\documentclass[prd,nofootinbib,onecolumn,10pt]{revtex4}

\pdfoutput=1

\usepackage{amssymb}
\usepackage{amsmath}
\usepackage{graphicx,subfigure}
\usepackage{longtable}
\usepackage{verbatim}
\usepackage{amsfonts}

\usepackage{graphicx,amsmath,amssymb,epsfig,verbatim,mathrsfs,array,layout,textcomp,latexsym,xspace}
\usepackage{multirow}

\allowdisplaybreaks[1]

\renewcommand{\{}{\left\lbrace}
\renewcommand{\}}{\right\rbrace}
\renewcommand{\[}{\left\lbrack}
\renewcommand{\]}{\right\rbrack}

\newcommand{\nn}{\nonumber}

\newcommand{\gev}{\ensuremath{\mathrm{\,Ge\kern -0.1em V}}\xspace}
\newcommand{\gevcc}{\ensuremath{{\mathrm{\,Ge\kern -0.1em V\!/}c^2}}\xspace}

\begin{document}

\begin{flushright}
DO-TH 15/05 \\
QFET-2015-17
\end{flushright}

\title{$\bar B \to  \bar K \pi \ell \ell$ in and outside the $\bar K^*$ window}

\author{Diganta Das}
\author{Gudrun Hiller}
\affiliation{Institut f\"ur Physik, Technische Universit\"at Dortmund, D-44221 Dortmund, Germany}
\author{Martin Jung}
\affiliation{
TUM Institute for Advanced Study, Lichtenbergstr. 2a, D-85747 Garching, Germany.}
\affiliation{
Excellence Cluster Universe, Technische Universit\"at M\"unchen, Boltzmannstr. 2, D-85748 Garching,
Germany.}

\begin{abstract}
We study the impact of $\bar{B}\to\bar{K}\pi\ell\ell$ decays on $\bar{B}\to\bar{K}^* (\to \bar K \pi) \ell\ell$, taking into
account the $\bar K^*$ at finite width. Interference effects can generically be sizable, up to  ${\cal{O}}(10 \%)$, but are reduced in several 
ratios of observables of the angular distribution. Information on strong phases  is central to control interference effects, which cannot be
removed by sideband subtractions. We point out ways to probe the strong phases; only a single one is required to describe leading effects in the region of low
hadronic recoil. We find that recent LHCb data on the $\bar{B^0}\to\bar{K}^{*0} \mu \mu$  angular observables  at low recoil are in
good agreement with the standard model.
\end{abstract}

\maketitle

\section{Introduction}

The semileptonic flavor-changing neutral-current decay $\Bar{B}\to \bar{K}^* \ell\ell$\footnote{CP-averaging is tacitly implied throughout this work.}
is one of the key modes at current and future high luminosity flavor facilities.
If  analyzed through the quasi-4-body $\Bar{B}\to \bar{K} \pi \ell\ell$ decays,
several  observables can be obtained that allow to precisely probe flavor physics in the standard model (SM) and  beyond.

With increasing experimental precision further backgrounds become of importance.
In fact the shape of the $\bar{B}\to\bar{K}^* (\to \bar{K} \pi) \ell\ell$ angular
distribution is distorted by non-resonant $\bar{B}\to\bar{K}\pi\ell\ell$ decays as well as resonances  
decaying to  $\bar K\pi $. 
Building on previous works~\cite{Das:2014sra} we study the impact of such backgrounds, taking into account the $\bar K^*$ at finite width.
Since the sole suppression of the non-resonant $\bar B \to \bar K \pi \ell \ell$ mode relative to $\bar B \to \bar K^* \ell \ell$  is due to phase
space, generically sizable interference effects of order $1/(4 \pi)$ are expected.  Non-resonant decays have dominant S- and P-wave components;  
the leading background stems therefore from P-P and S-P  interference. Only the latter can be separated from $\bar B \to
\bar K^* \ell \ell$ by 
its different angular structure.
The contamination from resonant $\bar K \pi$ contributions in an S-wave, as originating from $\bar{B}\to\bar\kappa(800) \ell\ell$ and 
$\bar{B}\to\bar{K}_0(1430)\ell\ell$ decays, has been previously considered in \cite{Becirevic:2012dp, Blake:2012mb, Matias:2012qz,Doring:2013wka}.

We work in the region of low hadronic recoil, which is advantageous due to the suppression of $1/m_b$ power corrections
\cite{Grinstein:2004vb,Bobeth:2010wg,Beylich:2011aq} and its direct accessibility to  form factor calculations based on lattice QCD
\cite{Horgan:2013hoa} and heavy hadron chiral perturbation theory (HH$\chi$PT), {\it e.g.,}  Refs.~\cite{Burdman:1992gh,Lee:1992ih}.

The relative strong phases  between the $\bar K^*$ and its backgrounds can in principle be probed experimentally using  interference.
An important feature of the $\Bar{B}\to\bar{K}\pi\ell\ell$ decay
is that it gives access to new combinations of short-distance coefficients that are not present in the $\bar K^*$ signal mode~\cite{Das:2014sra}. 
We discuss resulting opportunities for reducing the background and for probing new physics.

The paper is organized as follows:  
We give details on the finite width implementation  of the $\bar K^*$ in Sec.~\ref{sec:details}.  In Sec.~\ref{sec:interference} we estimate the
backgrounds to $\bar{B}\to \bar K^*  (\to \bar{K}\pi)\ell\ell$ for relevant observables. In Sec.~\ref{sec:pheno} we show how to probe strong phases
with ratios of angular observables and discuss SM tests and beyond the SM (BSM) searches with  SM-nulltests. In Sec.~\ref{sec:data} we compare SM
predictions to the latest preliminary LHCb findings for $\bar{B^0}\to\bar{K}^{*0} \mu \mu$ angular observables based on $3 \, {\rm fb}^{-1}$
\cite{LHCb:2015dla}. In Sec.~\ref{sec:conclusions} we conclude. In several appendices we give auxiliary information.

\section{Framework \label{sec:details}}

The $\bar B \to \bar K \pi \ell \ell$ decay amplitudes $H_{0,\parallel,\perp}^{L/R}$  factorize in the
operator product expansion (OPE) \cite{Buchalla:1998mt,Grinstein:2004vb} at leading order in $1/Q$, $Q=\{m_b, \sqrt{q^2}\}$, into universal 
short-distance coefficients $C_{\pm}^{L/R}$ and form factors (see Ref.~\cite{Das:2014sra} for details),
\begin{align} \label{eq:trans}
H_{0,\parallel}^{L/R} = C_{-}^{L/R}(q^2)  \cdot
\mathcal{F}_{0,\parallel} (q^2,p^2,\cos \theta_K)\;, 
\quad H_{\perp}^{L/R} = C_+^{L/R}(q^2)  \cdot \mathcal{F}_\perp (q^2,p^2,\cos \theta_K)\;.  
\end{align}
The contributions from various hadronic final states are contained in the generalized transversity form factors 
\begin{align} \label{eq:fullformfactor}
{\cal F}_0 & \equiv {\cal F}_0\left(q^2, p^2, \cos \theta_K\right)=F_0\left(q^2, p^2, \cos \theta_K\right)+\sum_R P^{0}_{J_R} (\cos\theta_K) \cdot F_{0 J_R}\left(q^2, p^2\right)  \, ,   \\
{\cal F}_i & \equiv  {\cal F}_i\left(q^2, p^2, \cos \theta_K\right)=F_i\left(q^2, p^2, \cos \theta_K\right)+\sum_R \frac{P^{1}_{J_R} (\cos\theta_K) }{\sin \theta_K}  \cdot F_{iJ_R}\left(q^2, p^2\right)  \, , \quad i= \parallel, \perp \, . \nonumber
\end{align}
Here, the first terms $F_{0,\|,\perp}$ on the right-hand 
side correspond to the non-resonant $\bar{B}\to\bar{K}\pi$
transversity form factors ({\it cf.} App.~\ref{app:B2Kpi}), whereas the
second terms containing the form factors $F_{(0,\|,\perp)R}$ belong to resonant states $R$ with spin $J_R$ decaying to $\bar{K}\pi$.
The $P^m_\ell$ are the associated Legendre polynomials.
In this work $q^2$ and $p^2$ denote the invariant mass squared of the dilepton- and $\bar K \pi$-system, respectively.
$\theta_K$ is the angle between the kaon and the $\bar B$ in the $\bar K \pi$ center-of-mass system.

In our numerical estimate we employ the $\bar{B}\to\bar{K}\pi$ form factors from HH$\chi$PT, given in Eq.~(\ref{eq:ffinput}).
Sizable uncertainties are present in these leading order results, already parametrically
from the HH$\chi$PT coupling constant, $g$ ($g^2$) in $F_{0,\parallel}$ ($F_\perp$) of $13\%$ ($26\%$), in addition to higher-order corrections. The expansion is expected to work better towards zero recoil. Alternative determinations for  $\bar{B}\to\bar{K}\pi$ form factors are desirable.

For $\bar{B}\to\bar{K}^*\ell\ell$ decays, it is useful to match the corresponding
P-wave contributions  $F_{(0,\|,\perp)P}=F_{(0,\|,\perp)P}(q^2, p^2)$ onto the common $\bar{B}\to\bar{K}^*$ transversity form factors $f_{(0,\|,\perp)}(q^2)$,  see App.~\ref{app:B2Kst} for details,
\begin{align}\label{eq:KstFull}
F_{0P} =&  - 3 f_0(q^2)  \,P^{BW}_{K^*}(p^2)\, e^{i \delta_{K^*}}\, , \quad F_{\parallel P} =  - 3  \sqrt{ \frac{ 1}{ 2}}  \, f_\parallel(q^2) \, P^{BW}_{K^*}(p^2)\,e^{i \delta_{K^*}}\, , \quad 
F_{\perp P} =  3 \sqrt{ \frac{ 1}{ 2}} \, f_\perp(q^2)  \, P^{BW}_{K^*}(p^2)\,e^{i \delta_{K^*}}\, .
\end{align}
Here we included factors $e^{i \delta_{K^*}}$ to account for a
relative strong phase $\delta_{K^*}$ between the $\bar K^*$ and the  non-resonant contributions. In general it can assume values between $-\pi$
and $+\pi$. 
Eq.~(\ref{eq:trans}) implies that there is only one universal strong phase for all transversity amplitudes  between the $\bar K^*$ and its
non-resonant background.
The strong phase should vary with $p^2$, and this could be taken into account given knowledge of the functional form.
By keeping  in this work $\delta_{K^*}$ constant in each $p^2$-integration window it becomes an effective $p^2$-bin specific phase.

As explicitly shown in  Ref.~\cite{Das:2014sra}, the 
non-resonant background dominates over the one from the scalar mesons $\kappa(800)$ and $\bar{K}_0(1430)$ in the $\bar B \to \bar K^*
\ell \ell$ dilepton mass distribution.
Since in addition the fraction of states with longitudinal polarization $F_L$ from purely non-resonant decays at low recoil is large, $\sim
0.5$~\cite{Das:2014sra}, the contributions from S-wave resonances are subdominant also in this observable. In the angular coefficients $I_{3
..9}$, S-wave contributions can be isolated with an angular analysis or are even absent. We therefore consider  the non-resonant $\bar K \pi$ decays
as an effective model for the background. We note, however, that this can be refined as resonance contributions can be modeled in a
straightforward manner, at the price of additional phases and parametric uncertainties.
In the remainder of this work we denote by $\bar B \to  \bar K \pi \ell \ell$ decays originating from the $\bar K^*$ as well as non-resonant modes,
including interference, unless  stated otherwise.

We incorporate the finite width of the  $\bar K^*$ by the usual Breit-Wigner (BW) lineshape,
\begin{align}
P_{K_J}^{BW}(p^2) =\sqrt{\frac{m_{K_J} \Gamma_{K_J}}{\pi} } \frac{1}{p^2-m_{K_J}^2+ i m_{K_J}  \Gamma_{K_J}(p^2)} \, ,\quad{\rm with}\quad\int d
p^2 |P_{K_J}^{BW}(p^2)|^2 = 1 \, ,
\end{align}
$m_{K_J}$ and $\Gamma_{K_J}$ being the mass and mean width of the resonance $K_J$ with spin $J$, respectively.
We further take into account the running width of the $\bar K^*$:
\begin{align} \label{eq:runningwidth}
 \Gamma_{K^*}(p^2)= \Gamma_{K^*}^0  \left( \frac{ p^*} {p^*_0}   \right)^3  \frac{m_{K^*}}{\sqrt{p^2}}\frac{1+(r_{BW} \,
 p^*_0)^2}{1+(r_{BW} \, p^* )^2}  \, ,\quad\rm{where}\\
 p^* =  \frac{\sqrt{\lambda_p}}{2 \sqrt{p^2}}\, , \quad p^*_0= p^*|_{p^2=m_{K^*}^2} \, , \quad \lambda_p=\lambda(p^2, m_K^2,m_\pi^2) \, ,
\end{align}
with the Blatt-Weisskopf parameter $r_{BW}$ (see Table \ref{tab:input} for numerical input) and the common phase space function
$\lambda(a,b,c)=a^2+b^2+c^2-2(ab+ac+bc)$.
Other $\bar K^*$-lineshapes may also be studied, however, in view of the current experimental precision and the form factor uncertainties we refrain
in this work from doing so. We remark that experimental information on the lineshape in the kinematical situation relevant here could be obtained from
angular studies in $\bar B_s \to \bar K^* ( \to \bar K \pi)  \ell \nu$ decays.

The invariant-mass cuts suitable for $\bar B \to \bar K^* \ell \ell$ experimental studies are taken from Ref.~\cite{Das:2014sra}, which we follow
closely: 
\begin{align} \nonumber
0.64 \, \mbox{GeV}^2 < p^2 < 1 \, \mbox{GeV}^2: & ~~~~P ~~(\bar K^* \mbox{signal  window) cut}  \, ,\\
p^2_{\rm min}=0.40  \, \mbox{GeV}^2 < p^2 < 1.44 \, \mbox{GeV}^2: & ~S+P  ~(\bar K^*\mbox{total window) cut} \, ,
\label{eq:cuts}
\end{align}
where $p^2_{\rm min}=(m_K+m_\pi)^2$.

Upon evaluation of the $\bar K^*$ at finite width  kinematics is affected; notably there will be events above the zero-width endpoint 
$q^2>(m_B-m_{K^*})^2$, where $\lambda_{K^*} \equiv \lambda(m_B^2, q^2, m_{K^*}^2)<0$. To take this fully into account would require, besides
enlarging the phase space, taking into account hadronic form factors computed at finite width as well, which are not available
presently, see, however, Ref.~\cite{Briceno:2014uqa}. For concreteness, we pursue the following phenomenological avenue: we use $\lambda(m_B^2, q^2,
p^2)$ instead of $\lambda_{K*}$ in the overall phase space factor (\ref{eq:NKstar}) of the $\bar K^*$-contribution, keep $\lambda_{K*}$ elsewhere, in
particular in $\bar B \to \bar K^*$ form factors (\ref{eq:Kstff}), and have the plots end at  $\lambda_{K*}=0$ above which the rate dies out anyway.
The effects from different treatments around the endpoint are negligible in view of other uncertainties.

\section{$\bar{B}\to \bar{K}\pi \ell\ell$ distributions  \label{sec:interference}}

In this section we study the impact of non-resonant $\bar{B}\to \bar{K}\pi \ell\ell$ decays  on the $\bar{B}\to \bar K^*  (\to \bar{K}\pi)\ell\ell$
analysis. The background induces in general a shift and a phase-related uncertainty in the observables. 
 We work out the interference effects on the dilepton mass distributions (Sec.~\ref{sec:dileptonmass}), on the fraction of  longitudinally polarized $\bar K^*$, $F_L$, 
(Sec.~\ref{sec:FL})  and on the angular observables (Sec.~\ref{sec:ang}).
Auxiliary information on  the full angular distribution of $\bar B \to \bar K \pi \ell \ell$ decays  has been collected in App.~\ref{sec:primer}, and is based on
Ref.~\cite{Das:2014sra} to which we refer for further details.

\subsection{Dilepton spectrum \label{sec:dileptonmass}}

\begin{figure}[ht]
\centering{
\includegraphics[width=0.3\textwidth]{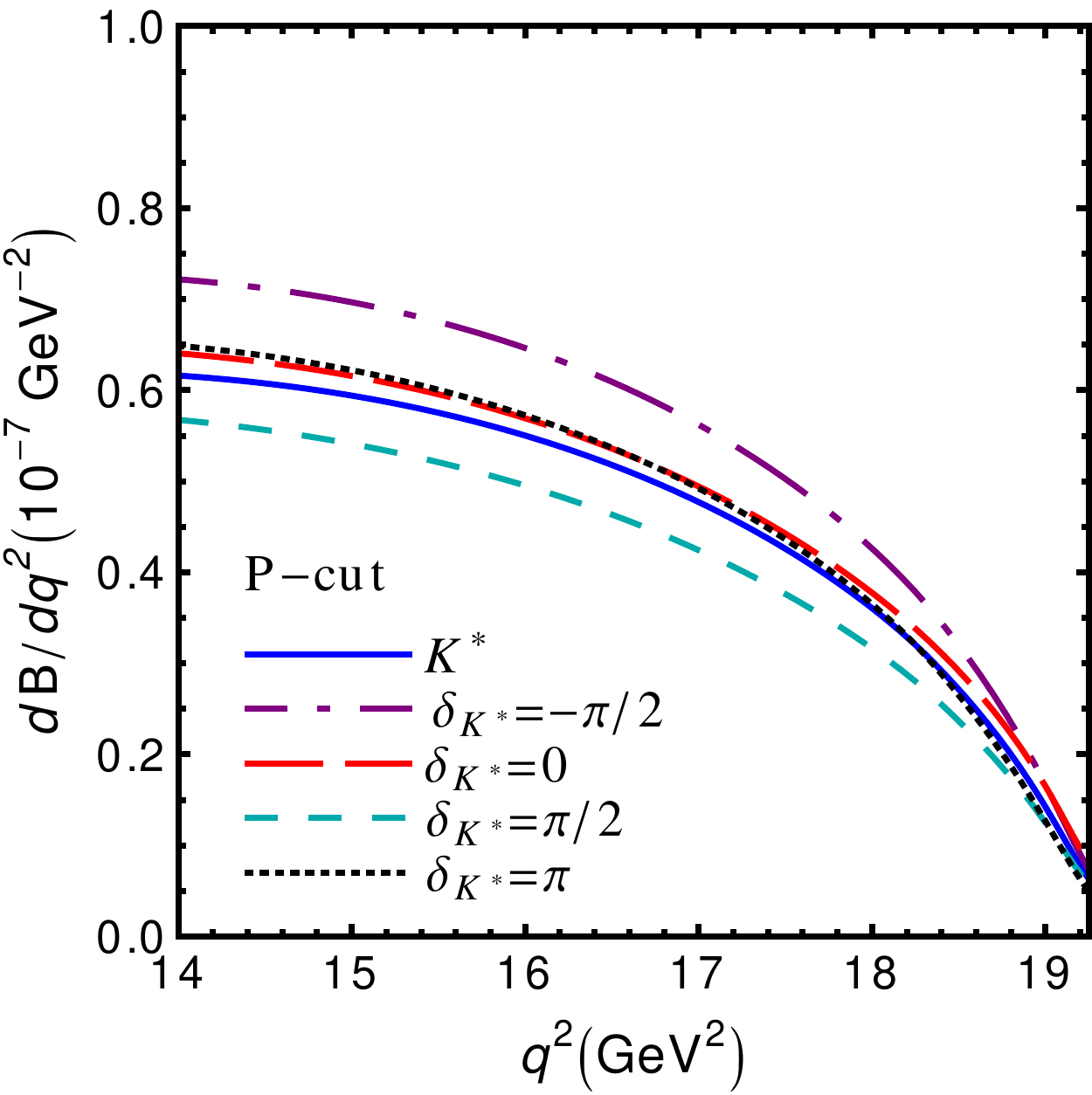}\qquad
\includegraphics[width=0.3\textwidth]{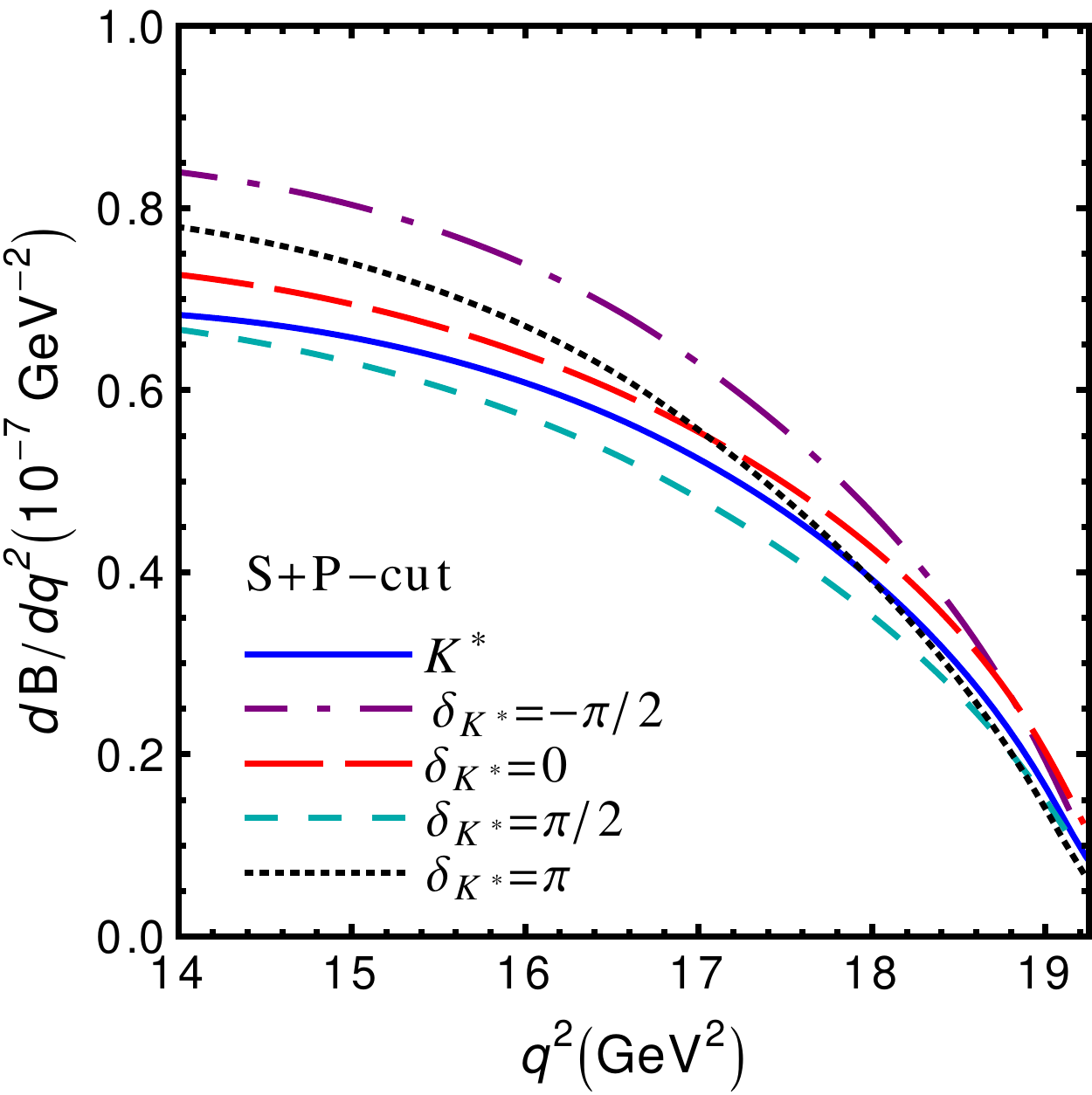}\qquad
\includegraphics[width=0.3\textwidth]{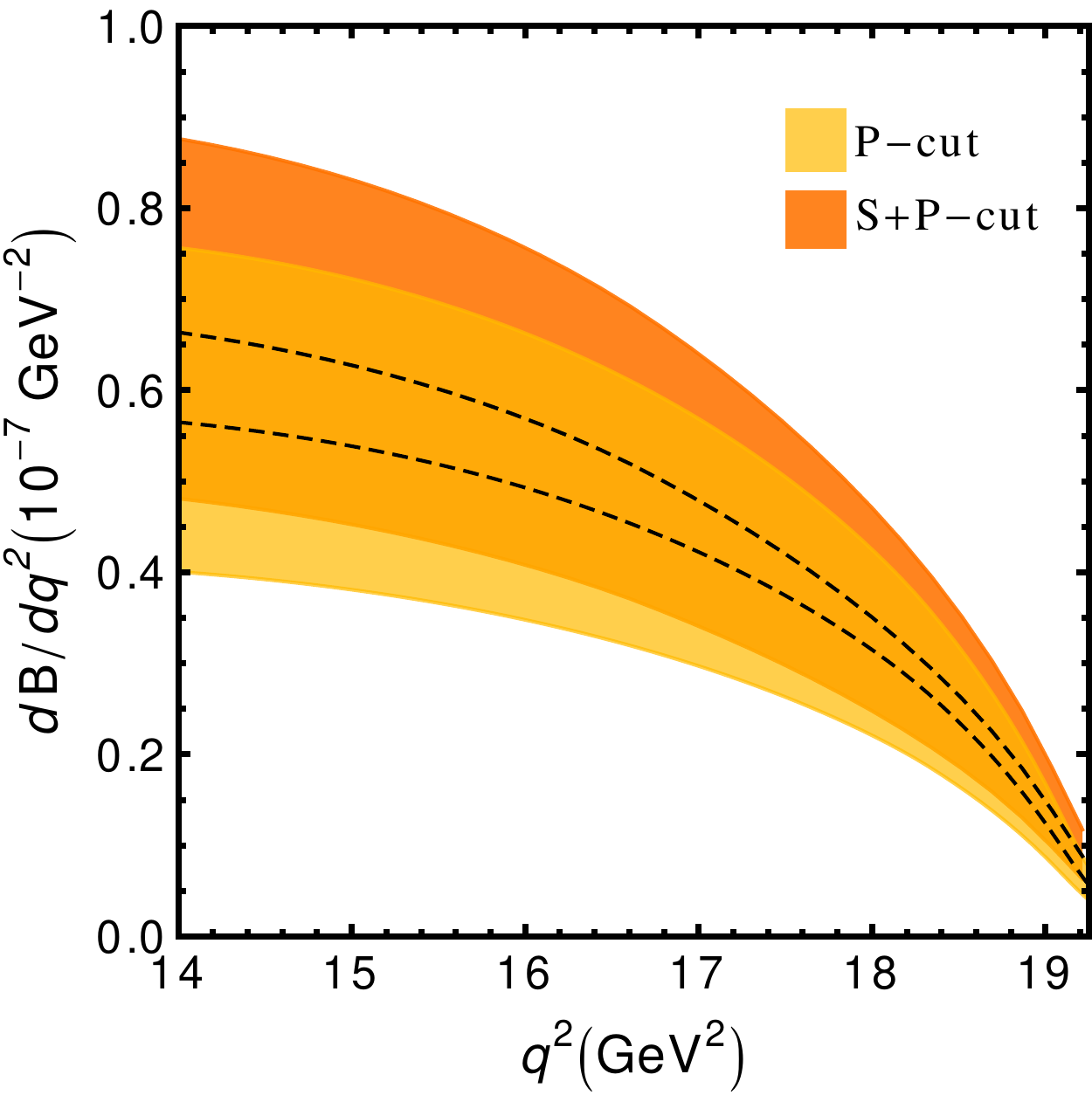}
}
\caption{\small The SM differential branching fraction $d{\cal{B}}(\bar B \to \bar K \pi \ell \ell)/dq^2$ in the P-cut 'signal' window (left)
and the S+P total window (mid) for central values of input and for different values of the strong phase. The solid blue curve corresponds to the
resonant $K^*$ contribution. All curves are with running width in the Breit-Wigner propagator for the $K^*$, \emph{c.f.} Eq.~\eqref{eq:runningwidth}.
In the plot to the right the bands correspond to the uncertainties coming from form factors and parametric inputs for fixed strong phase
$\delta_{K^*}=\pi/2$.}
\label{fig:br}
\end{figure}

\begin{figure}[ht]
\centering{
\includegraphics[width=0.3\textwidth]{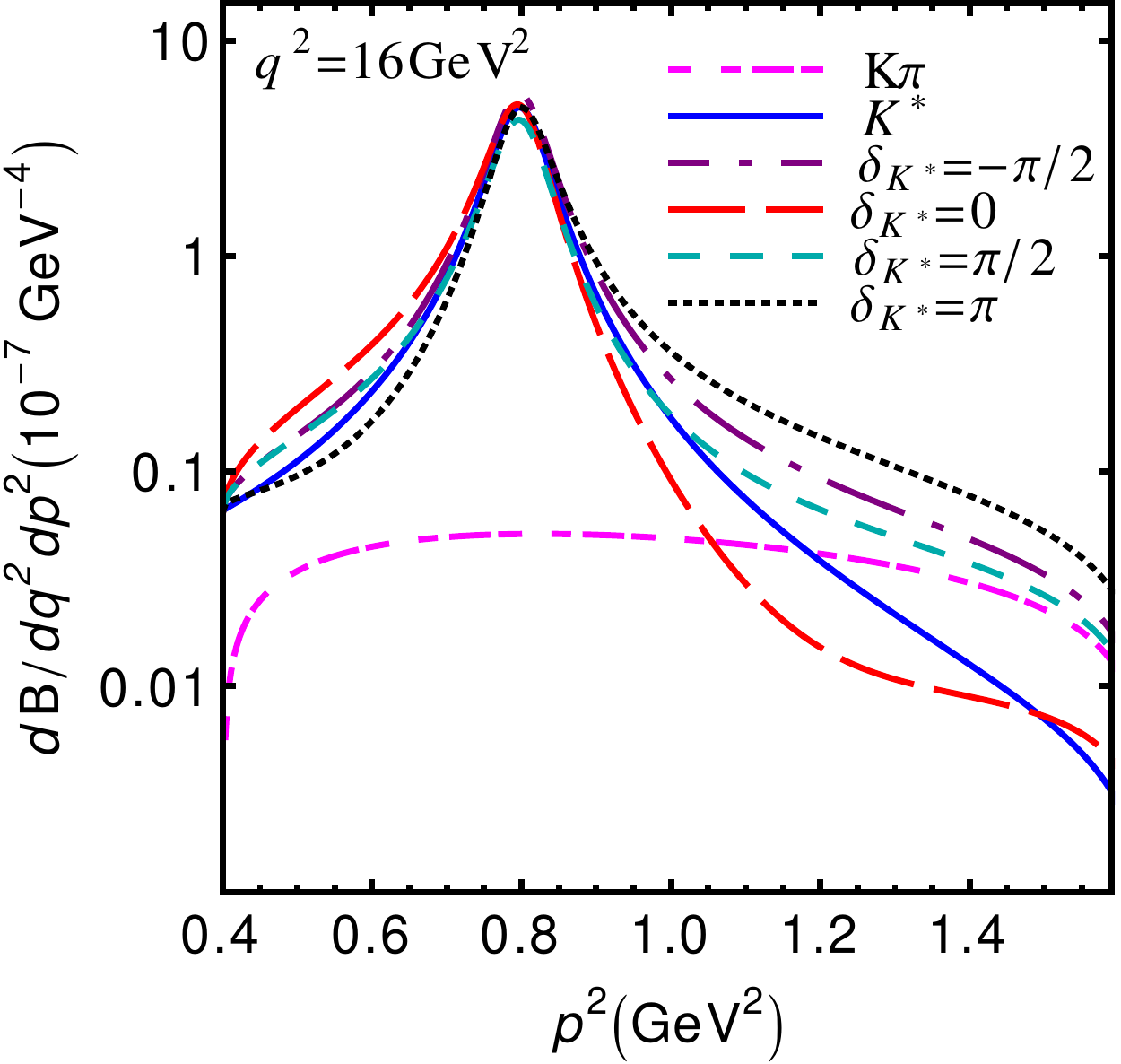}\qquad\qquad
\includegraphics[width=0.3\textwidth]{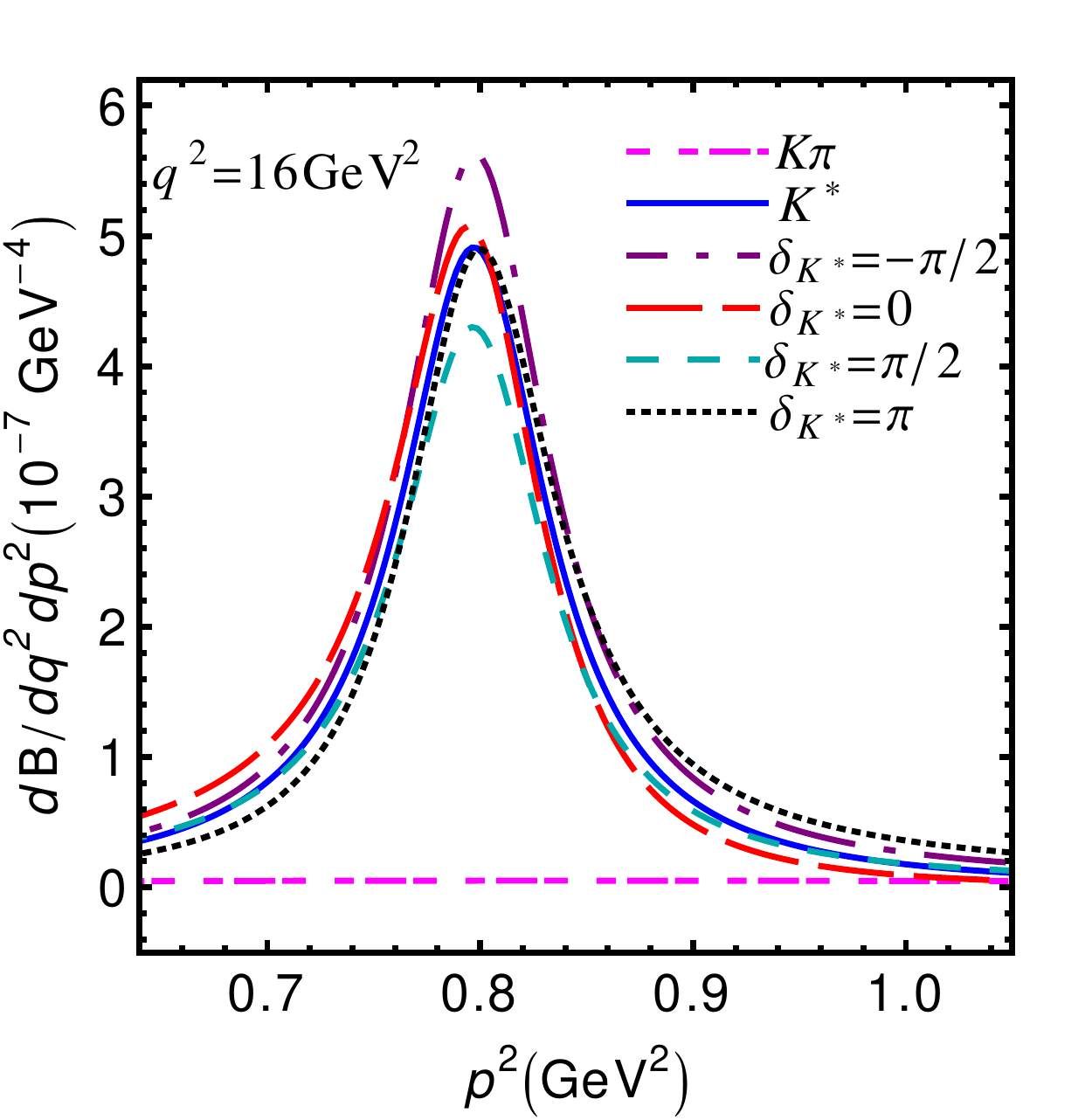}

}
\caption{\small $d^2{\cal{B}}(\bar B \to \bar K \pi \ell \ell)/dq^2dp^2$ at $q^2=16 \, \mbox{GeV}^2$ in the SM for central values of input and
different values of the strong phase. The solid blue curve corresponds to the resonant $K^*$, the magenta curve to the purely non-resonant
contribution. The plot to the right is a zoom of the left one around the $\bar K^*$-window.
}
\label{fig:brp2}
\end{figure}

In Fig.~\ref{fig:br} we show the influence of the interfering non-resonant contribution on the SM differential branching fraction
$d{\cal{B}}(\bar B \to \bar K \pi \ell \ell)/dq^2$ in the P-wave 'signal' window (left) and the S+P total window (mid); in the panel on the
right the uncertainties from form factors and parametric inputs are illustrated. 
In Fig.~\ref{fig:brp2} we show similarly $d^2{\cal{B}}(\bar B \to \bar K \pi \ell \ell)/dq^2dp^2$ for fixed $q^2=16 \,\mbox{GeV}^2$, with a zoom into
the $K^*$ signal region on the right. For $p^2 \gtrsim 1 \, \mbox{GeV}^2$ the non-resonant branching ratio becomes comparable to the $\bar K^*$ one.
We note that the numerical difference between using constant or running width (\emph{c.f.}~Eq.~\eqref{eq:runningwidth}) amounts to less than a few
percent.

\begin{figure}[ht]
\centering{
\includegraphics[width=0.3\textwidth]{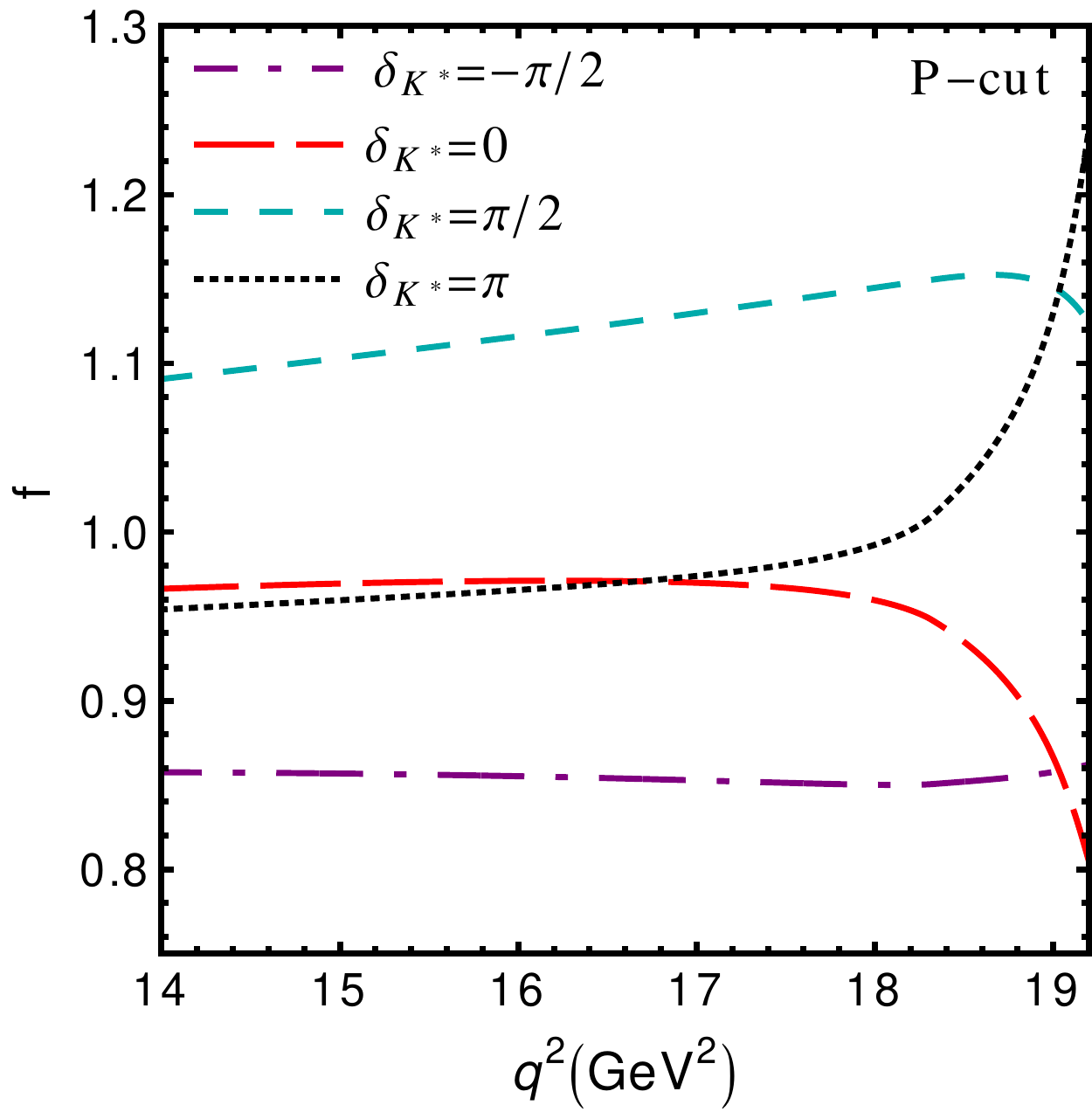}\qquad\qquad
\includegraphics[width=0.3\textwidth]{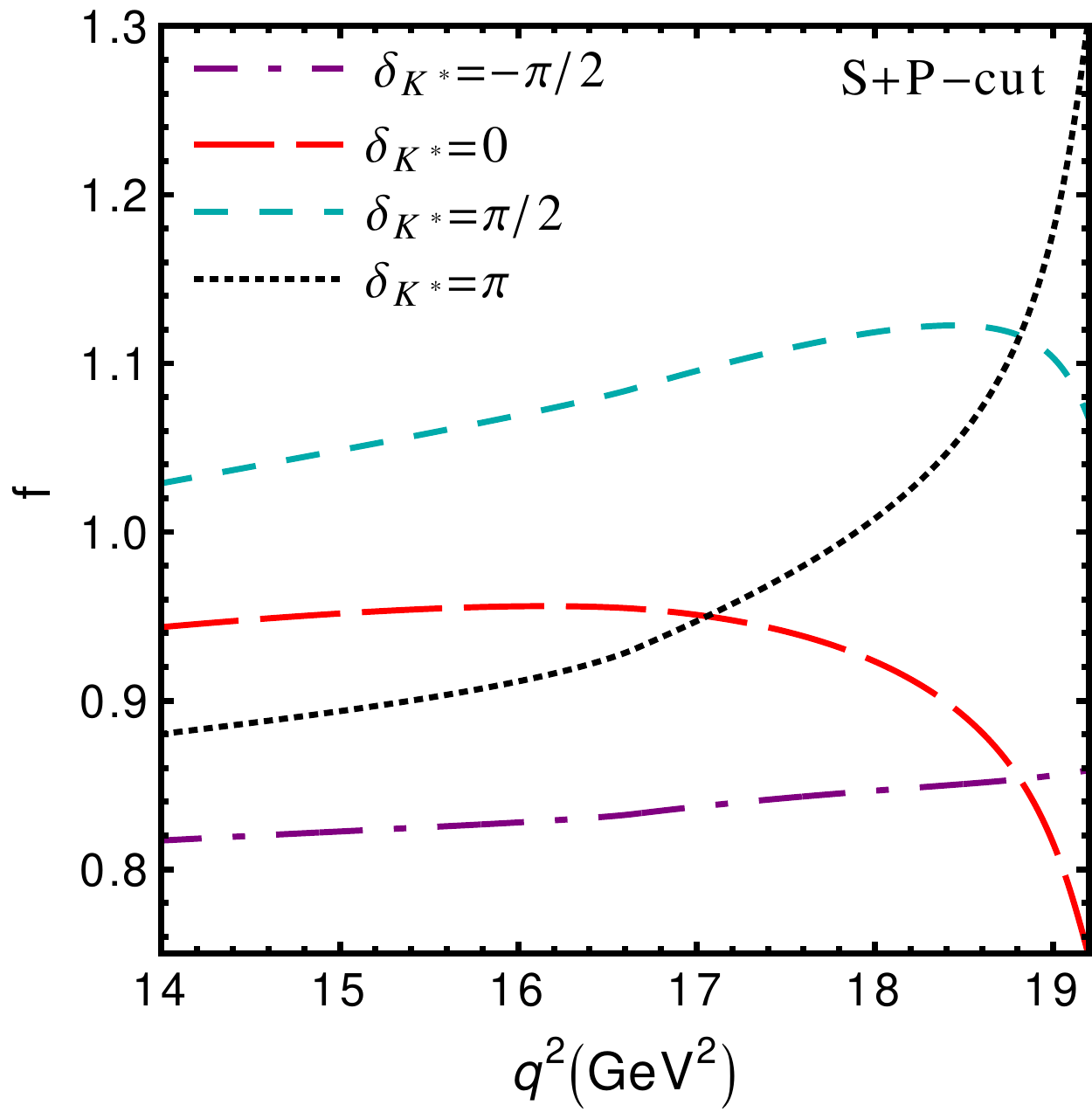}
}
\caption{\small 
The correction fraction defined in Eq.~(\ref{eq:ratio}) in the P signal cut window (left) and S+P cut window (right) in the SM for different values
of the relative strong phase.
\label{fig:breps}}
\end{figure}

The spread induced by varying the  relative strong phase is considerable. We quantify this
in Fig.~\ref{fig:breps}, where we show the fraction of resonant $(\bar K^*)$ to all events,
\begin{align} \label{eq:ratio}
f=\frac{ \int dp^2 d^2{\cal{B}}(\bar B \to \bar K^* (\to \bar K \pi) \ell \ell)/dq^2 dp^2}{ \int d p^2d^2{\cal{B}}(\bar B \to \bar K \pi \ell \ell)/dq^2 dp^2} \, ,
\end{align}
that is,  the denominator includes resonant $\bar K^*$ and non-resonant decays and their interference. The correction amounts to up to
$15\%$, depending on $\delta_{K^*}$, and can be even larger in the very $\bar B\to \bar K^* \ell \ell$ endpoint region. Since the sole suppression of 
the non-resonant $\bar B \to \bar K \pi \ell \ell$ mode relative to $\bar B \to \bar K^* \ell \ell$ decays  is due to phase space, effects of order
$1/(4 \pi)$ are actually expected.
This is presently within the uncertainties of the $\bar B \to \bar K^* \ell \ell$ branching ratio which amount to about 30 
percent from form factors and parametric input, see plot to the right in Fig \ref{fig:br}. Nevertheless, it stresses the importance of (angular)
observables with less sensitivity to hadronic physics. We discuss examples for the latter in the next sections.
Notably, even very rough bounds on the strong phase would reduce the uncertainties related to interference. We study this further in
Section \ref{sec:strong}.

\subsection{Longitudinal polarization fraction \label{sec:FL}}

\begin{figure}[ht]
\centering{
\includegraphics[width=0.3\textwidth]{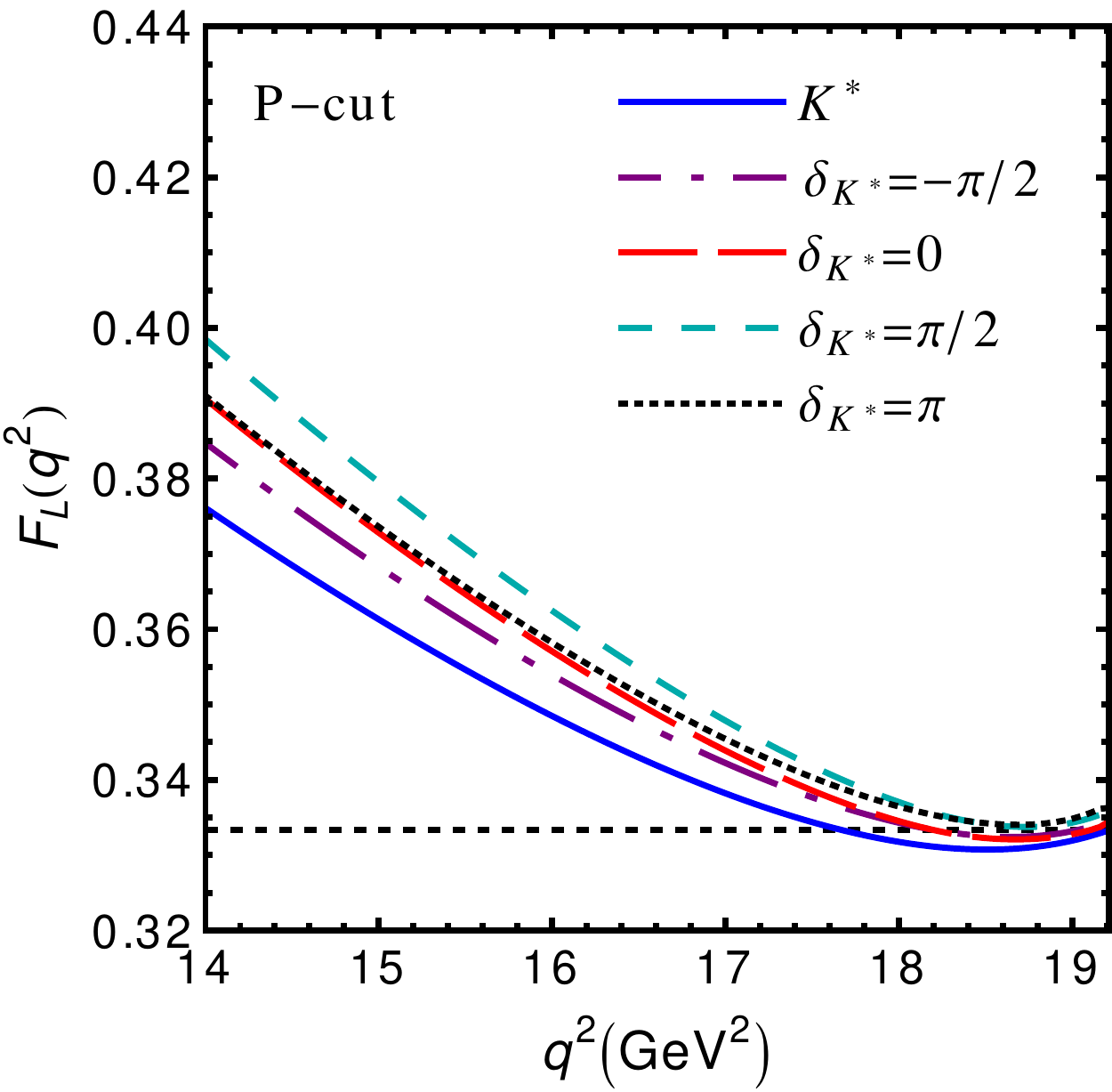}\qquad
\includegraphics[width=0.3\textwidth]{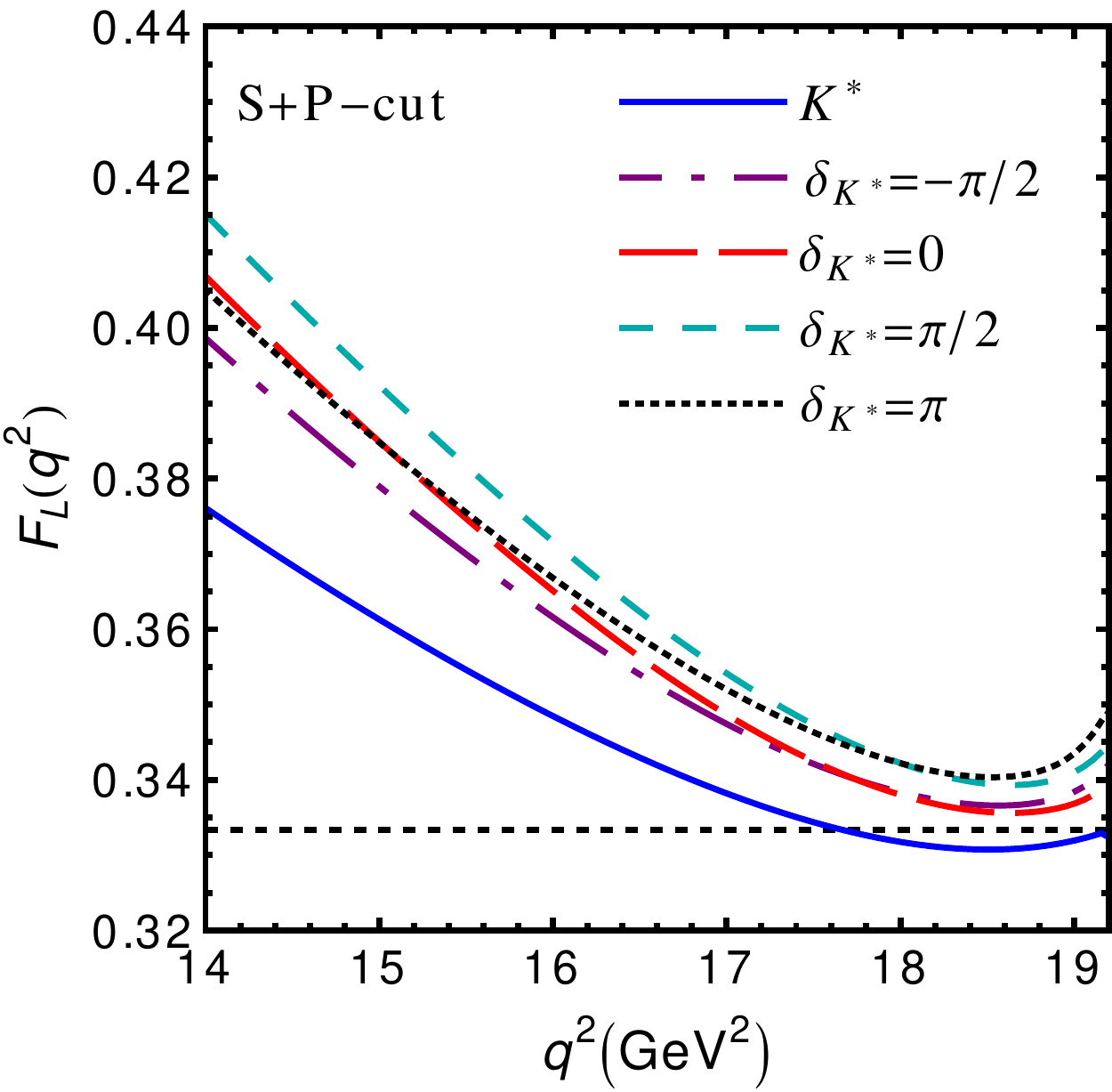}\qquad
\includegraphics[width=0.3\textwidth]{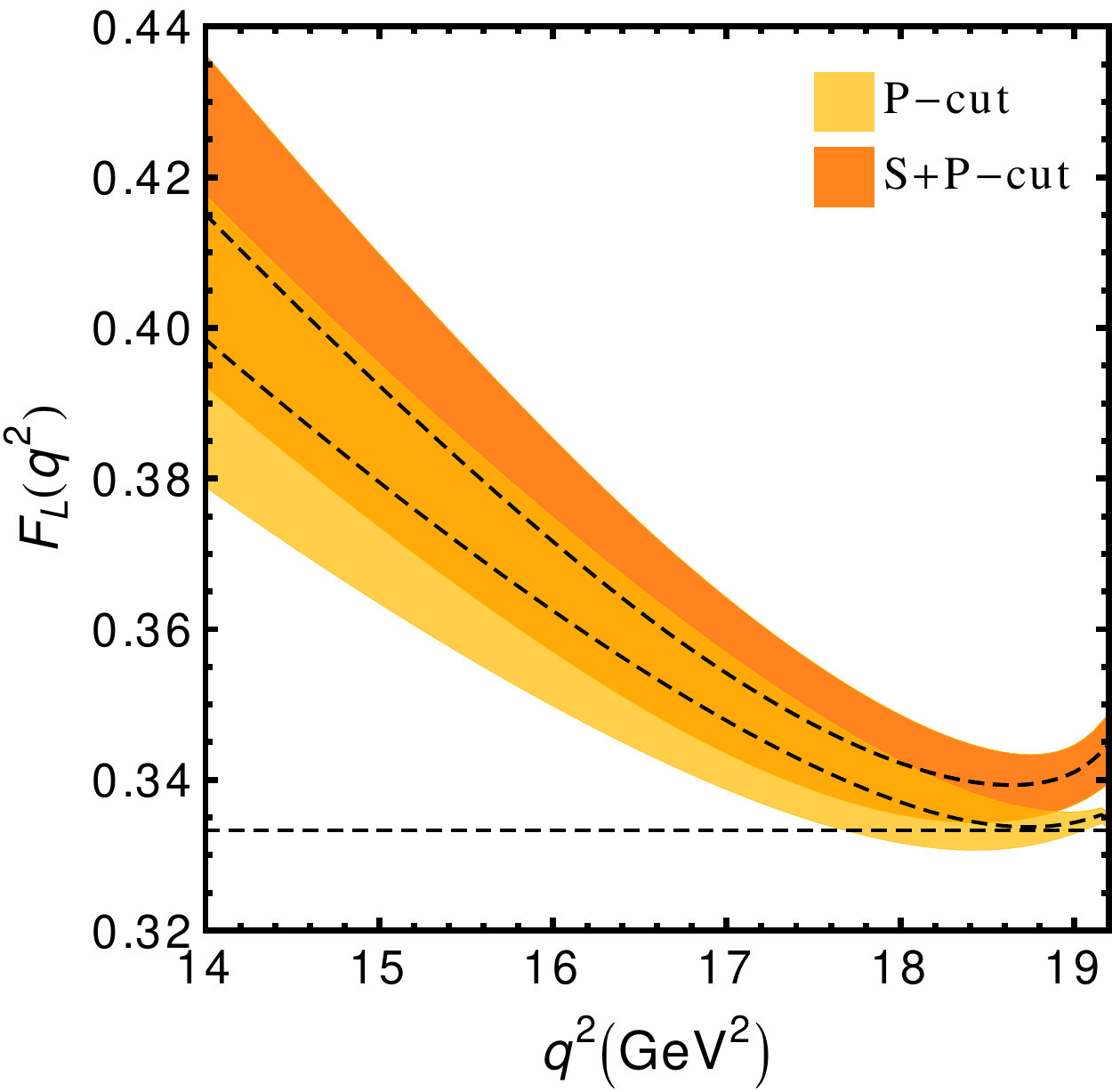}
}
\caption{\small The longitudinal polarization fraction $F_L$ in the P-cut 'signal' window (left)
and S+P total window (mid) in the SM
for central values and
for different values of the strong phase.
The solid blue curve corresponds to the resonant $K^*$ contribution, 
the horizontal dotted lines to the endpoint prediction $F_L=1/3$. In the plot to the right the bands correspond to the uncertainties coming from form
factors and parametric inputs for fixed strong phase $\delta_{K^*}=\pi/2$.
}
\label{fig:FL}
\end{figure}

 The longitudinal polarization fraction, given by
\begin{align}
F_L =\frac{d \Gamma(\bar B \to \bar K \pi \ell \ell)/dq^2 \big |_{{\cal{F}}_{\perp, \parallel}=0}}{d \Gamma (\bar B \to \bar K \pi \ell \ell)/dq^2}\,,
\end{align}
is shown in Fig.~\ref{fig:FL}. The $\bar K \pi$ background shifts $F_L$ to larger values by about 6\% in the P-cut and
$11\%$ in the S+P-cut window, while the uncertainty from the strong phase is only up to $2\%$ in the P-cut and $3\%$ in the S+P-cut window. The shift
remains strictly positive even when including the hadronic uncertainties and its size is larger than the present experimental uncertainty for this
observable, see Sec.~\ref{sec:data}. The inclusion of this effect is therefore important when interpreting the available data.

\subsection{Angular observables  \label{sec:ang}}

The angular coefficients $I_i$ are observables of  the $\bar B \to \bar K \pi \ell \ell$ angular distribution, see Appendix \ref{sec:primer} for a brief overview.
It is useful to define integrated angular coefficients $\hat I_i=\hat I_i(q^2)$ as follows:
\begin{align} \nonumber
\hat I_i & = \int d p^2  \int_{-1}^{+1} d \cos \theta_K I_i(q^2,p^2, \cos \theta_K) \, , \quad  i=3,6,9 \, ,\\
\hat I_i  &= \int d p^2   \[ \int_{0}^{+1} - \int_{-1}^0 \] d \cos \theta_K I_i(q^2,p^2, \cos \theta_K) \, , \quad  i=4,5,7,8  \, .
\end{align}

 The relations between the $ \hat I_i$ and the coefficients of the pure P-wave analysis, $J_i, J_{ic}$ 
({\it cf.}  Ref.~\cite{Das:2014sra}) read:
  \begin{align} \nonumber
\hat I_i&= \frac{4}{3} J_i + \mbox{pure D-waves and higher} \, , \quad  i=3,6,9 \, ,\\
 \hat I_i&= \frac{2}{3} J_{ic} + \mbox{D-waves and higher} \, , \quad  i=4,5,7,8  \, .
 \end{align}
 The first equation holds up to pure D-wave contributions and higher ones. The second equation receives in addition corrections from S-D wave interference.
 We recall that the leading contributions of the background are in S- and P-wave. The dominant effect in the $\bar K^*$-signal window is hence P-P interference.
 For the $J_i, J_{ic}$ we follow here the conventions spelled out in  \cite{Das:2014sra}. The relation to the commonly used ones  \cite{Bobeth:2008ij,Bobeth:2010wg} (BHP) read
 $J_i=3/4 J_i^{BHP}$ for $i=3,6,9$ and $J_{ic}=3/2 J_{i}^{BHP}$ for $i=4,5,7,8$.
 The $J_i$ are building blocks for further observables, often designed to have specific features such as  reduced hadronic uncertainties.

To discuss the  shift and induced uncertainties related to the interfering backgrounds
we define correction fractions 
\begin{align} \label{eq:eps}
\epsilon_i =\frac{ \hat  I_i(\bar B \to \bar K^* (\to \bar K \pi) \ell \ell)}{ \hat I_i(\bar B \to \bar K \pi \ell \ell)} \,  ,   \quad i=3,4,5,6\, .
\end{align}
As in Eq.~(\ref{eq:ratio}), the denominators of the $\epsilon_i$ include both $\bar K^*$ and (interfering) non-resonant contributions, whereas in the
numerator only the $\bar K^*$ is included, \emph{c.f.} Eq.~(\ref{eq:KstFull}).  The $\epsilon_i$ depend on the cut in $p^2$; we employ an
identical one for both numerator and denominator.

\begin{figure}[ht]
\centering{
\includegraphics[width=0.30\textwidth]{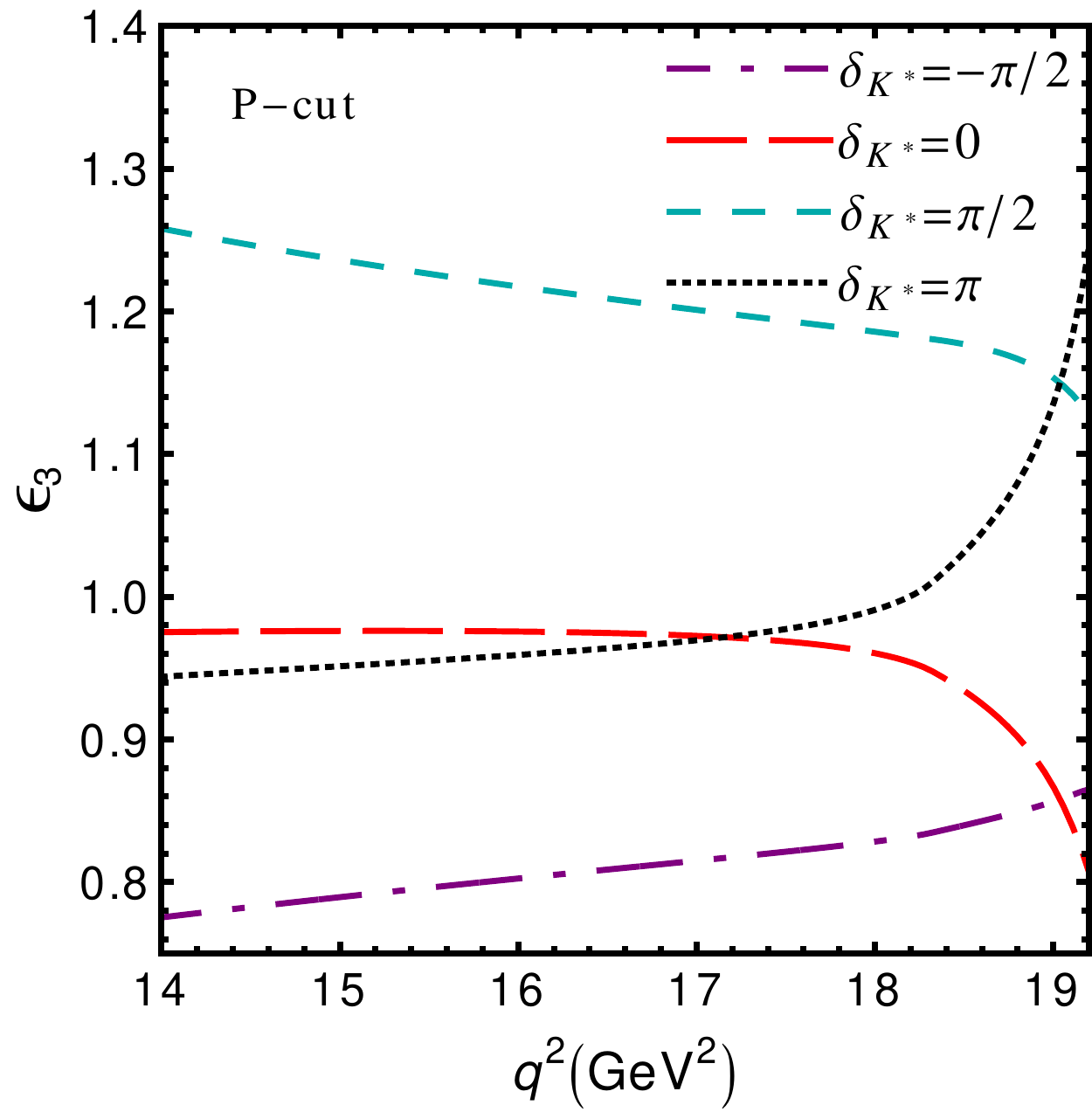}\qquad\qquad
\includegraphics[width=0.30\textwidth]{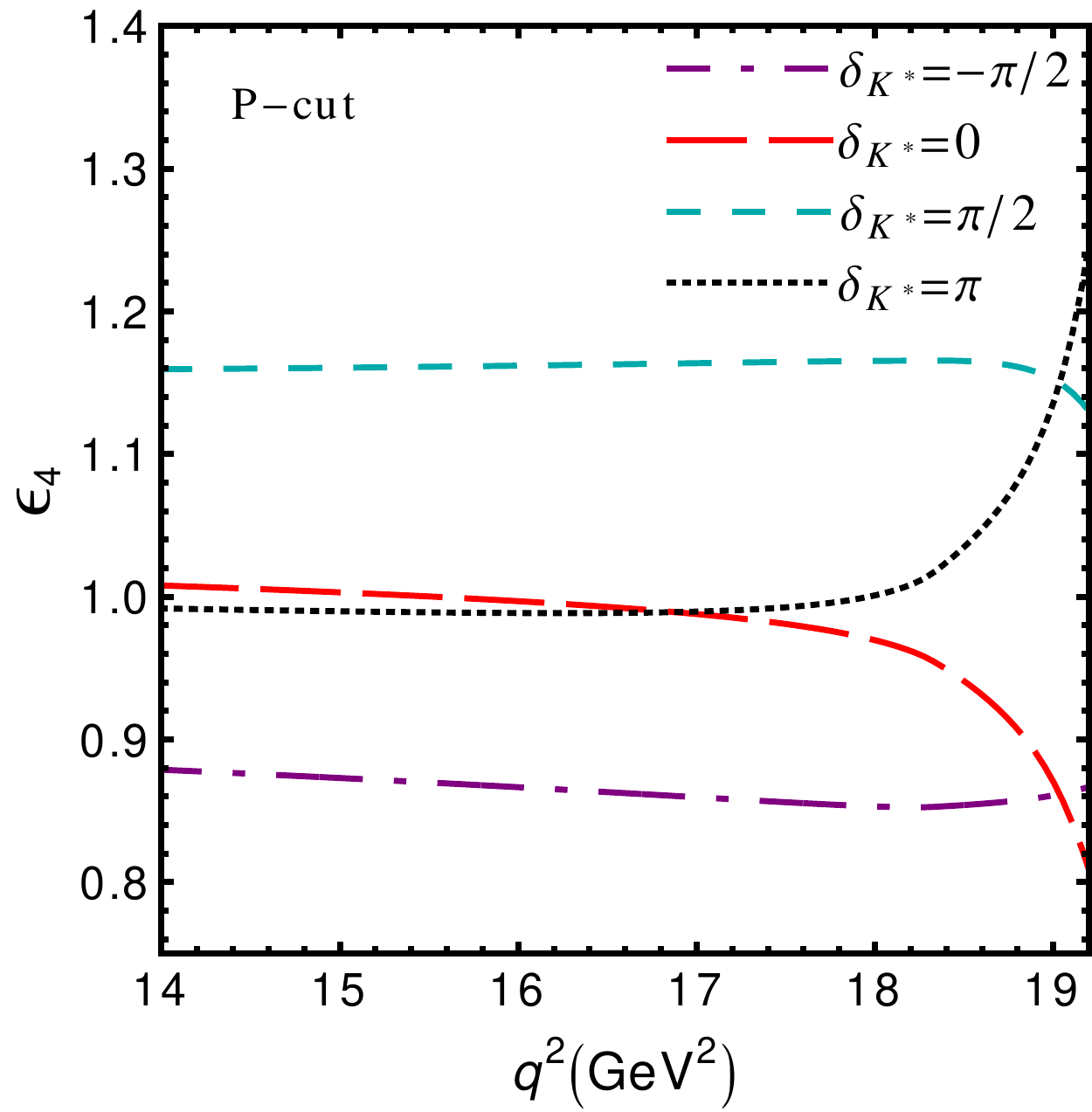} \\
\includegraphics[width=0.30\textwidth]{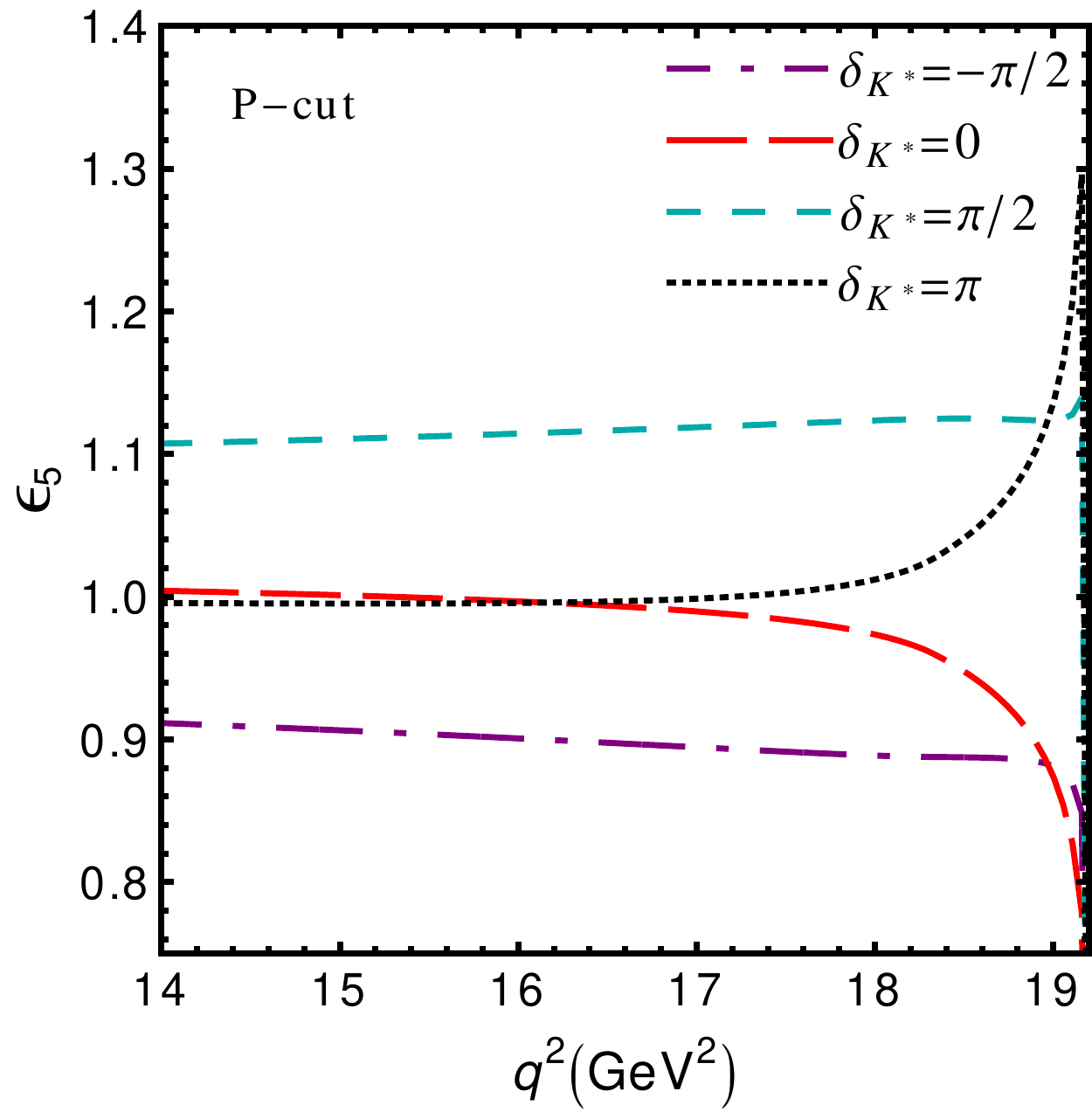}\qquad\qquad
\includegraphics[width=0.30\textwidth]{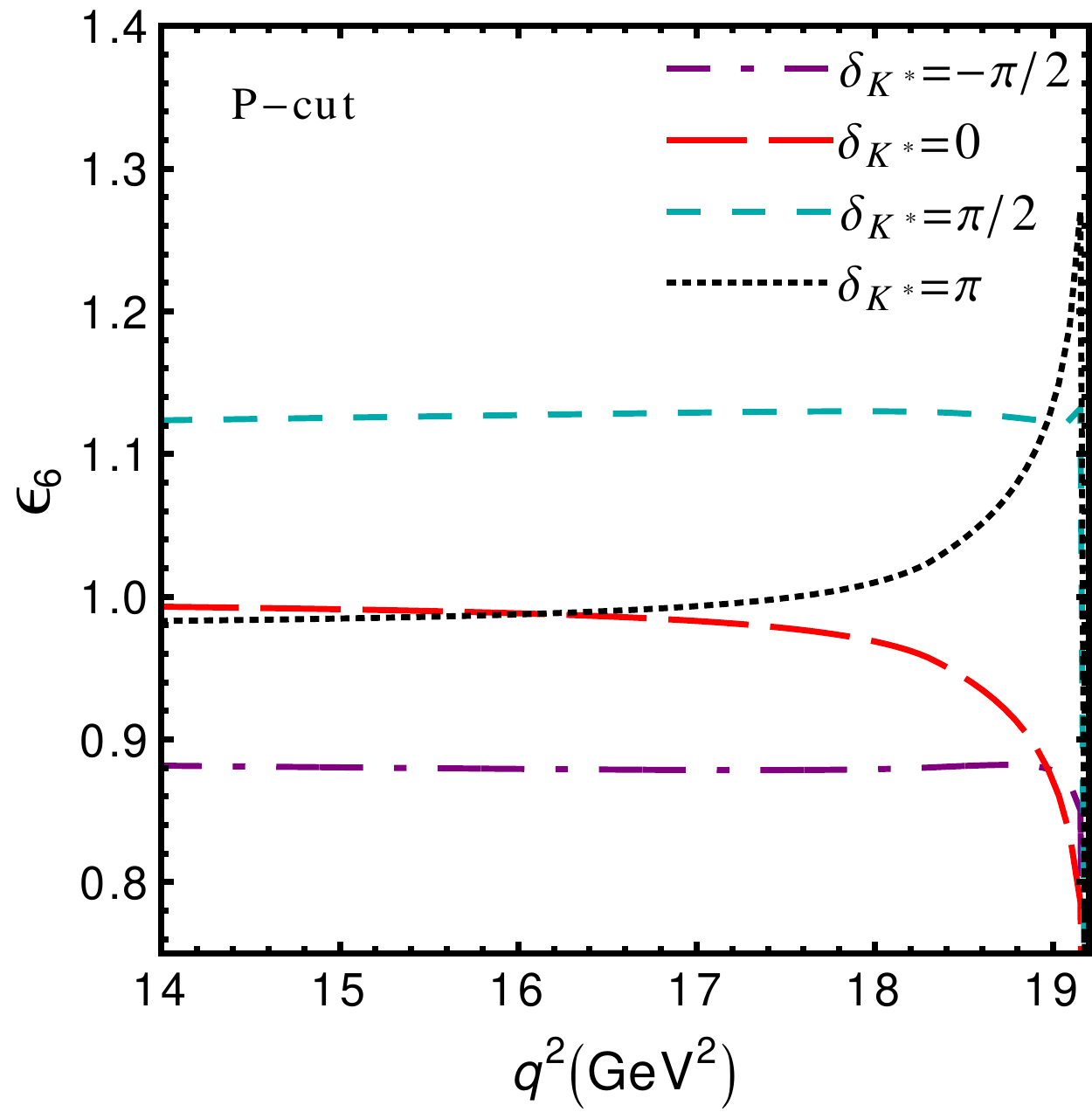}
}
\caption{\small The correction fractions  $\epsilon_{3,4,5,6}$, \emph{cf.} Eq.~(\ref{eq:eps}), in the P-signal-cut window in the SM basis for central
values and for different values of the strong phase.}
\label{fig:afbeps}
\end{figure}

\begin{figure}[ht]
\centering{
\includegraphics[width=0.3\textwidth]{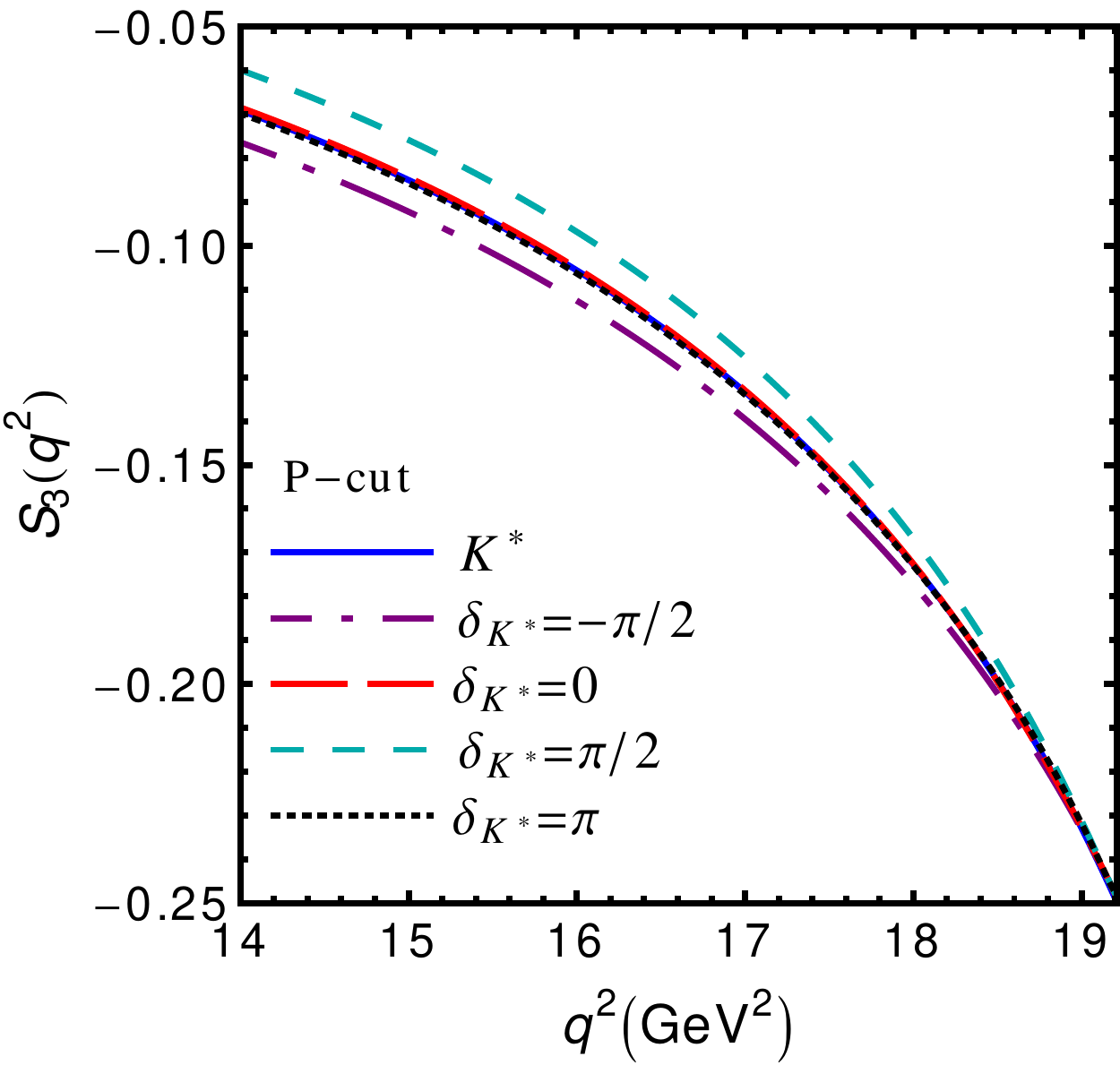}\qquad
\includegraphics[width=0.3\textwidth]{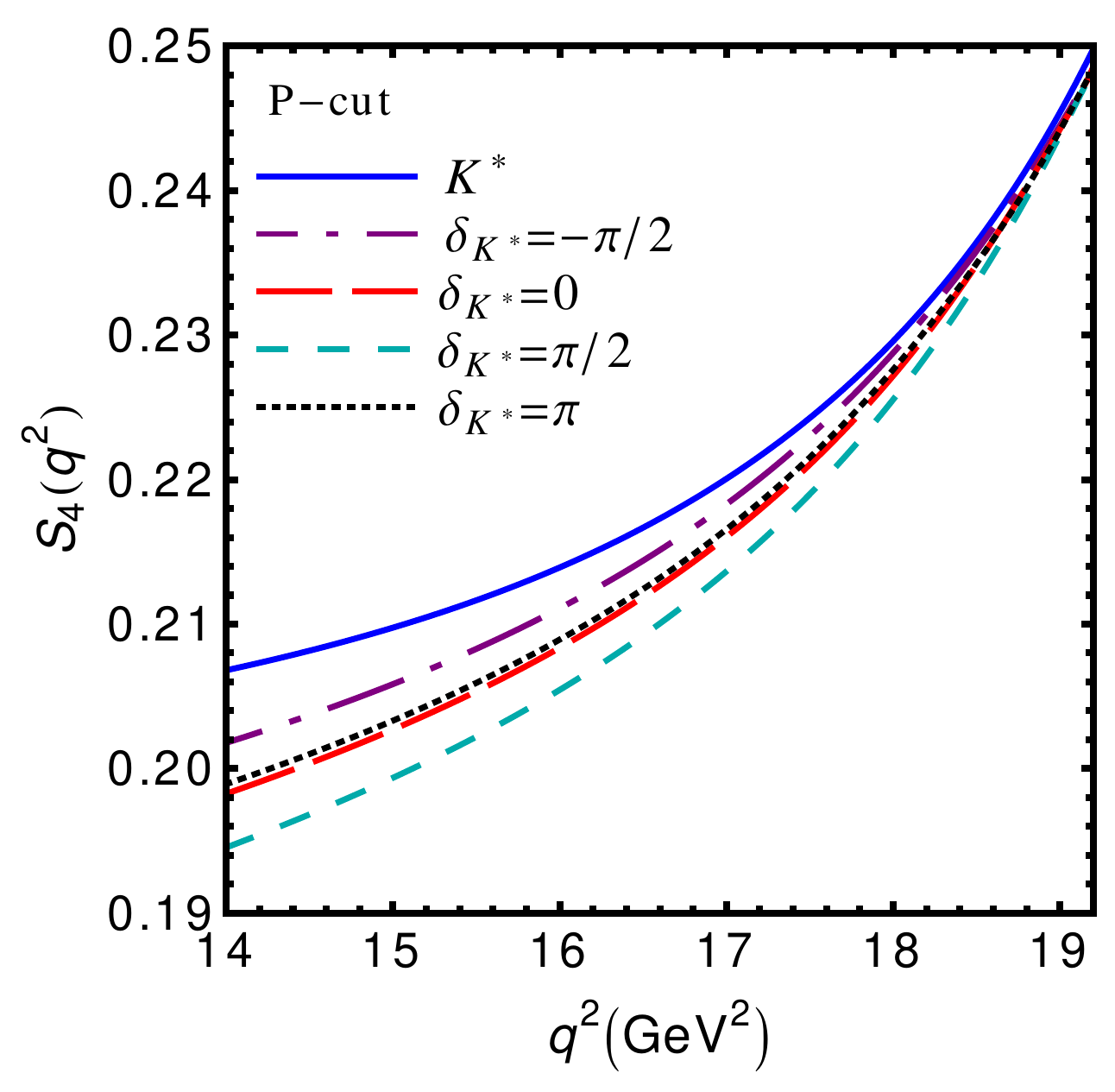}\qquad
\includegraphics[width=0.3\textwidth]{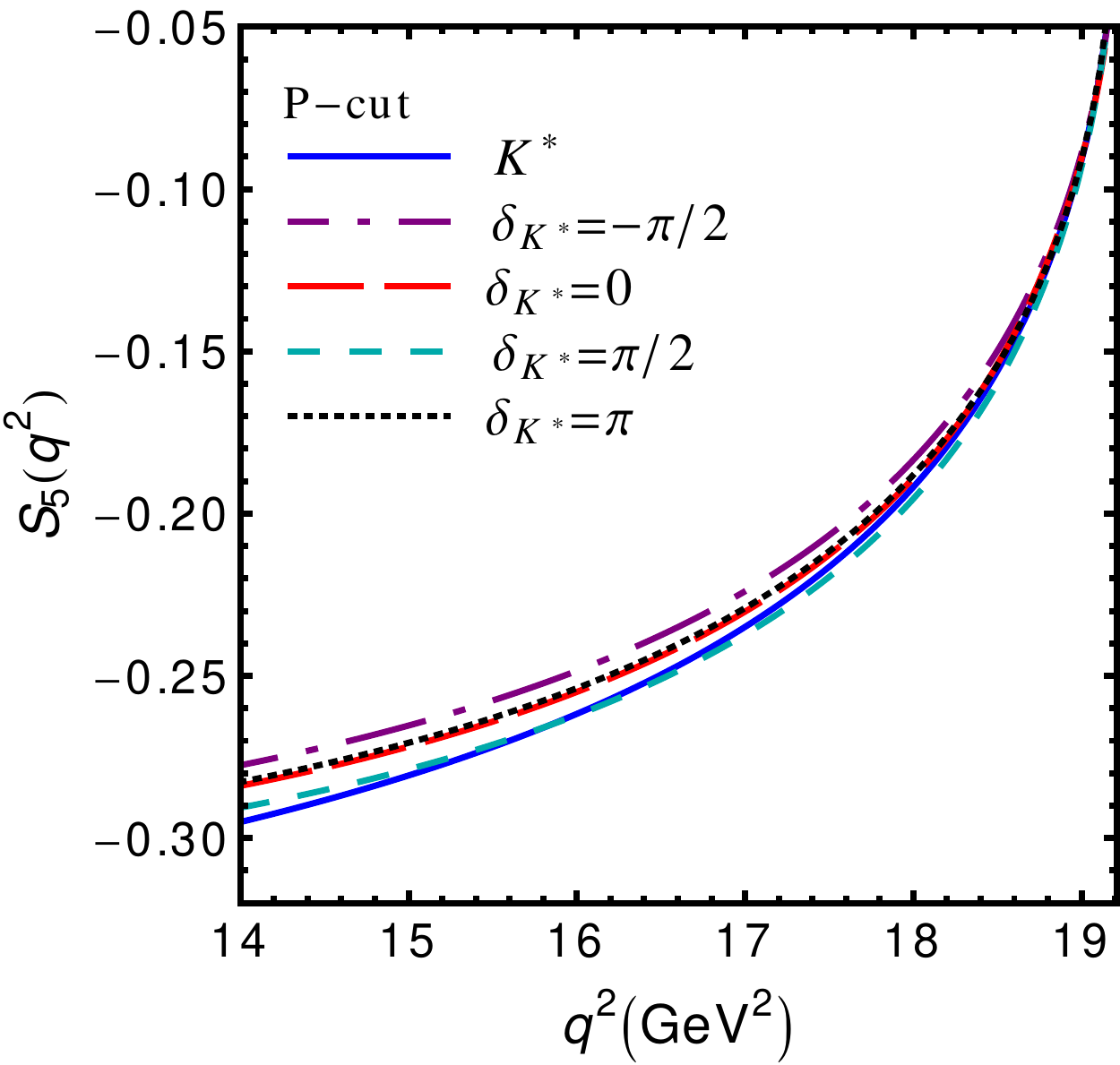}
\includegraphics[width=0.3\textwidth]{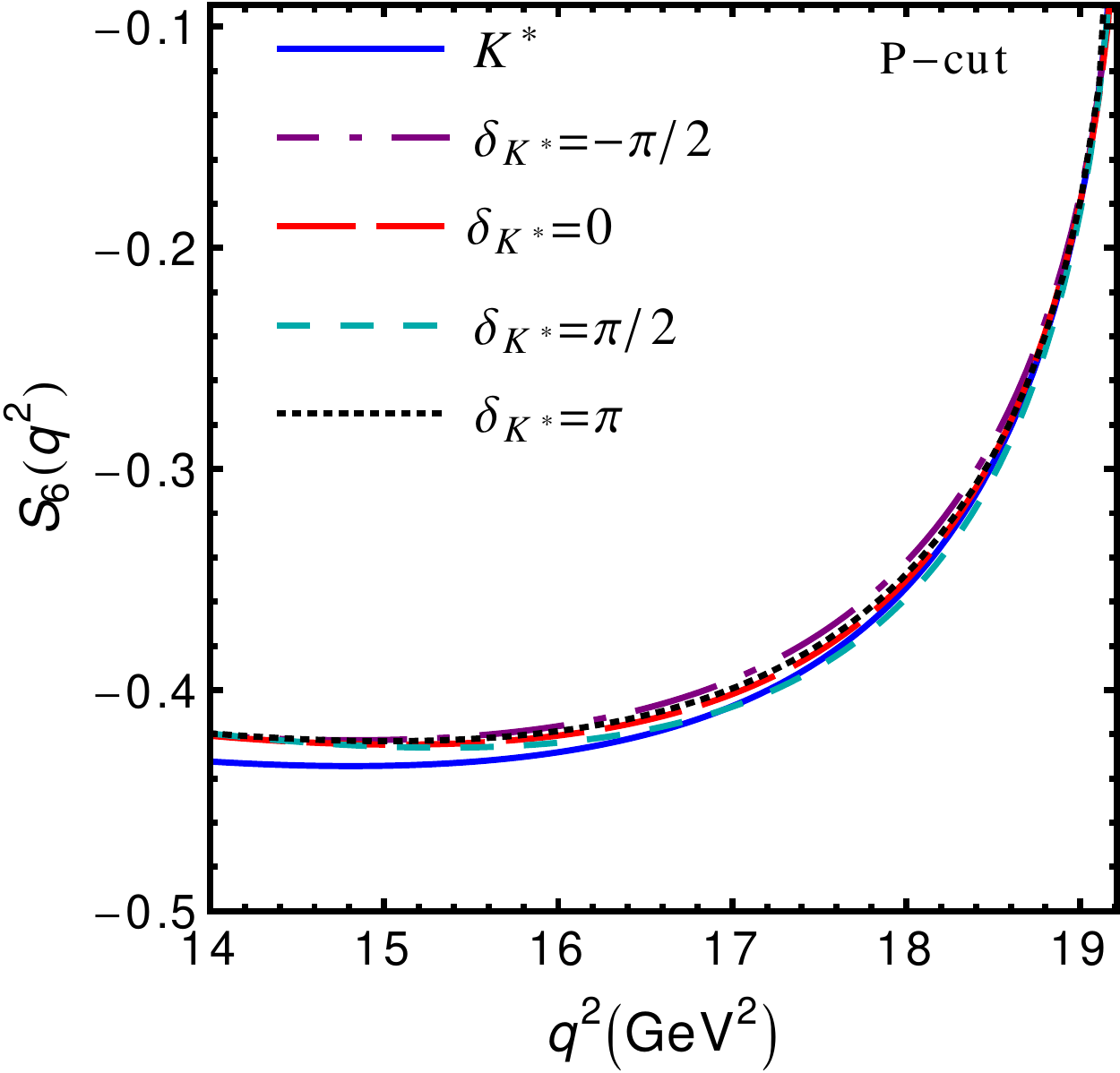}\qquad\qquad
\includegraphics[width=0.3\textwidth]{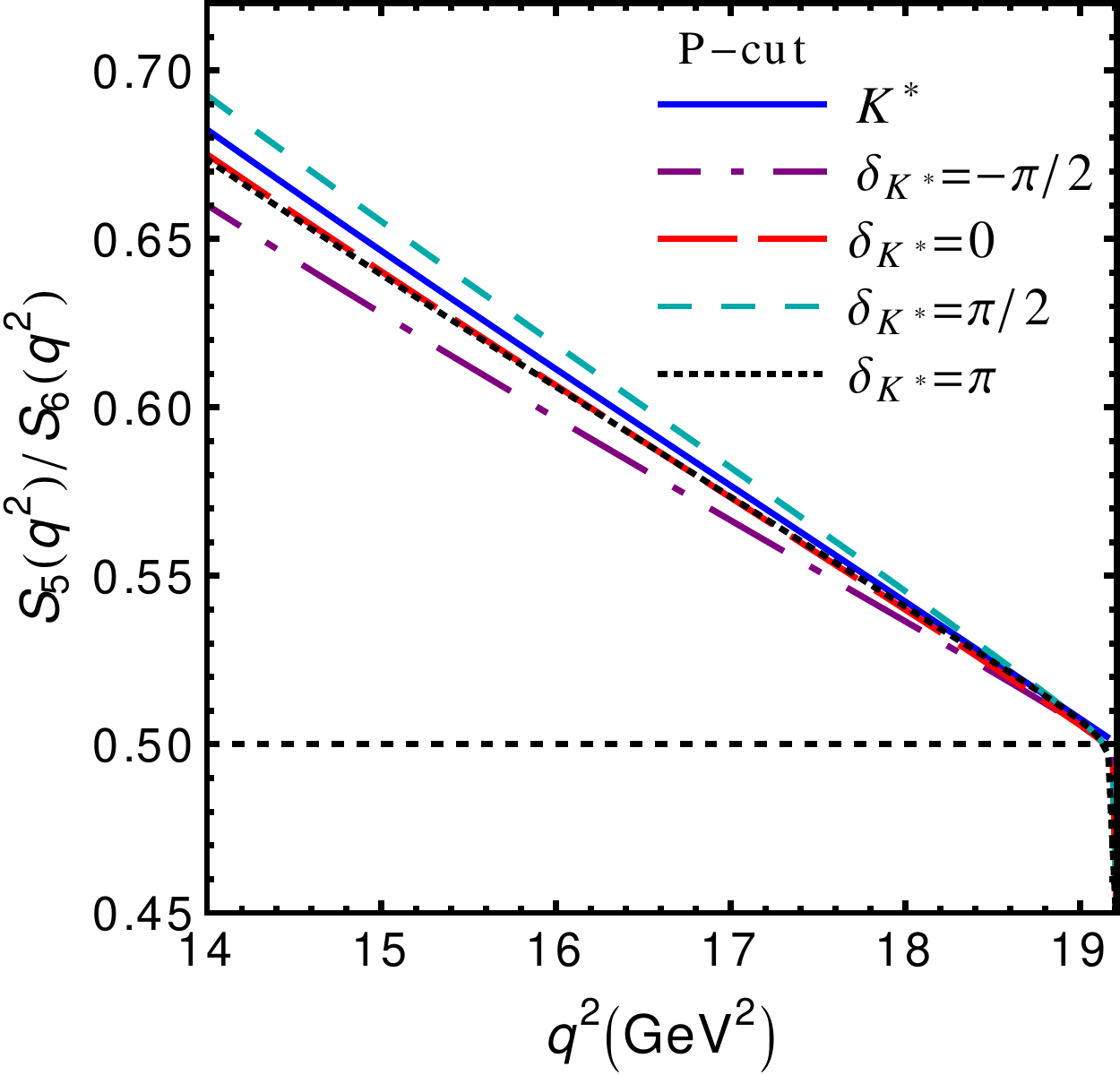}
}
\caption{\small $S_{3,4,5,6}$ as given in Eq.~(\ref{eq:Si})  and $S_5/S_6$ in P signal-cut window in the SM  for central values and different values
of the strong phase.
The solid blue curve corresponds to the resonant $K^*$ contribution. 
}
\label{fig:S345}
\end{figure}

As shown in Fig.~\ref{fig:afbeps}, the corrections are up to 15\% 
for $I_{5,6}$, 20\% 
for $I_4$ and 30\% 
for $I_3$. The corresponding effects in
the S+P cut window are very similar and not shown. The qualitative dependence on the strong phase is in all cases similar and follows the one of the
branching ratio, shown in Fig.~\ref{fig:breps}, which drastically reduces the net effect in ratios, as a feature of a universal strong phase.
We show this explicitly for  the observables
\begin{align} \label{eq:Si}
 S_i \equiv \frac{\hat I_i(\bar B \to \bar K \pi \ell \ell)}{d \Gamma (\bar B \to \bar K \pi \ell \ell)/dq^2} \, ,
\end{align}
where we adopt the notation from $\bar B \to\bar K^* \ell \ell$ studies \cite{Altmannshofer:2008dz} to allow for easier comparison.
Interference affects the observables $S_5$ and $A_{\rm FB} =S_6$ very little, followed by $S_4$ and then $S_3$,  
as can be seen in Fig.~\ref{fig:S345} for the P-cut window. The uncertainties on the $S_i$ from the strong phase are up to 14\% on
$S_3$, while they do not exceed a few percent for $S_{4-6}$. 
The largest shift from interference receives $S_4$, which gets suppressed by about $5\%$ 
over the whole low recoil region. The shifts
in the other angular observables are  smaller and vary with $q^2$.

\begin{figure}[ht]
\centering{
\includegraphics[width=0.3\textwidth]{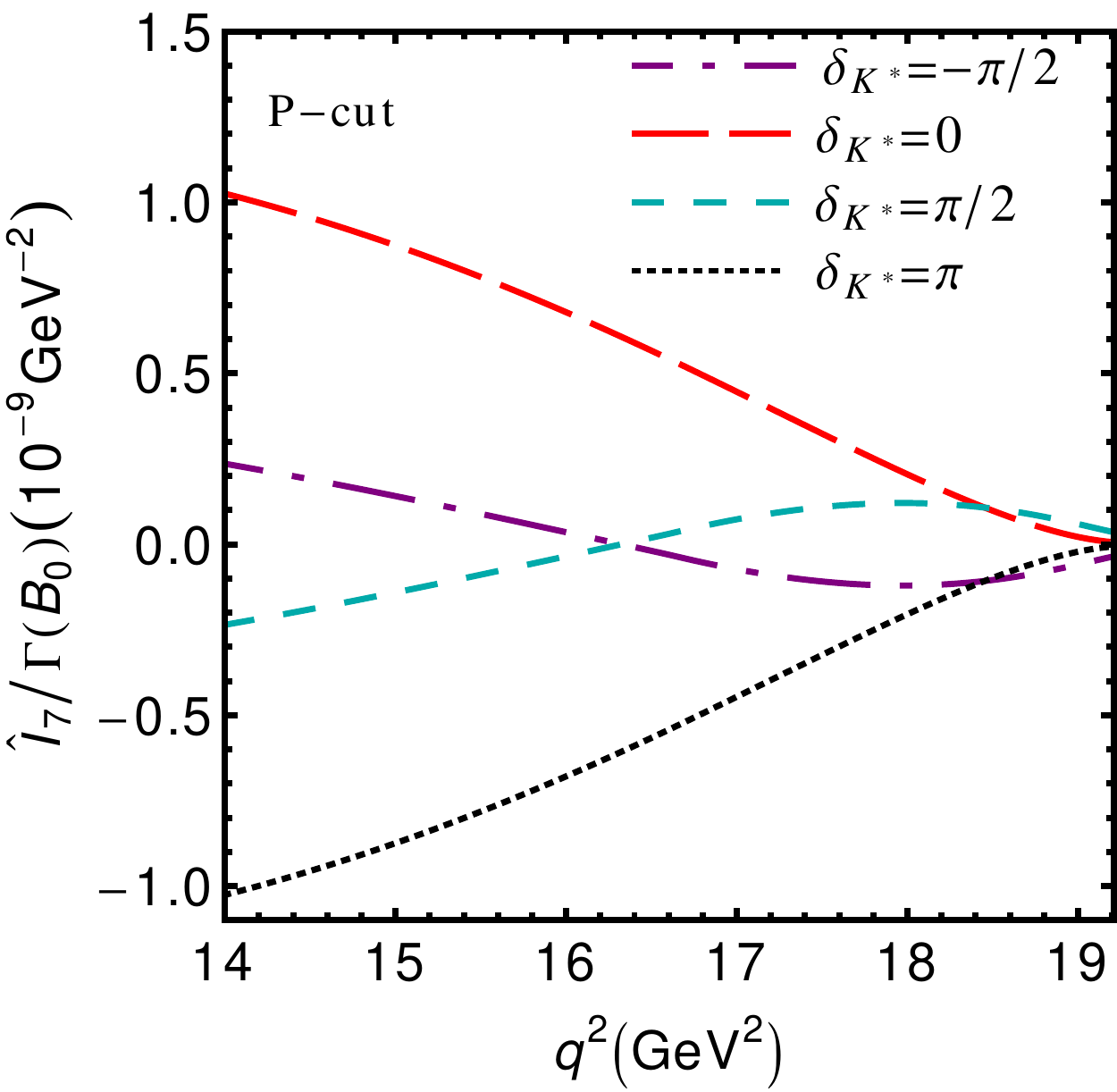}\qquad
\includegraphics[width=0.3\textwidth]{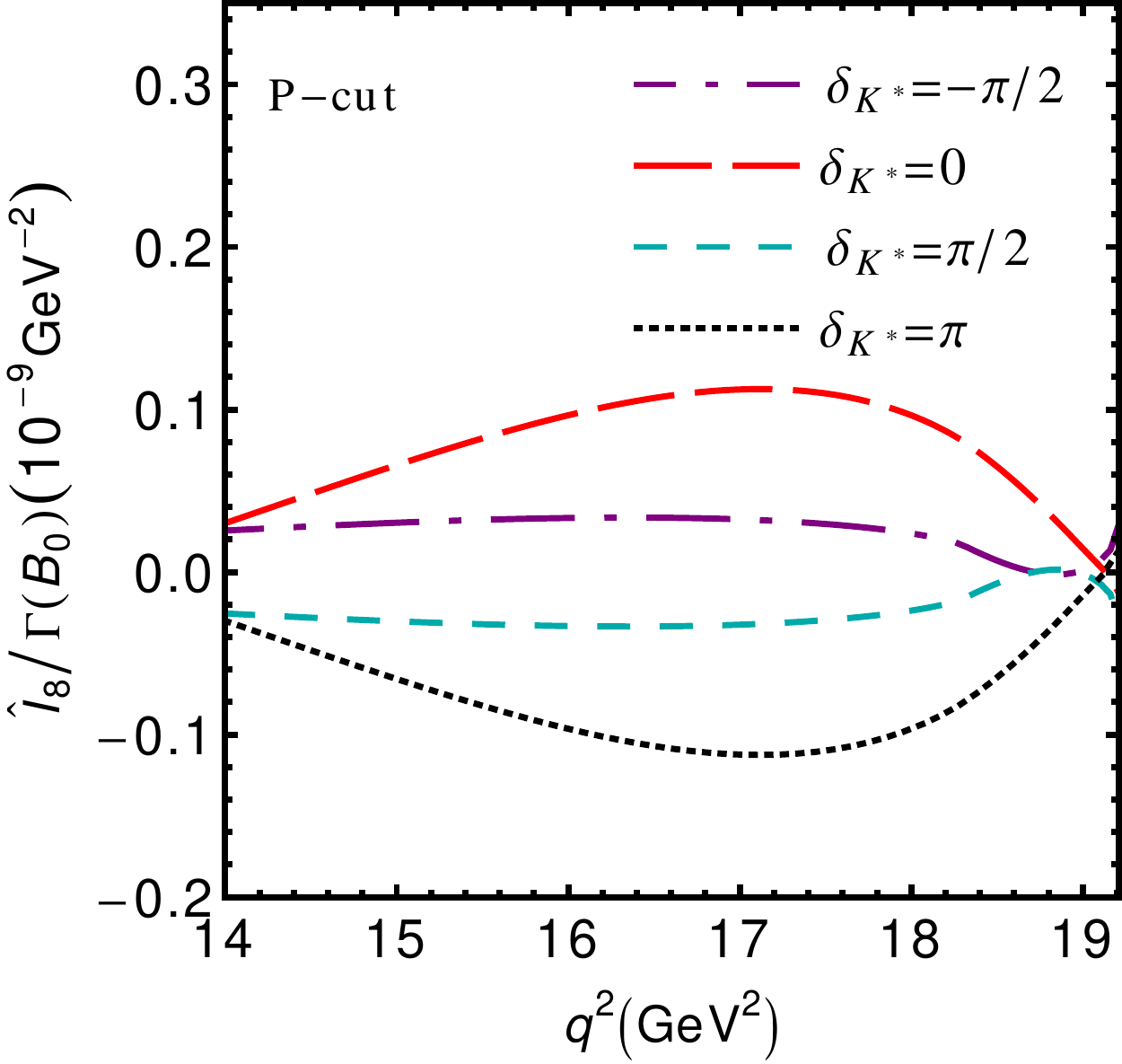}\qquad
\includegraphics[width=0.3\textwidth]{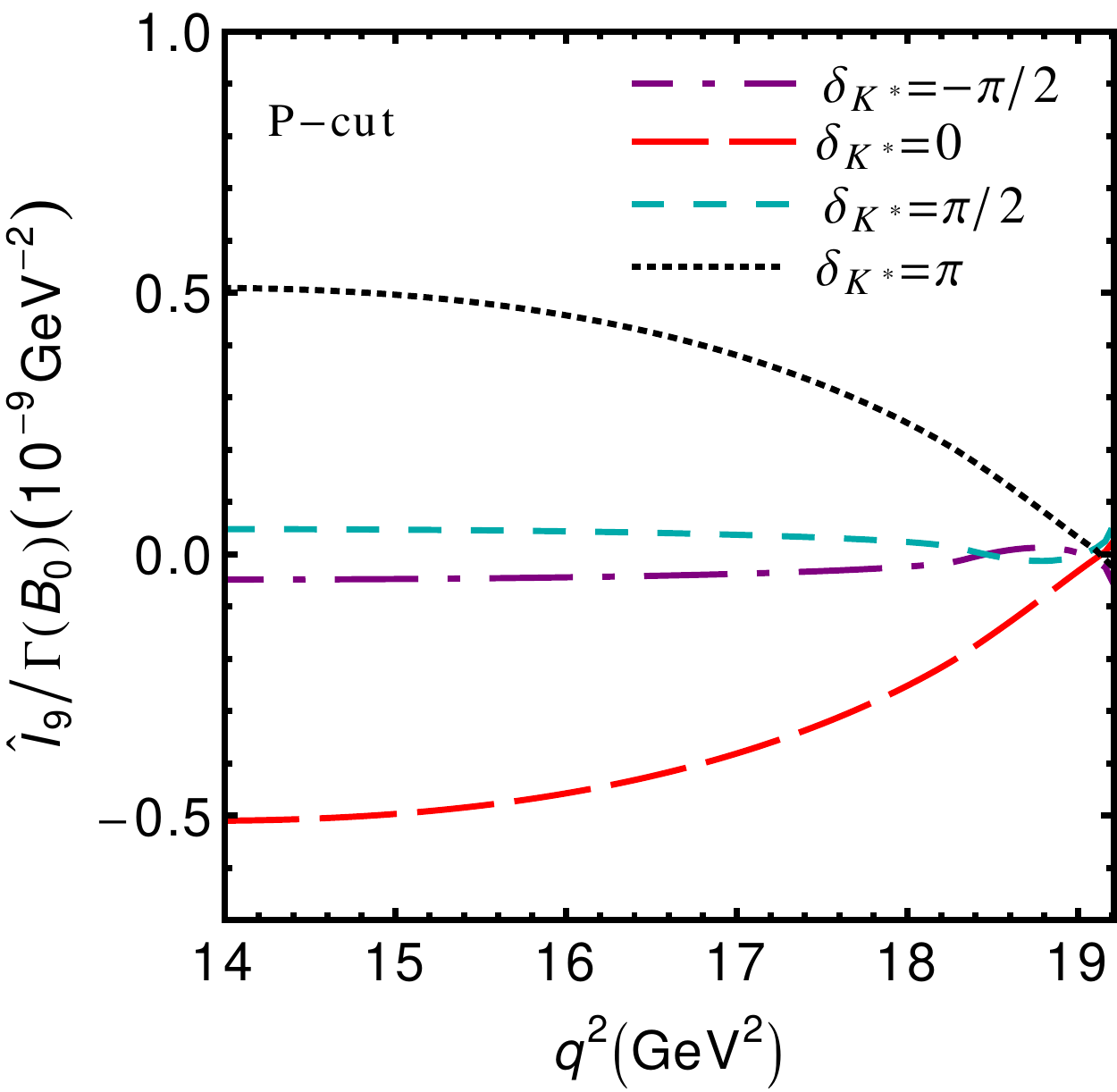}\\
\includegraphics[width=0.3\textwidth]{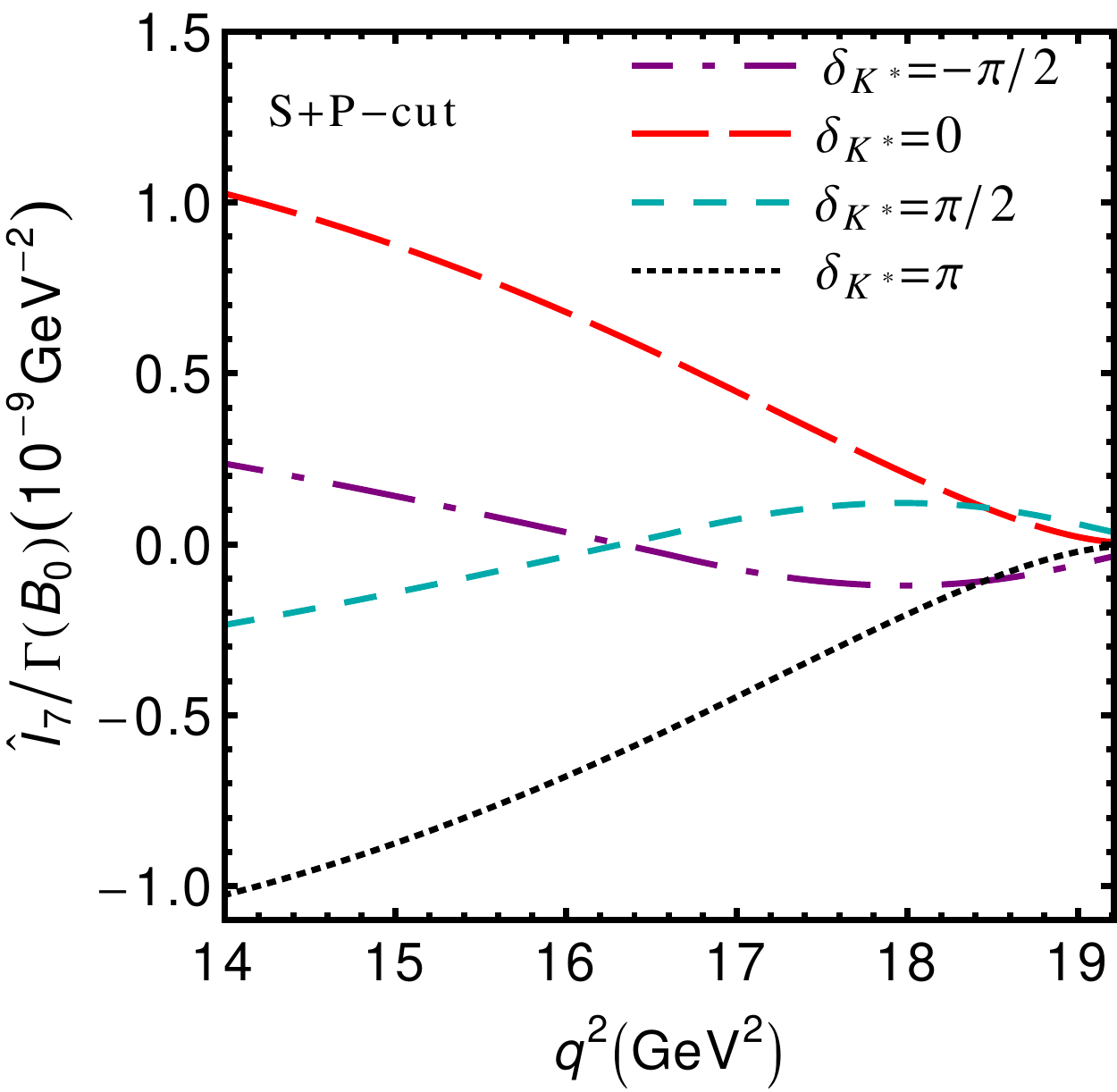}\qquad
\includegraphics[width=0.3\textwidth]{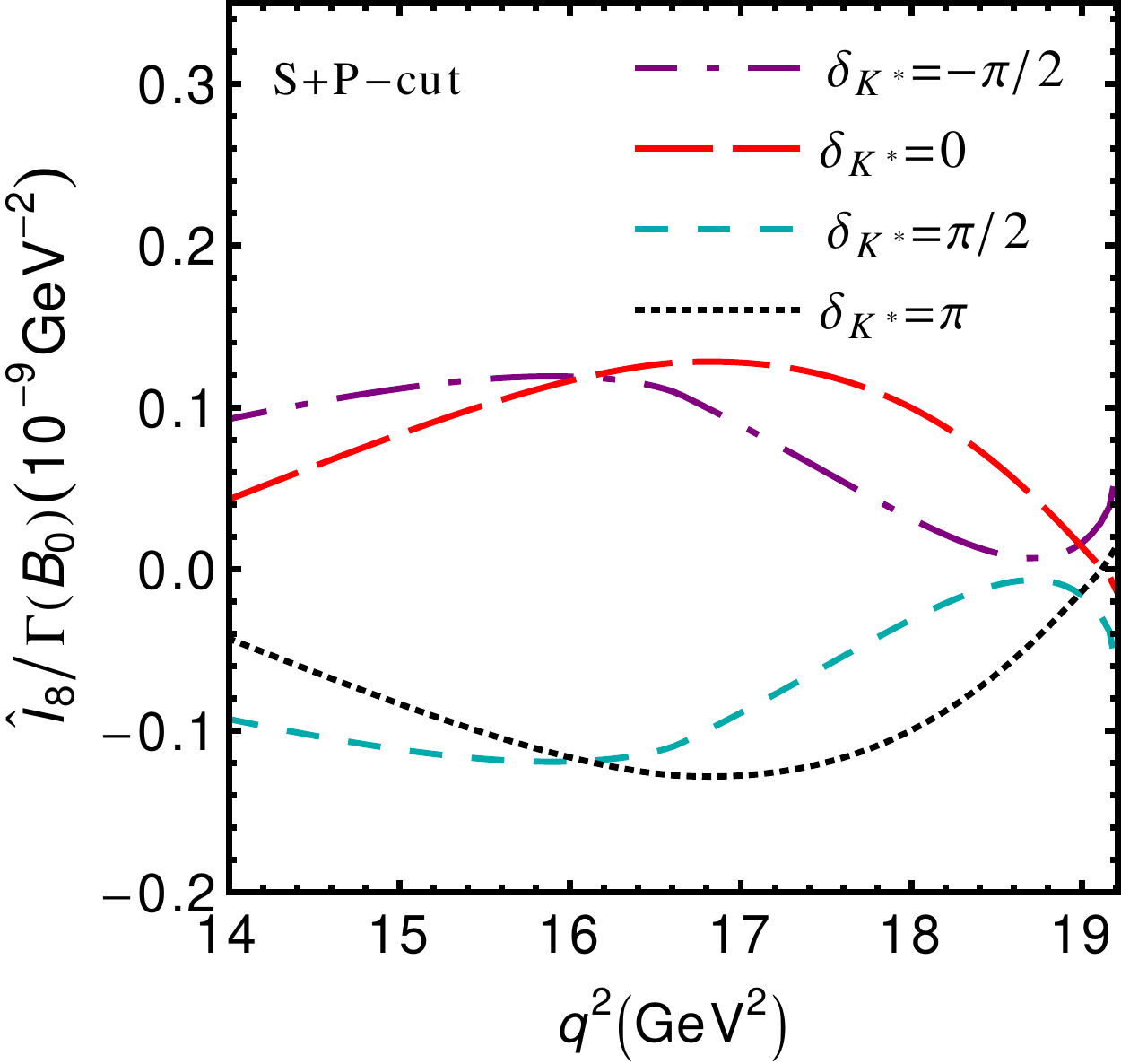}\qquad
\includegraphics[width=0.3\textwidth]{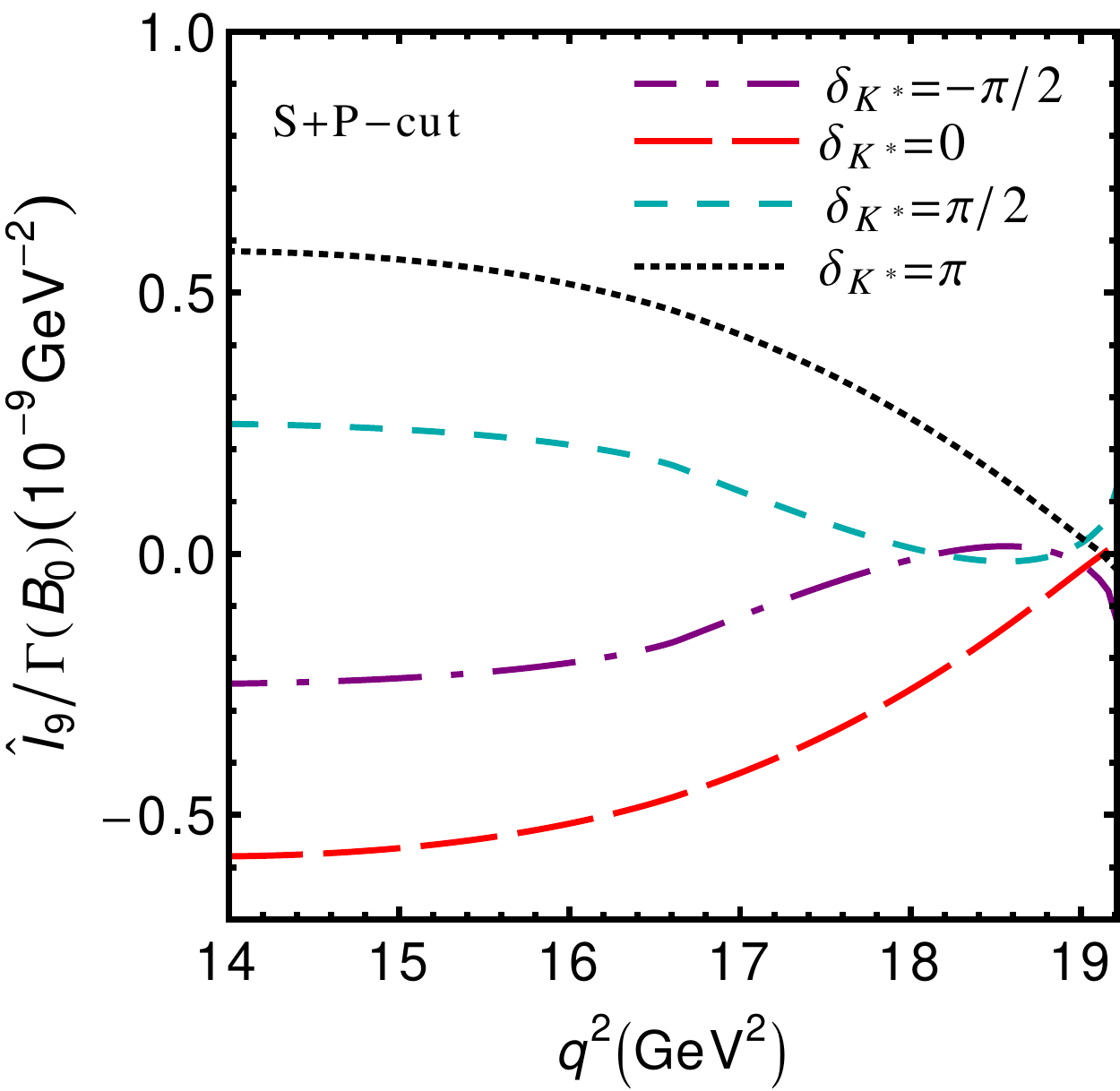}
}
\caption{\small The angular coefficients $\hat I_{7,8,9}/\Gamma(B)$ in the SM for different values of the relative strong
phase in the P-cut window (upper row) and S+P cut window (lower row). \label{fig:789}}
\end{figure}

$I_{7,8,9}$  are SM nulltests of the $\bar K^*$ distribution. This follows from universality  {\it i.e.}, $\bar K^*$-polarization independence of
the short-distance coefficients of the leading-order low-recoil OPE \cite{Bobeth:2010wg}, which extends to the case with CP violation. For $I_7$ this
is even true in the more general SM+SM'basis given in Eq.~(\ref{eq:Iope})  \cite{Bobeth:2012vn}.
However, interference of the $\bar K^*$ with the non-resonant $\bar K \pi$ contribution induces small backgrounds, see Fig.~\ref{fig:789}, where 
$\hat I_{7,8,9}$ are shown in the SM, normalized to the mean total width $\Gamma(B)$ after P-cut and S+P-cut integration. Comparing their size to
the dilepton spectrum shown in Fig.~\ref{fig:br}, the effect is at most of the order of a few percent and largest for $\hat I_7$,
followed by $\hat I_9$.
The induced  values for $|\hat I_{7,8,9}^{\rm SM}|$ are largest for $\delta_{K^*}$ near  $0$ and $\pi$. On the contrary, the  largest
interference effects in the dilepton spectrum and other ${\rm Re}$-type observables like $I_{3,4,5,6}$  are assumed at   $\delta_{K^*} \sim \pm
\pi/2$.
As a result, strong-phase-related uncertainties do not cancel efficiently in ratios $S_{7,8,9}$, and remain sizable, at ${\cal{O}}(1)$.

\section{Phenomenlogy \label{sec:pheno}}

The unknown strong phase $\delta_{K^*}$ implies a sizable uncertainty in the SM predictions, see Figs.~\ref{fig:br}-\ref{fig:789}, which is very
difficult to control theoretically. We therefore start this section by discussing opportunities from $I_{7,8,9}$ for probing strong phases
(Sec.~\ref{sec:strong}), before discussing BSM physics (Sec.~\ref{sec:bsm}).

\subsection{Probing strong phases \label{sec:strong}}
\begin{figure}[ht]
\centering{
\includegraphics[width=0.3\textwidth]{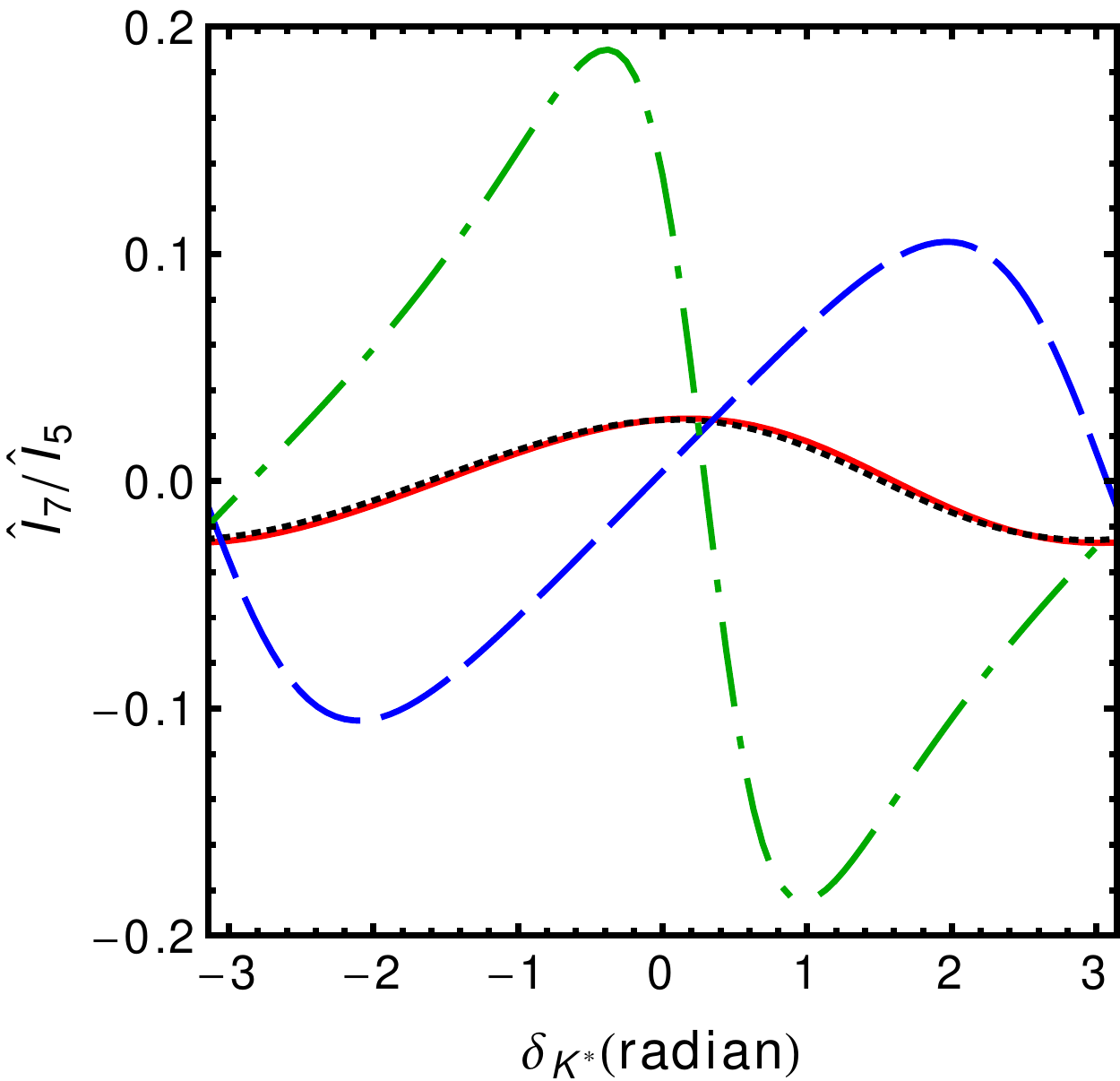}\qquad
\includegraphics[width=0.3\textwidth]{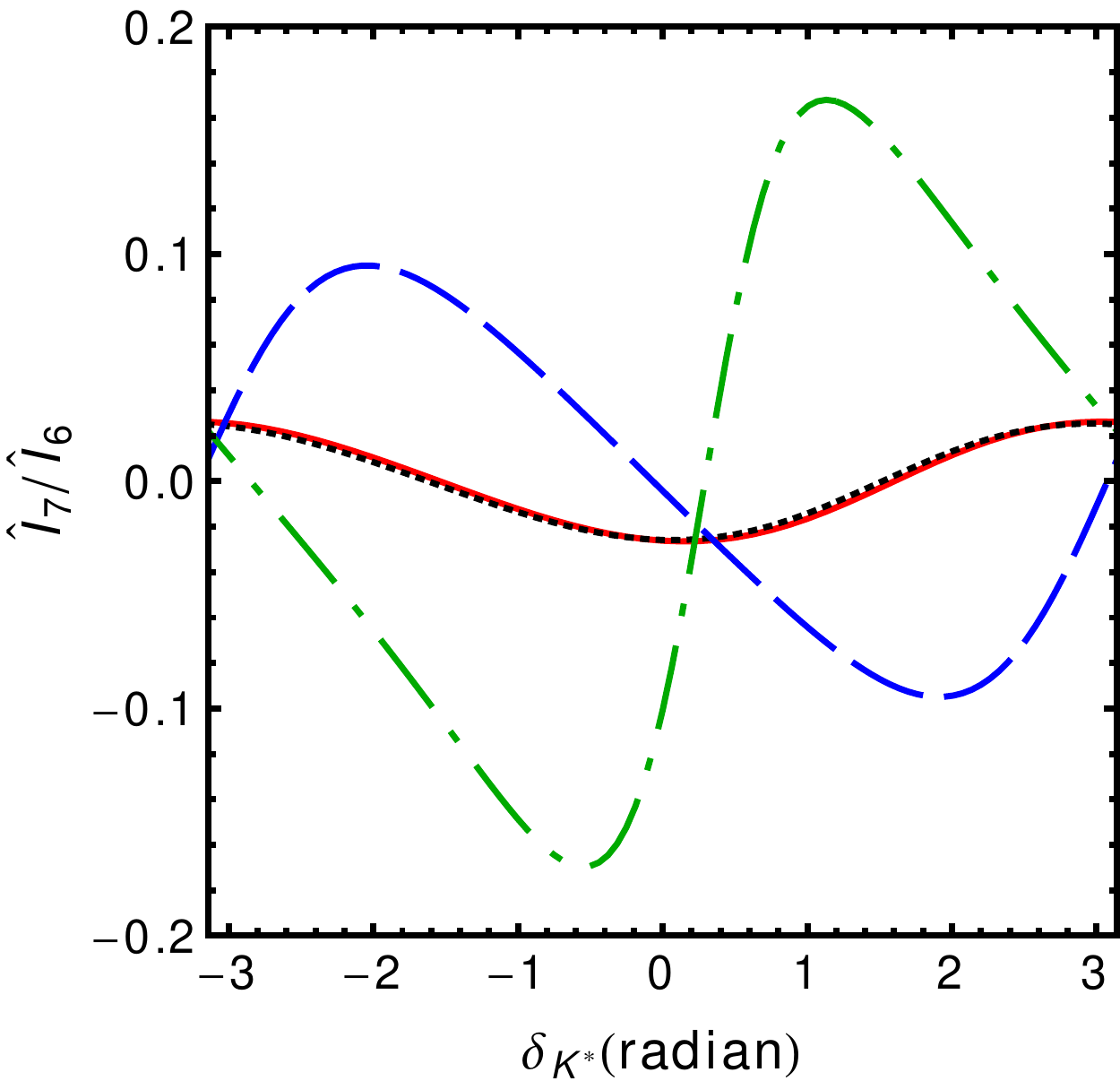}\qquad
\includegraphics[width=0.3\textwidth]{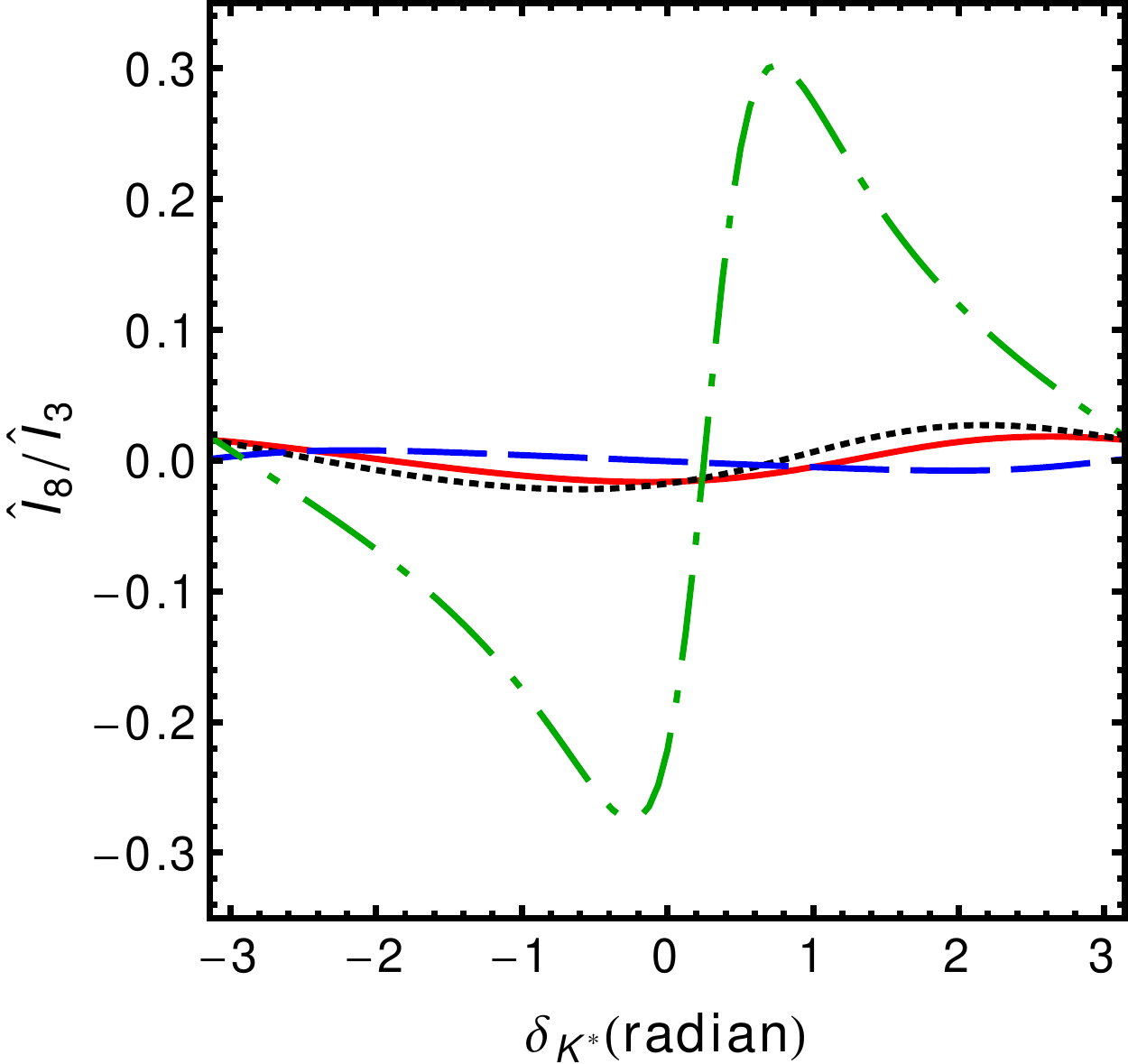}
\includegraphics[width=0.3\textwidth]{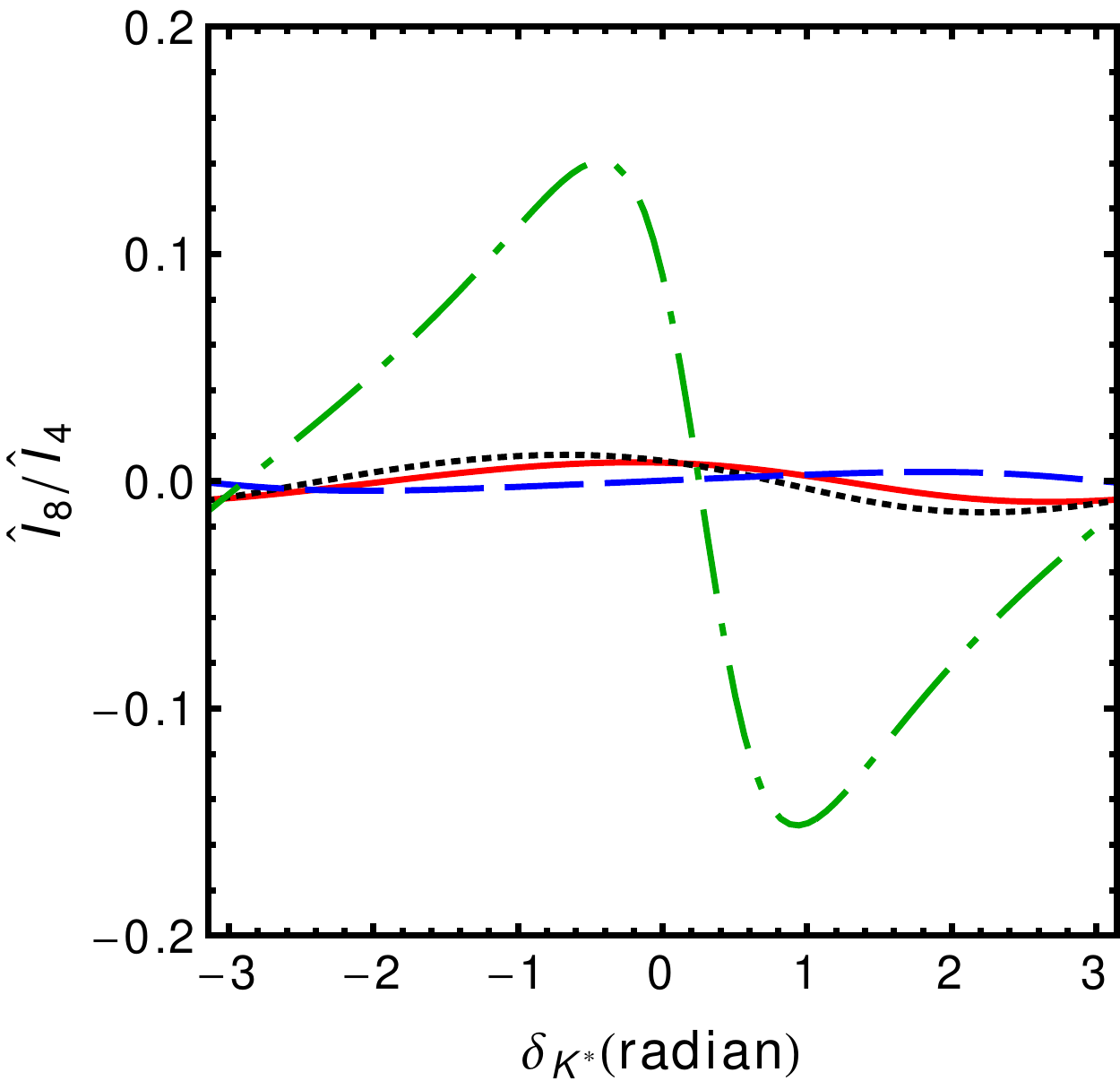}\qquad
\includegraphics[width=0.3\textwidth]{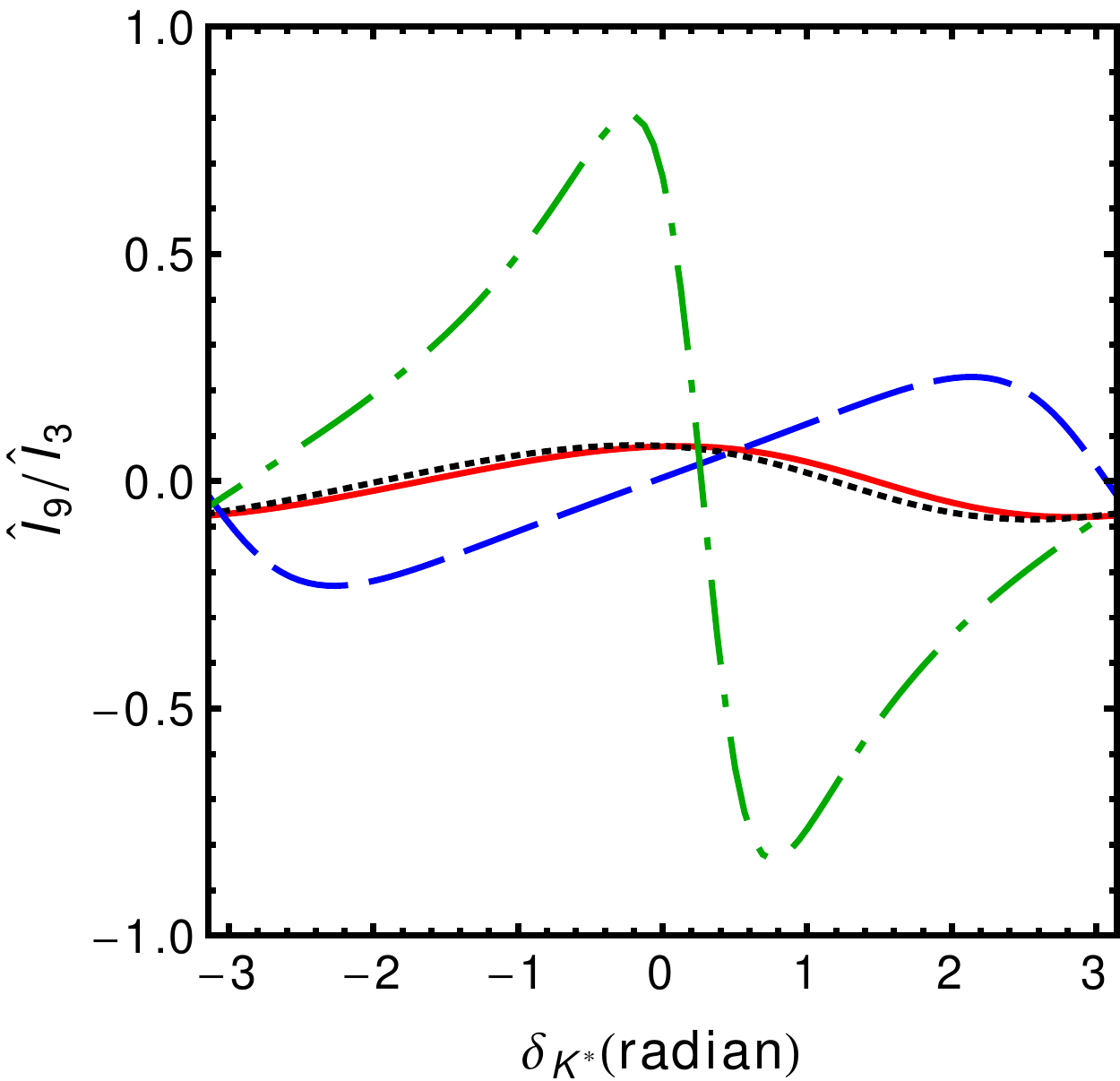}\qquad
\includegraphics[width=0.3\textwidth]{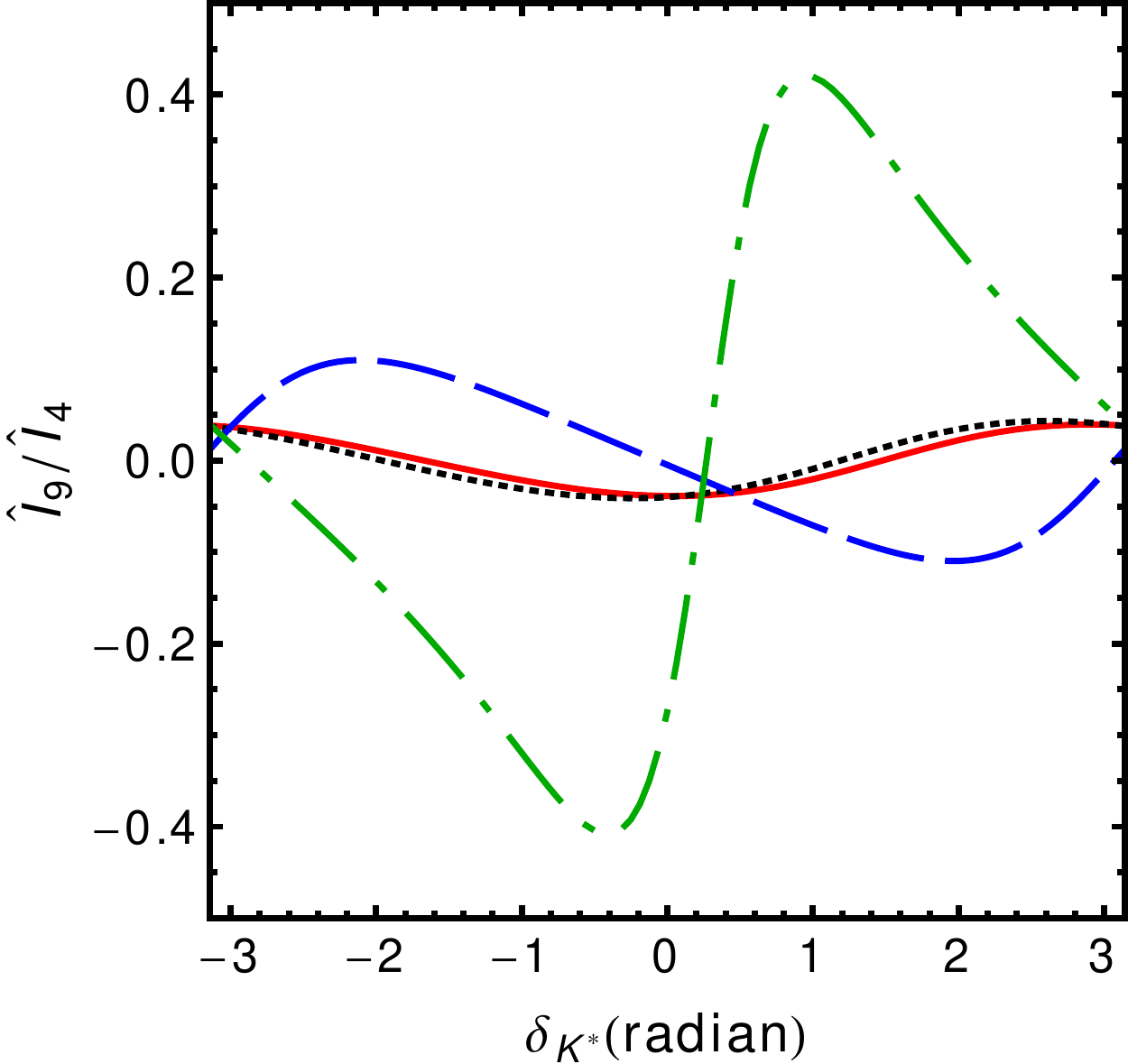}
}
\caption{\small 
Short-distance-free ratios in the SM basis as in Eq.~(\ref{eq:strongphaseratios}) versus  $\delta_{K^*}$ at $q^2=16 \, \mbox{GeV}^2$ 
in the P-cut signal window (red solid), the  S+P-cut total window (black dotted), below the signal window, $p^2_{\rm min} < p^2 < 0.64 \,
\mbox{GeV}^2$ (blue dashed), and above the signal window, $0.9 \, \mbox{GeV}^2<p^2 < 2 \, \mbox{GeV}^2$ (green dash-dotted).
\label{fig:delP}}
\end{figure}

The ratios 
\begin{align}  \label{eq:strongphaseratios}
\frac{\hat I_7}{\hat  I_5}\, , \frac{\hat  I_7}{\hat  I_6}\, , \frac{\hat  I_8}{\hat  I_3}\, , \frac{\hat  I_8}{\hat  I_4}\, , \frac{\hat  I_9}{\hat  I_3}\, , \frac{\hat  I_9}{\hat  I_4} \, , S_8 \, , S_9\, ,
\end{align}
are all short-distance-free in the SM basis, \emph{cf.} Eq.~(\ref{eq:IopeSM}). They can be used to obtain information on the strong phase between the
resonant and non-resonant contributions to $\bar B\to \bar K^*\ell\ell$ decays, since for these ratios the dependence on the strong phase fully
remains,  as discussed in the previous section. The functional dependence on the strong phase $\delta_{K^*}$ varies with the $p^2$-cuts as detailed in
App.~\ref{app:epsilon}. We see this leading behaviour explicitly in Fig.~\ref{fig:delP}. It is evident that the observables are sensitive to the
strong phase, which can be measured in any of the ratios up to a twofold ambiguity, which could be resolved with a second measurement of a ratio with
a different numerator.
As stressed already in Sec.~\ref{sec:details}, $\delta_{K^*}$ depends on $p^2$. Hence, phases extracted using different $p^2$-cuts are in general not
the same.

The observables in Eq.~\eqref{eq:strongphaseratios} can be larger outside the $\bar K^*$-window, where signal and interfering background become
more comparable. This is especially the case in the $p^2$-region above the $\bar K^*$, where more phase space is available away from the
$\bar K^*$-peak than below. Being outside the $\bar K^*$-window comes, however, at the price of fewer events. It would require experimental
simulations to estimate the ideal $p^2$ cuts for maximal sensitivity; however, note again that the strong phase is expected to vary over $p^2$. Theory
uncertainties from ratios of the form factors $F_{nr}$ and $F_{K^*}$ apply.

\subsection{Beyond the Standard Model \label{sec:bsm}}

While a  complete exploration of the BSM sensitivity of all angular observables in the full basis in Eq.~(\ref{eq:7910}) is beyond the scope of this
work, here we concentrate on $I_{7,8,9}$ because  {\it i)} they are SM nulltests of $\bar B \to \bar K^* ( \to \bar K \pi) \ell \ell$ decays and  {\it
ii)} they involve new combinations of short-distance coefficients   \cite{Das:2014sra}. Specifically, $\delta \rho$ and ${\rm Re} \rho_2^-$ can
in $\bar B \to \bar K \pi \ell \ell$ only be accessed with $I_7$ and $I_{8,9}$, respectively, see Eq.~(\ref{eq:Iope}). Note that $\delta\rho$ and
${\rm Re}\rho_2^-$ can also be probed with $\Lambda_b \to \Lambda \ell \ell$ decays \cite{Boer:2014kda}. A closer look exhibits that there arise new
constraints only, if interference arises between \emph{i)} primed and unprimed Wilson coefficients and \emph{ii)} contributions from operators with vector and axial vector structure. Given the presence of $C_{9,10}$ in the SM, this requires at least one coefficient $C^\prime_i \neq 0$ from new
physics.
Consequently, we focus on exploring the sensitivity to primed operators. We recall that in this paper we do not consider  CP violation, as it is
consistent with semileptonic and radiative $b \to s$ data to do so and small in the SM. Larger effects in $I_{8,9}$ are of course possible with CP
violation, which is driven by ${\rm Re}({\cal{F}} {\cal{F}}^*)$ rather than its imaginary part; however, this is a feature that can already be probed
with $\bar B \to \bar K^* \ell \ell $ decays \cite{Bobeth:2008ij,Bobeth:2012vn}.
The assumption of negligible CP-violation can be checked by measuring CP-asymmetries \cite{Bobeth:2008ij}.

Integrating $\hat I_{7,8,9}$ in the SM over high $q^2$, $15 \, \mbox{GeV}^2 \leq q^2 \leq 19.2 \, \mbox{GeV}^2$, in the P-cut window, we find,
roughly, {\it cf} (\ref{eq:imregions}), 
\begin{align}
\int d q^2 \hat I_{(7,8,9)}^{\rm \, SM}(q^2)/\Gamma(B) \simeq (+1.7, +0.4, -1.4) \cdot 10^{-9} \, \cos \delta_{K^*} \, . 
\end{align}
Despite the uncertainty from the unknown strong phase, all these observables remain small in the SM; a measurement of a larger value
would indicate a non-vanishing BSM contribution.

\begin{figure}[ht]
\centering{
\includegraphics[width=0.3\textwidth]{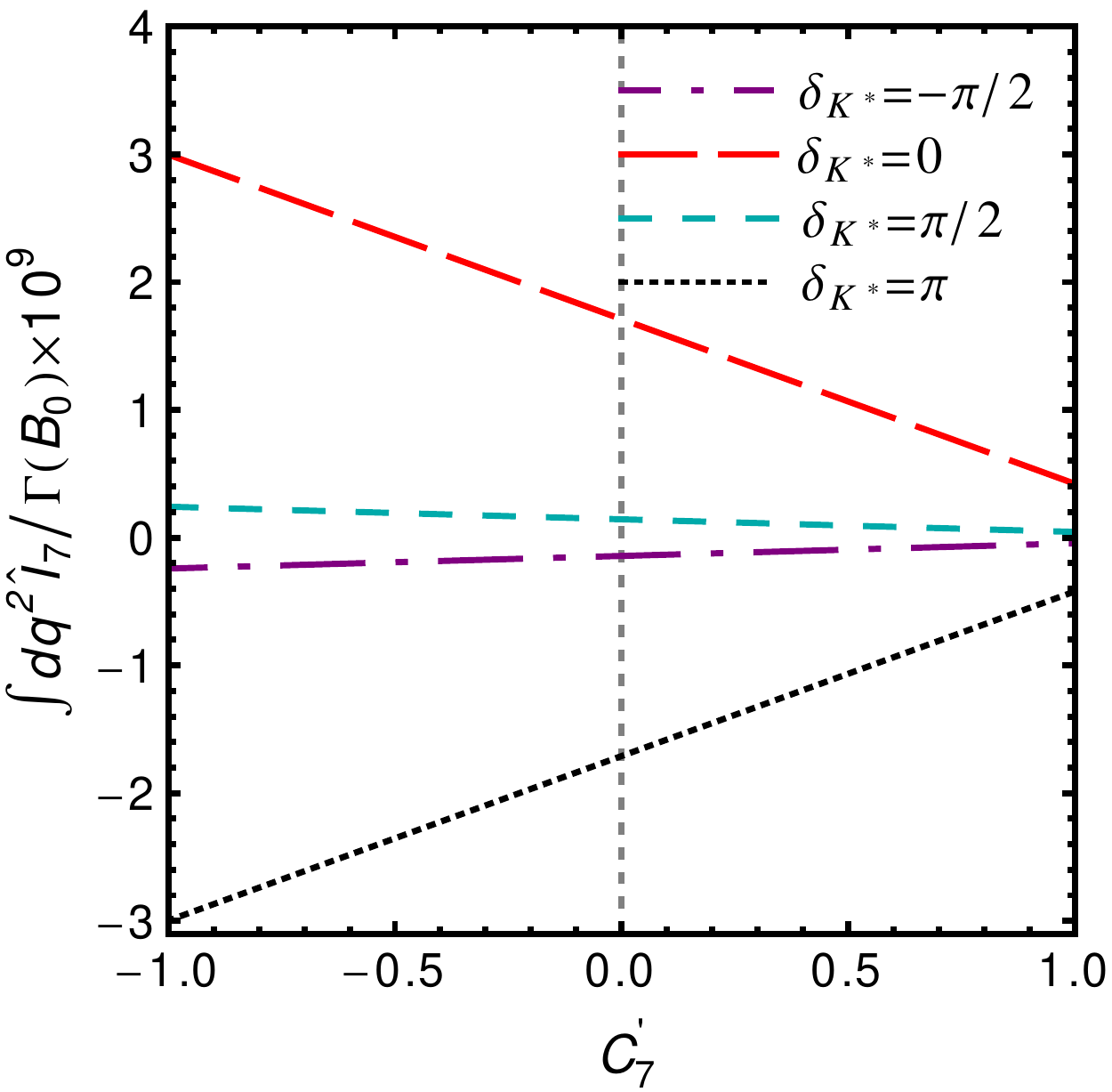}\qquad
\includegraphics[width=0.3\textwidth]{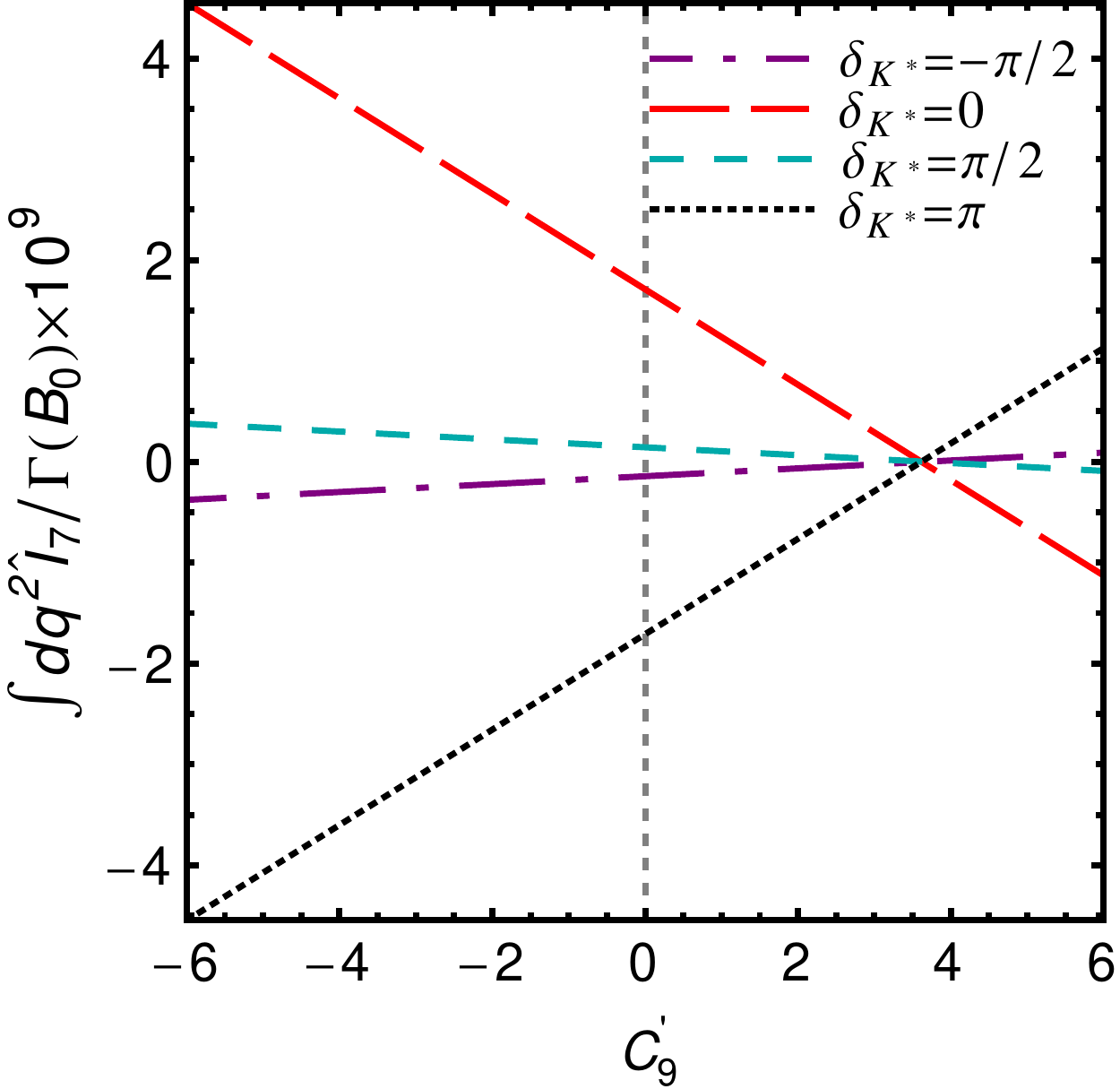}\qquad
\includegraphics[width=0.3\textwidth]{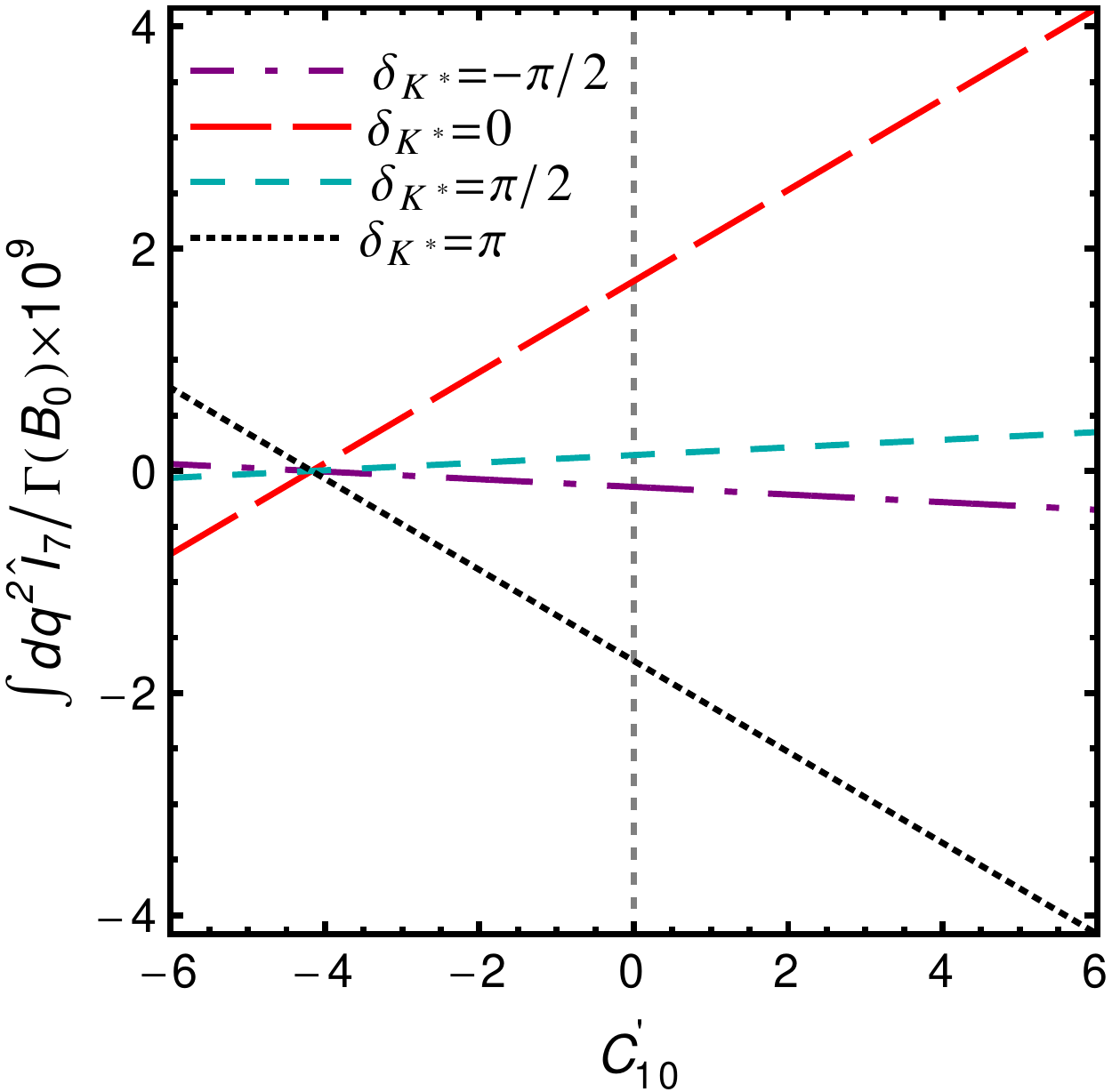} 
\includegraphics[width=0.3\textwidth]{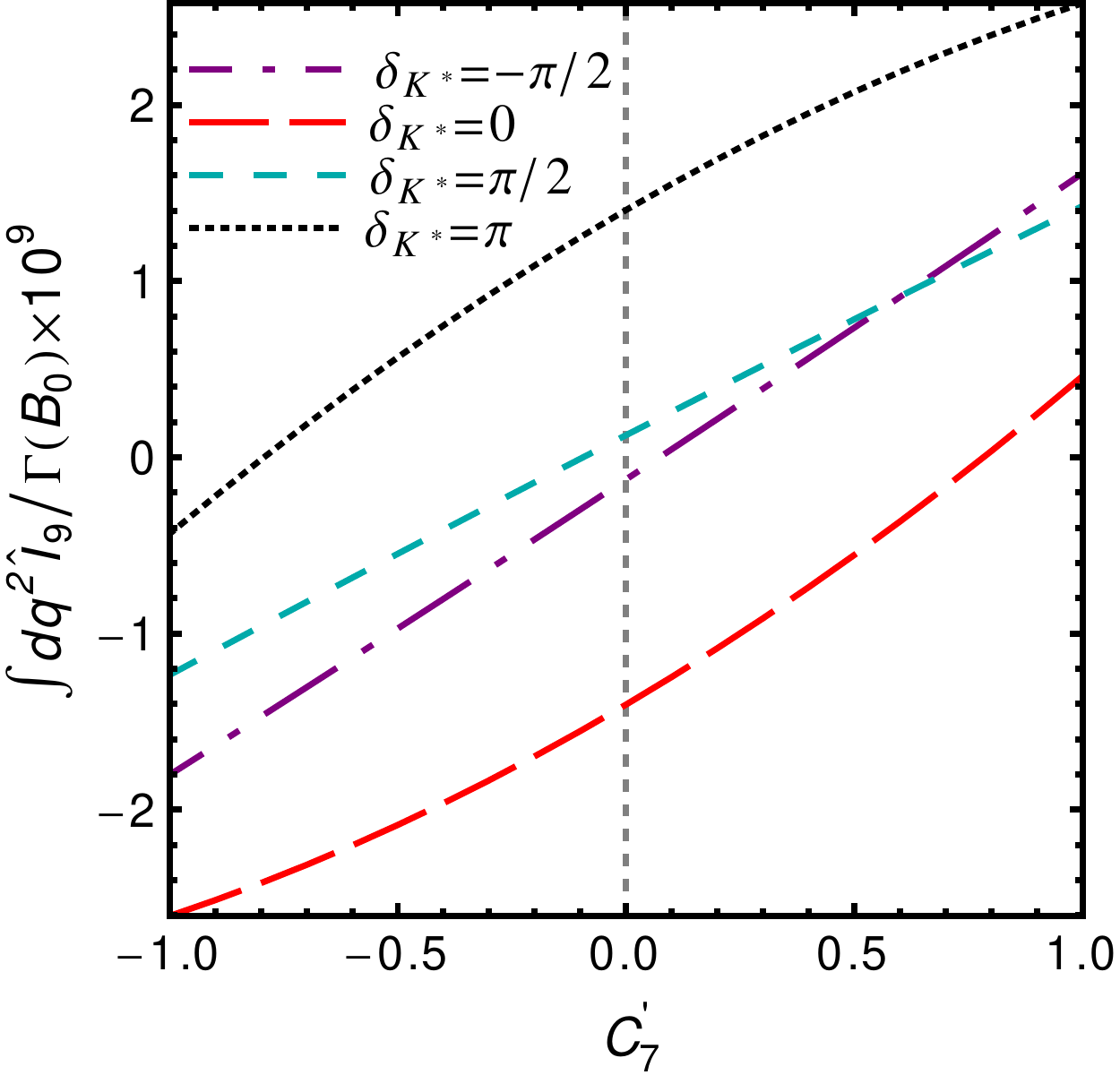}\qquad
\includegraphics[width=0.3\textwidth]{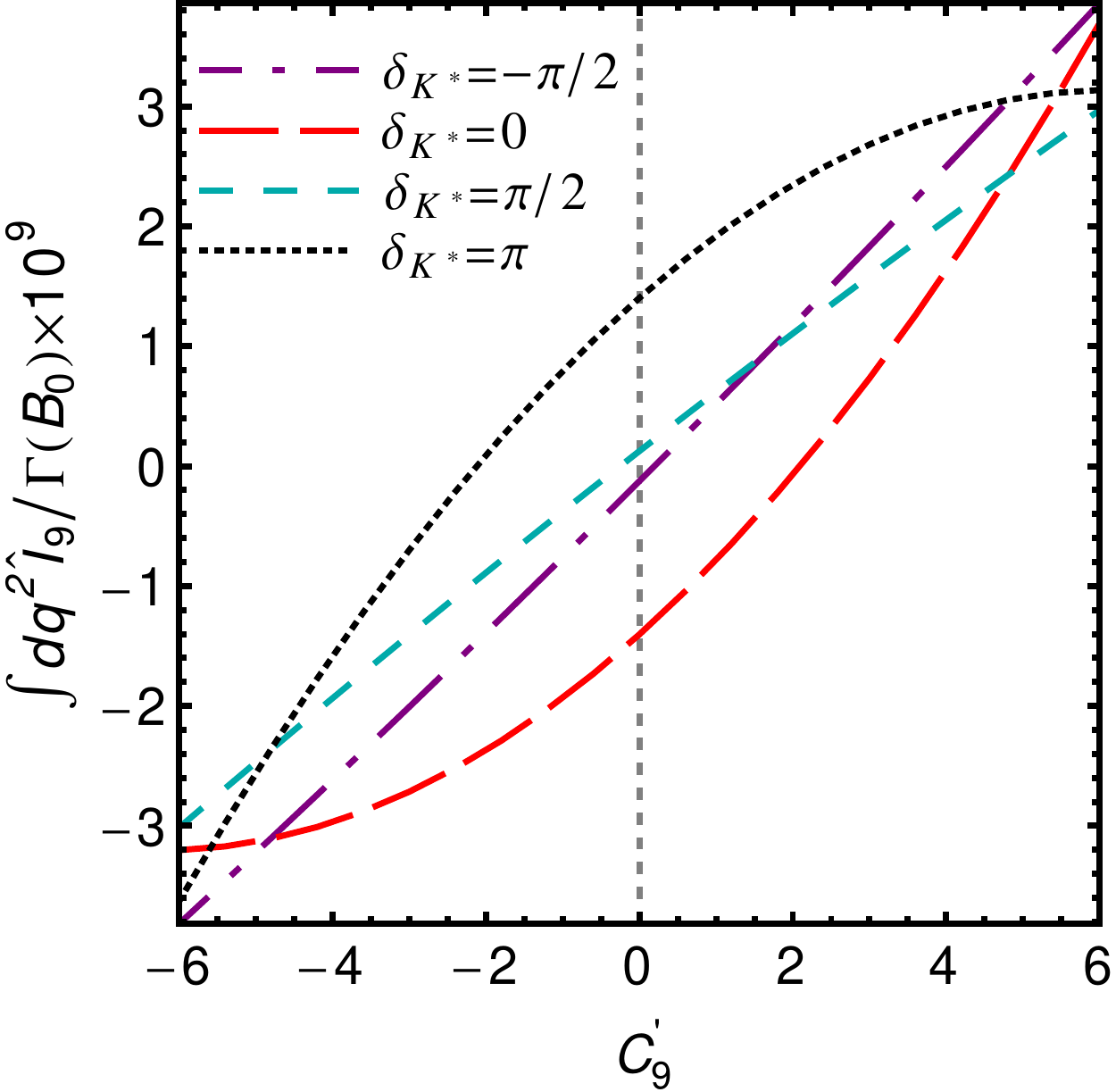}\qquad
\includegraphics[width=0.3\textwidth]{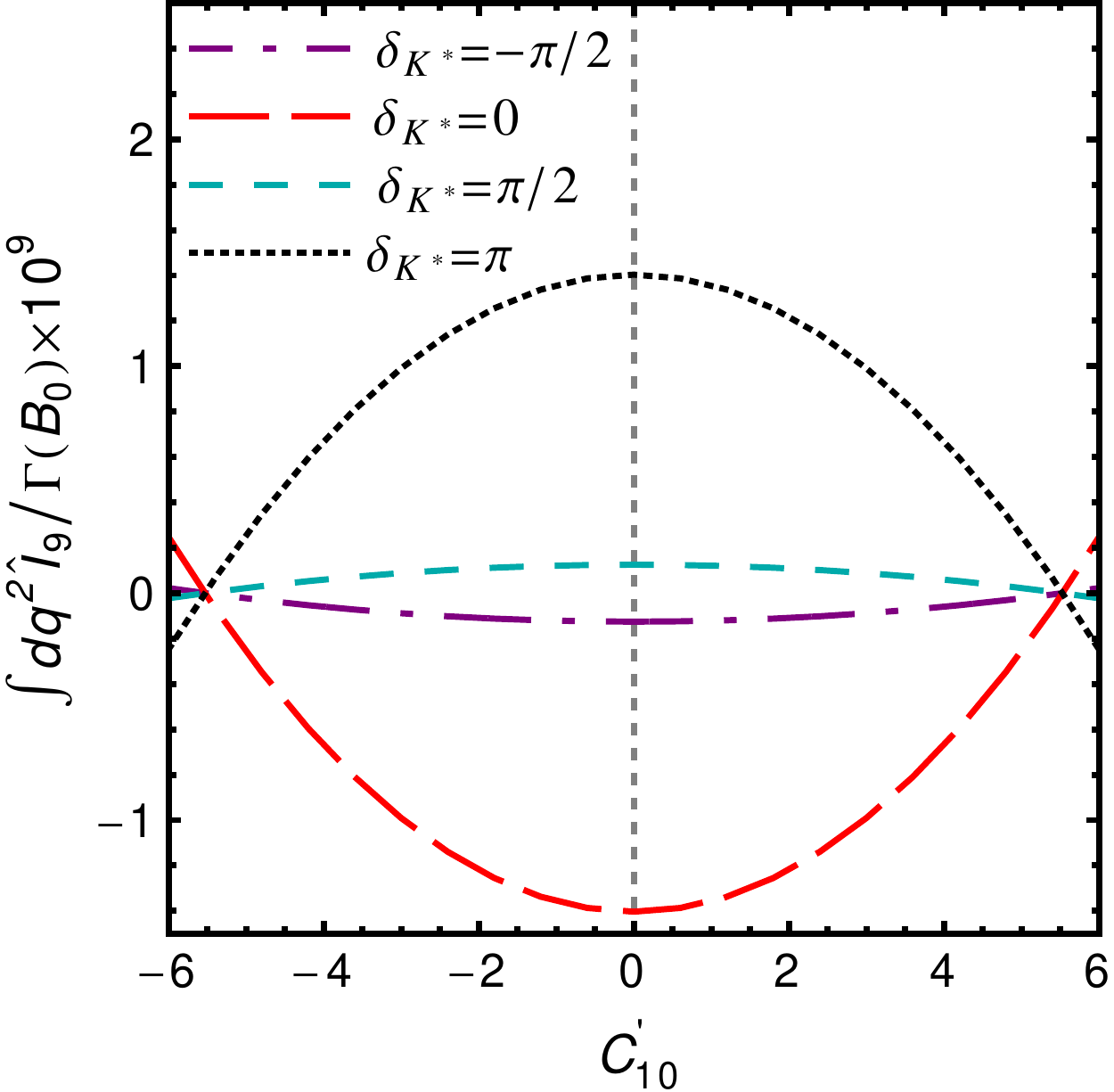}
}
\caption{\small The angular coefficients $\int d q^2 \hat I_{7,9}/\Gamma(B)$ integrated over high-$q^2$, $15 \, \mbox{GeV}^2 \leq q^2 \leq 19.2 \,
\mbox{GeV}^2$, in the P-cut window with one right-handed BSM Wilson coefficient switched on while all others assume SM values. The vertical line
corresponds to the SM.
\label{fig:np} }
\end{figure}

New physics effects are exemplified in Fig.~\ref{fig:np}. Since $I_8$ and $I_9$ involve the same short-distance physics, we only show
one of them, $I_9$, which can be  larger in magnitude. Note, that in the presence of right-handed currents there is even in the CP-limit a non-zero ${\rm Im}
\rho_2^+$  induced by quark loops, so the dependence of $I_{8,9}$ on $C_{7,9}^\prime$ is tilted relative to  $C_{10}^\prime$. Current low recoil data
constrain $|S_{7,8,9}|$ to be below the ${\cal{O}}(5-10\%)$  level \cite{LHCb:2015dla}, about to probe BSM effects.

\begin{figure}[ht]
\centering{
\includegraphics[width=0.3\textwidth]{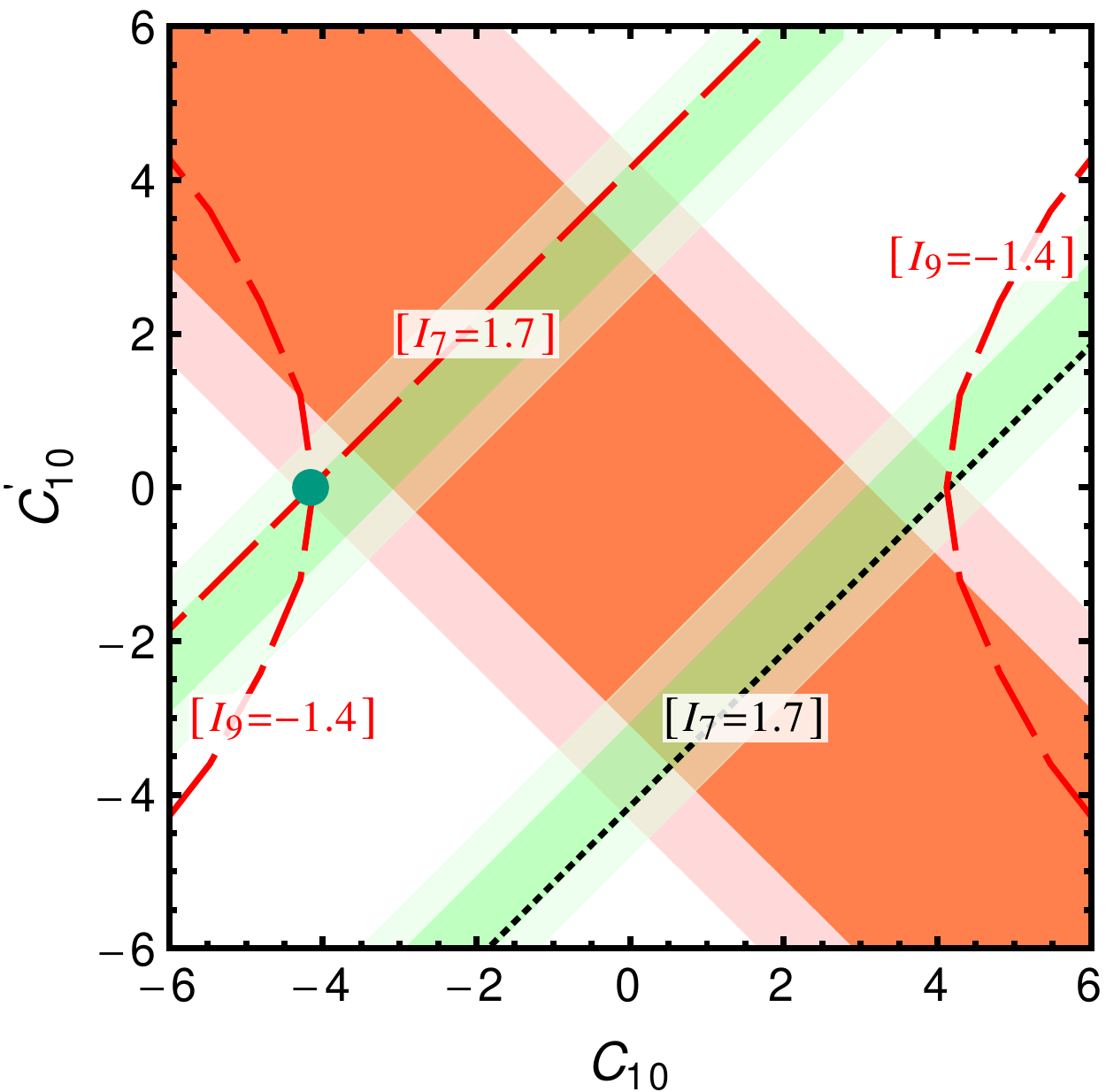}\qquad
\includegraphics[width=0.3\textwidth]{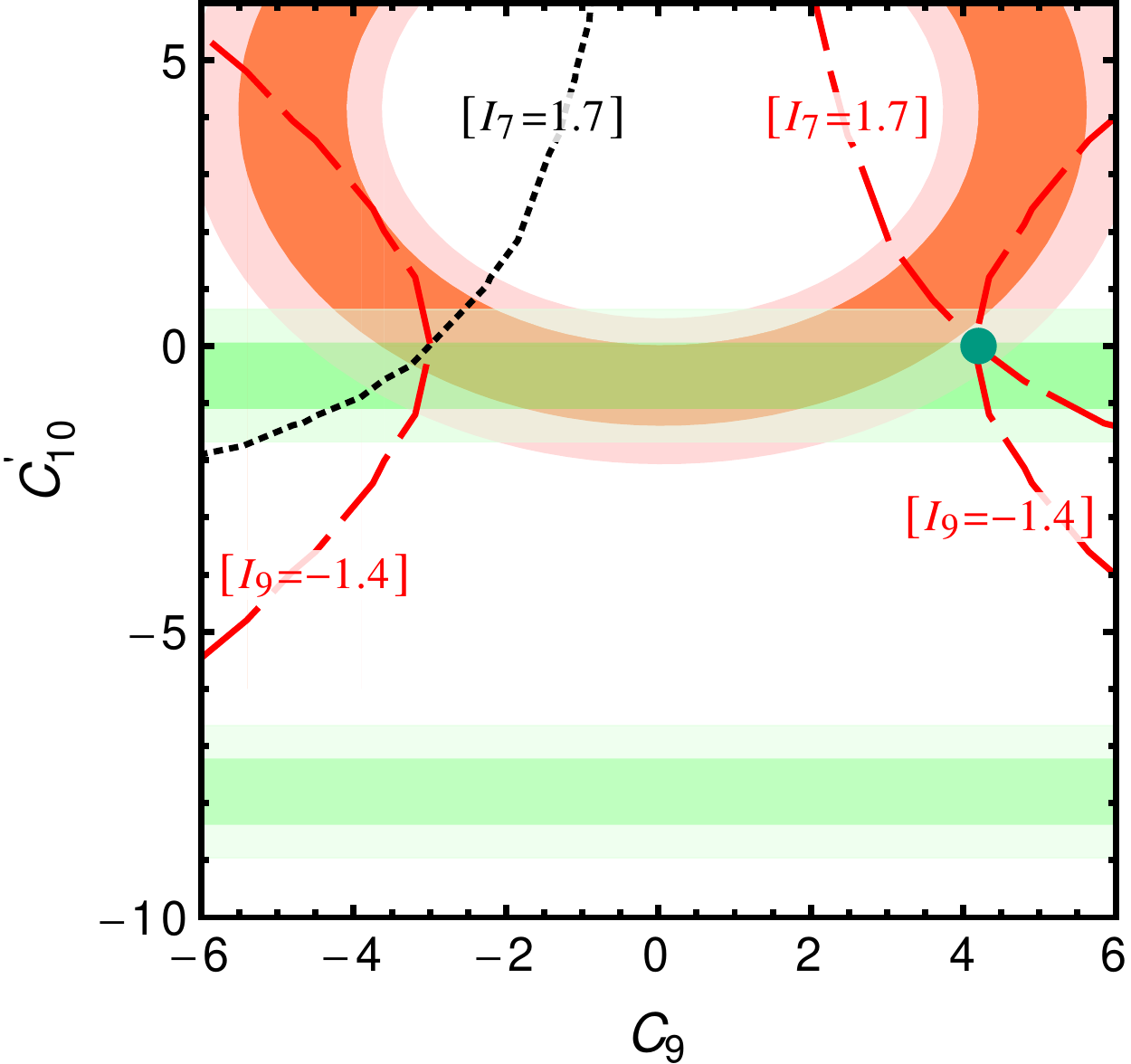}\qquad
\includegraphics[width=0.3\textwidth]{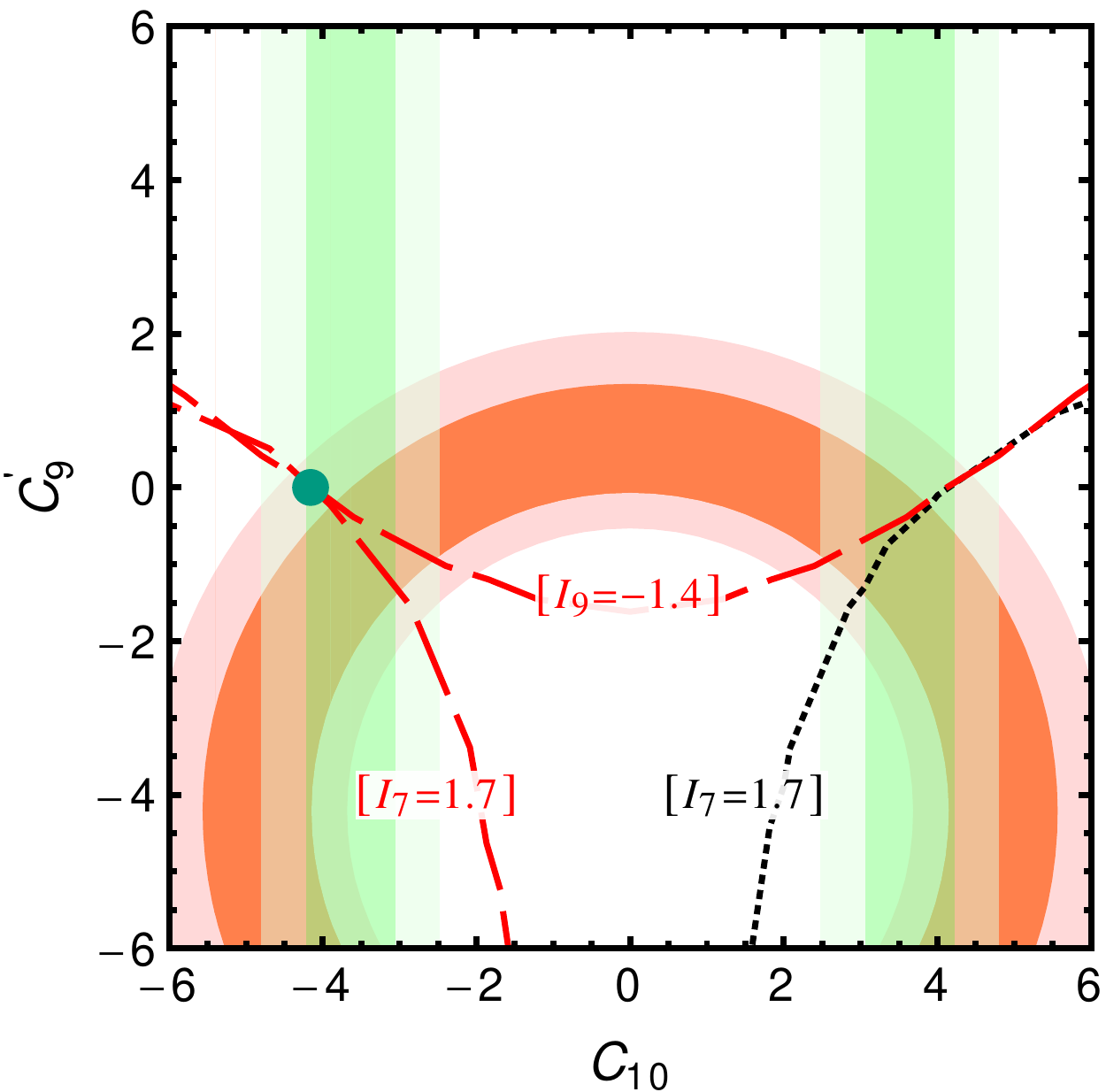}
}
\caption{\small Contours from hypothetical measurements for the high-$q^2$-integrated observables $\hat I_{7,9}/\Gamma(B^0)$ (in units of
$10^{-9}$) for central values of input versus constraints from present measurements for $\mathcal B(B_s\to\mu^+\mu^-)$~\cite{CMS:2014xfa} (green)
and $\mathcal B(B\to K\mu^+\mu^-)|_{q^2\in [1.1,6]{\rm GeV^2}}$~\cite{Aaij:2014pli} (orange) in different $C-C^\prime$ planes (Wilson coefficients that are not shown assume SM values). The
darker and lighter areas correspond to $1\sigma$ and $2\sigma$, respectively. 
The contours correspond to conic sections given by the bilinears in Eq.~\eqref{eq:effW}. The red dashed (black dotted) curve corresponds to
$\delta_{K^*}=0\, (\pi)$. The green blob corresponds to the SM.
\label{fig:contours}}
\end{figure}

In Fig.~\ref{fig:contours}  we show the resulting contours from a hypothetical SM-like measurement of high-$q^2$-integrated $\hat
I_{7,9}/\Gamma(B^0)$ in comparison with those from other observables. They demonstrate the complementarity with other observables as well as the need
to get contraints on the strong phase: without knowledge of the latter, the whole area between $I_{7,9}$-contours for $\delta_{K^*}=0,\pi$ remains
viable. Nevertheless, even without this knowledge measurements will allow to exclude a significant part of the parameter space. Of course the
determination of these parameters eventually requires a global fit to all $|\Delta B|=|\Delta S|=1$ processes with the strong phase (at low recoil)
as additional parameter.

\section{Comparison of $\bar B \to \bar K^* \mu \mu$ data with SM predictions\label{sec:data}}

In Table \ref{tab:compare} we compare our SM predictions including finite width effects to the recent LHCb data on $\bar B \to \bar K^* \mu \mu$
angular observables~\cite{LHCb:2015dla}. In the latter study S-wave backgrounds have been considered by including S-wave observables explicitly as
nuisance parameters.
We give therefore additional values in parentheses with the S-wave component, present only in ${\cal{F}}_0$, removed by replacing $F_0 \to F_0-
\int d \cos \theta_{K} F_0/2$, \emph{cf.} Eq.~\eqref{eq:Fi}. We use the most reliable bins in the low recoil region: the one with the largest
$q^2$-interval to allow for maximal smearing, $15 < q^2 < 19 \, \mbox{GeV}^2$, and the one closest to the endpoint, with highest momentum transfer and
furthest away from the $c \bar c$-threshold, $17 < q^2 < 19 \, \mbox{GeV}^2$. Note the different conventions between the $S_i$ used in this work, Eq.~(\ref{eq:Si}), and LHCb~\cite{LHCb:2015dla}. For the SM predictions, as usual numerator and denominator are $q^2$-integrated before dividing them. In the long run this
may not be necessary, as it may be feasible to extract amplitudes without binning \cite{Egede:2015kha}.
\begin{table}[ht]
\centering{
\begin{tabular}{c|c|c||c|c|c}
         &    LHCb$\[15,19\]^{a,b} $ & SM$\[15,19\]$ &   LHCb$
\[17,19\] ^{a,b} $ & SM$\[17,19\]$ & endpoint  \\\hline
         $F_L$ & $0.344 \pm 0.031$ & $0.351(0.342)\pm 0.010 \pm 0.003$ &
$0.354
\pm 0.054$ & $0.338(0.333)\pm 0.006 \pm 0.002$ & $1/3$  \\
  $A_{\rm FB}=S_6$ &  $-0.355 \pm 0.029$ & $-0.391(-0.396)\pm 0.016 \pm
0.005$ &
$-0.306 \pm 0.049$ &  $-0.349(-0.351) \pm 0.015 \pm 0.007$& $0^c$ \\
                    $S_3$ & $-0.122 \pm 0.026$ & $-0.129(-0.131)\pm 0.009
\pm
0.007$ & $-0.145 \pm 0.062$ & $-0.167(-0.169) \pm 0.007 \pm 0.005$ & $-1/4$\\
           $S_4$ & $0.214 \pm 0.029$ & $0.215(0.218)\pm 0.005 \pm 0.002$
& $0.202
\pm 0.052$ &  $0.226(0.227) \pm 0.003 \pm 0.002$ &  $+1/4$\\
            $S_5$ & $-0.244 \pm 0.029$ & $-0.230(-0.233)\pm 0.009 \pm
0.006$ &
$-0.245 \pm 0.050$&  $-0.191(-0.193) \pm 0.008 \pm 0.006$& $0^c$\\
             $S_5/S_6$ & $0.687 \pm 0.093^d$ & $0.588(0.591)\pm 0.008\pm
0.009$ &
$0.800 \pm 0.195^d $& $0.548(0.550) \pm 0.004 \pm 0.005$& $1/2$\\
\end{tabular}
}
\caption{
Angular observables as measured by LHCb  \cite{LHCb:2015dla} for $15 < q^2 < 19 \, \mbox{GeV}^2$ and $17 < q^2 < 19 \, \mbox{GeV}^2$, and SM
predictions including $\bar K \pi$-interference using $\bar K^*$ finite width and with P-signal cut (\ref{eq:cuts}). The first uncertainty in the
SM predictions corresponds to the one from form factors and parametric input whereas the second one is due to the strong phase varied within
$[-\pi,\pi]$; for the values in parantheses the S-wave contributions have been subtracted. 
Endpoint values \cite{Hiller:2013cza} refer to $\bar B \to \bar K^* \ell \ell$ decays, $q^2_{\rm end}=19.2 \, \mbox{GeV}^2$.
$^a$Uncertainties added in quadrature and symmetrized. 
$^b$Value adopted to the definitions used in this work, see Eq.~(\ref{eq:Si}).
$^c$The observable is proportional to the transverse perpendicular amplitude, which goes with a non-negligible slope to zero.
$^d$Correlations  included.
 \label{tab:compare}}
\end{table}
In both bins the data are in good agreement  with the SM predictions. The largest deviations are in
$S_5/S_6$ and $S_5$ in the $17-19~{\rm GeV^2}$ bin, at around 1.3 $\sigma$ and 1.1 $\sigma$.
Both $F_L$ and the branching ratios which appear in the denominators of the $S_i$, drop mildly when S-wave contributions are removed. The
corresponding SM branching ratios read ${\cal{B}}[17,19]=[0.71 (0.70) \pm 0.195 \pm 0.10] \cdot 10^{-7}$ and ${\cal{B}}[15,19]=[1.83(1.81) \pm 0.50
\pm 0.25] \cdot 10^{-7}$. The agreement with the SM is similarly good in both cases.

The good agreement in the observables which can be used to extract form factor ratios ($F_L, S_3, S_4$) \cite{Hambrock:2013zya} suggests
that within present uncertainties the low $q^2$ OPE appears to work, specifically that the binning is sufficient and non-universal $\bar c
c$-effects are sufficiently small. Note that the ``$P_4^\prime$-anomaly'', related to $S_4$, which was present in LHCbs 1$\rm{fb}^{-1}$ data
\cite{Aaij:2013qta} is gone, as also the agreement in the next-to endpoint bin is good, $S_4^{\rm SM}[15,17]=0.208(0.212) \pm 0.005 \pm 0.003$ versus
$S_4^{LHCb}[15,17]=0.250 \pm 0.049$. Comparing this observable to its value in bins involving higher $q^2$ values, we confirm that the S-wave
component is larger further away from the zero recoil endpoint \cite{Das:2014sra}.

The agreement with the zero recoil predictions is very good for $F_L$ and $S_4$, within $1 \sigma$, and good for the ratio $S_5/S_6$, with $1.5
\sigma$, followed by  $S_3$, with $1.7 \sigma$. As the endpoint relations are based on Lorentz invariance a discrepancy could
indicate a statistical fluctuation or unaccounted backgrounds. The measured central value is too large for $S_5/S_6$ and too small for $|S_3|$.
Both appear to favor $\delta_{K^*} \sim \pi/2$, a choice that also reduces the branching ratio, see Fig.~\ref{fig:br}. This could bring the data closer to SM
predictions with  lattice form factors \cite{Horgan:2013pva}, result in a suppression of $I_7$ and, further assuming negligible CP violation,
$I_{8.9}$ as well.
In addition to the branching ratio data from LHCb with $3~\rm{fb}^{-1}$, data for a smaller bin at the endpoint
could shed further light on this.

As has been pointed out in Ref.~\cite{Hiller:2013cza}, the slope towards zero recoil in $S_5$ and $S_6$ is a probe of BSM physics.
We note here that the slopes are essentially unaffected by interference, see Fig.~\ref{fig:S345}.

\section{Conclusions \label{sec:conclusions}}

We present a model-independent analysis of the impact of $\bar{B}\to\bar{K}\pi\ell\ell$ backgrounds
on the various observables of the benchmark mode $\bar{B}\to\bar{K}^* (\to \bar K \pi) \ell\ell$ at low hadronic recoil, taking into account the $\bar
K^*$ at finite width. Depending on the relative strong phase between the $K^*$ and the non-resonant contribution,
the differential branching ratio receives $\pm 14\%$ corrections in the $\bar K^*$-signal window. The effect of the interfering background is less
significant for several ratios of observables; the remaining uncertainties from the strong phase induced in $F_L, A_{\text{FB}}, S_{4,5}$ are of
the order of  a few percent. $S_3$ benefits less from cancellations and  receives uncertainties of $14\%$. In addition, noticable shifts of $5\%$
towards smaller values exist in $S_4$  and of $6\%$ towards larger values in $F_L$, see Figs.~\ref{fig:FL} and \ref{fig:S345}. Backgrounds to the
SM nulltests $S_{7,8,9}$ arise again at the percent level; larger values remain indications for new physics. In these ratios
sizable uncertainties from the strong phases persist.
Turning this around, the sensitivity of certain angular observables to strong phases, as shown in Fig.~\ref{fig:delP}, can be used to obtain phase information from
data. This method is independent of the underlying model as long as contributions from right-handed currents can be neglected.

Comparison to recent data on $\bar B \to \bar K^* \mu \mu$ angular observables \cite{LHCb:2015dla} in the low recoil region exhibits good
agreement, within $\lesssim (1-2) \,  \sigma$ from the SM expectations, see Table \ref{tab:compare}. Barring tuning, this suggests that within
uncertainties the low-recoil OPE works, specifically that the binning is sufficient and non-universal $\bar c c$-effects are sufficiently small. Data
on the $q^2$-distributions with finer binning as in $B \to K \mu \mu$ \cite{Aaij:2013pta} could shed further light on this matter.
While the agreement with the zero recoil predictions is good, the values for $S_3$ and $S_5/S_6$ slightly hint at a value for the strong phase
around $\delta_{K^*}\approx\pi/2$, which could also improve the consistency between the SM predictions and data for the branching ratio. 

It is clear that interference effects become of  importance for  future high precision studies. It is also evident  that there are sizable
uncertainties to the estimates presented in this work. Our study can be improved in several ways, mainly by including more precise  $\bar
B \to \bar K \pi$ form factors.
This should go in parallel with the experiments, as there is considerable feedback from data expected \cite{Hambrock:2013zya}. One should
also consider the strong phase as a parameter in the $|\Delta B|=|\Delta S|=1$ global fits.

Several features discussed in this work are not limited to the low recoil region: The generic size of interference is order $1/(4 \pi)$, and the different dependence
on the strong phase of ${\rm Re}$-type observables, $d \Gamma /d q^2$ and $I_{3,4,5,6}$, and ${\rm Im}$-type observables, $I_{7,8,9}$, with large net
interference effects in ratios between observables from the different sectors and a reduction of sensitivity for ratios within the same
sectors.
Another generic point is that the non-resonant interference could be probed by comparing  $\bar B \to \bar K^* \ell \ell$  to $\bar B \to \bar K \ell
\ell$ decays, as in the latter the interference is absent. Similarly, interference effects are suppressed in $\bar B_s \to \phi \ell \ell$ due to the
$\phi$'s narrow width~\cite{Das:2014sra}. Agreement of the fits in the individual sectors would support that interference effects are  not maximal,
constraining the strong phases.

Testing the SM with $|\Delta B|=|\Delta S|=1$ processes  has become a precision program and requires global fits. Here, investigations of sub-sectors
such as large versus low recoil data or exclusive versus inclusive modes provide ways to check for systematic uncertainties in theory
and experiment \cite{Bobeth:2010wg}.
Our  analysis shows that  presently $\bar B \to \bar K^* \mu \mu $ decays at low recoil are in agreement with the SM.

\begin{acknowledgments}

GH gratefully acknowledges the hospitality of the LPT Orsay group, where parts of this work have been carried out. We would like to thank Niklas
Bonacker, Dennis Loose, Ismo Toijala and Danny van Dyk for useful comments about EOS.
This work is supported in part by the DFG Research Unit FOR 1873 ``Quark Flavor Physics and Effective Field Theories'', the ERC Advanced Grant
project ``FLAVOUR'' (267104) and the DFG cluster of excellence ``Origin and Structure of the Universe''.

\end{acknowledgments}

%
\appendix

\section{ The $\bar B \to \bar K \pi \ell \ell$ angular distribution \label{sec:primer}}

The $\bar B \to \bar K \pi \ell \ell$ angular distribution, with the angles $\theta_\ell, \theta_K, \phi$ defined as in \cite{Bobeth:2008ij},
can be written as
  \begin{eqnarray}\label{eq:d5Gamma}
d^5\Gamma 
&=&\frac{1}{ 2  \pi} \left[ \sum c_i(\theta_\ell,\phi) I_i \left(q^2,p^2,\cos \theta_K\right) \right] dq^2dp^2d\cos\theta_Kd\cos\theta_\ell d\phi\,,
\end{eqnarray}
where 
\begin{align}
c_1 & =1, \quad c_2=\cos 2\theta_\ell, \quad c_3=\sin^2\theta_\ell\cos 2\phi, \quad c_4=\sin 2\theta_\ell \cos \phi, \quad c_5=\sin\theta_\ell\cos\phi, \nonumber \\ c_6& =\cos\theta_\ell, \quad c_7=\sin\theta_\ell\sin\phi, \quad c_8=\sin 2\theta_\ell\sin\phi, \quad c_9=\sin^2\theta_\ell\sin2\phi \, .
\label{eq:ci}
\end{align}
At leading order in the low recoil OPE, the angular coefficients $I_i$ factorize into form factors and short-distance coefficients: 
\begin{align} \nonumber
I_1 & = \phantom{-}\frac{1}{8} \left[   |{\cal F}_0 |^2 \rho_1^- +\frac{3}{2} \sin^2 \theta_K  \{  |{\cal F}_\parallel |^2 \rho_1^-+  |{\cal F}_\perp |^2 \rho_1^+\} \right]  \, ,  \\ \nonumber
I_2 & =- \frac{1}{8} \left[   |{\cal F}_0 |^2 \rho_1^- -\frac{1}{2} \sin^2 \theta_K  \{  |{\cal F}_\parallel |^2 \rho_1^-+  |{\cal F}_\perp |^2 \rho_1^+\} \right]   \, ,  \\ \nonumber
I_3 & = \phantom{-}\frac{1}{8}    \left[ |{\cal F}_\perp |^2 \rho_1^+ -  |{\cal F}_\parallel |^2 \rho_1^-  \right]  \sin^2 \theta_K  \, ,  \\ \nonumber
I_4 &= - \frac{1}{4} {\rm Re}({\cal F}_0 {\cal F}_\parallel^*) \,\rho_1^-  \sin \theta_K  \, ,  \\
\label{eq:Iope}
I_5 &=  \phantom{-}\left[{\rm Re}({\cal F}_0 {\cal F}_\perp^*)  {\rm Re} \rho_2^++ {\rm Im} ({\cal F}_0 {\cal F}_\perp^*) 
{\rm Im} \rho_2^-  \right] \sin \theta_K  \, ,  \\ \nonumber
I_6&= - \left[{\rm Re}({\cal F}_\parallel {\cal F}_\perp^*) {\rm Re} \rho_2^+ + {\rm Im} ({\cal F}_\parallel {\cal F}_\perp^*) {\rm Im} \rho_2^-  \right]\sin^2 \theta_K  \, ,   \\ \nonumber
I_7&=   {\rm Im} ({\cal F}_0 {\cal F}_\parallel^*)\,\delta \rho\,  \sin \theta_K \, ,   \\ \nonumber
I_8&=  \frac{1}{2} \left[ {\rm Re}({\cal F}_0 {\cal F}_\perp^*) {\rm Im} \rho_2^+  -  {\rm Im} ({\cal F}_0 {\cal F}_\perp^*) {\rm Re} \rho_2^-  \right] \sin \theta_K \, ,   \\ \nonumber
I_9&=  \frac{1}{2} \left[{\rm Re}({\cal F}_\perp {\cal F}_\parallel^*) {\rm Im} \rho_2^+ + {\rm Im} ({\cal F}_\perp {\cal F}_\parallel^*) {\rm Re} \rho_2^- \right]\sin^2 \theta_K   \, ,  \nonumber
\end{align}
where the short-distance coefficients read
\begin{align}
\rho_1^\pm & = \left| C_9^{\rm eff} \pm  C_9^\prime + \kappa  \frac{2 m_b m_B}{q^2  }
(C_7^{\rm eff} \pm  C_7^\prime)\right|^2 + |C_{10}  \pm  C_{10}^\prime|^2 \, , \nn\\
\delta \rho & = {\rm Re}\left[ \left(C_9^{\rm eff} -  C_9^\prime + \kappa  \frac{2 m_b m_B}{q^2  } 
(C_7^{\rm eff} -  C_7^\prime) \right)\left(C_{10}  -  C_{10}^\prime\right)^* \right]  \, , \nn\\
\label{eq:effW}
{\rm Re} \rho_2^+ & ={\rm Re} \left[\left( C_9^{\rm eff} + \kappa  \frac{2 m_b m_B}{q^2  }
C_7^{\rm eff} \right)C_{10}^*  - \left(C_9^\prime + \kappa  \frac{2 m_b m_B}{q^2  }  C_7^\prime\right)C_{10}^{\prime *} \right]  \, , \\
{\rm Im} \rho_2^+ & = {\rm Im}\left[ C_{10}^{\prime} C_{10}^* + 
\left(C_{9}^{\prime}+\kappa  \frac{2 m_b m_B}{q^2} C_7^{\prime}\right)\left(C_9^{\rm eff} + \kappa  \frac{2 m_b m_B}{q^2  } C_7^{\rm eff}\right)^{\! *}
\right]  \, , \nn\\
{\rm Re} \rho_2^- & = \frac{1}{2} \left[  |C_{10}|^2  -| C_{10}^\prime|^2+ \left|C_{9}^{\rm eff}+\kappa  \frac{2 m_b m_B}{q^2  }C_{7}^{\rm eff}\right|^2  -\left| C_{9}^\prime+\kappa  \frac{2 m_b m_B}{q^2  }C_7^{\prime}\right|^2 
\right]  
\, , \nn\\
{\rm Im} \rho_2^- & = {\rm Im} \left[C_{10}^\prime  \left(C_9^{\rm eff}  + \kappa  \frac{2 m_b m_B}{q^2  } 
C_7^{\rm eff} \right)^* - C_{10}  \left(C_9^\prime + \kappa  \frac{2 m_b m_B}{q^2  }  C_7^\prime\right)^* \right]  \,\nn ,
\end{align}
 and the generalized
transversity form factors are given in Eq.~(\ref{eq:fullformfactor}). The Wilson  coefficients $C_{7,9,10}^{(\prime)}$ correspond to the
low energy Hamiltonian
${\cal{H}}_{\rm eff}=-4 G_F/\sqrt{2} \, V_{tb} V_{ts}^* \, \alpha_e/(4 \pi) \sum ( C_i  {\cal{O}}_{i}+ C_i^\prime  {\cal{O}}_{i}^\prime)$, 
where
\begin{align} \nonumber
  {\cal{O}}_{7} & = \frac{m_b}{e} \bar{s} \sigma^{\mu\nu} P_{R} b  F_{\mu\nu}\,, \quad 
   &{\cal{O}}_{7}^\prime &= \frac{m_b}{e} \bar{s} \sigma^{\mu\nu} P_{L} b  F_{\mu\nu}\,,
\\ \label{eq:7910}
  {\cal{O}}_{9} & =  \bar{s} \gamma_\mu P_{L} b \, \bar{\ell} \gamma^\mu \ell \,, \quad
   &{\cal{O}}_{9}^\prime  &=  \bar{s} \gamma_\mu P_{R} b \, \bar{\ell} \gamma^\mu \ell \,,
\\
  {\cal{O}}_{10} & = \bar{s} \gamma_\mu P_{L} b \, \bar{\ell} \gamma^\mu \gamma_5 \ell \,,
  \quad  &{\cal{O}}_{10}^\prime  &=  \bar{s} \gamma_\mu P_R b \, \bar{\ell} \gamma^\mu \gamma_5\ell \,.
\nonumber
\end{align}
The effective coefficients $C_{7,9}^{\rm eff}$ equal $C_{7,9}$ up to contributions from 4-quark operators.
In our analysis we neglect the mass of the leptons and the strange quark.
 
Assuming only operators already present in the SM, which we term "SM basis", corresponds to no right-handed currents, $C_{7,9,10}^\prime=0$. In this
limit Eq.~(\ref{eq:Iope}) simplifies  to 
\begin{align} \nonumber
I_1 & = \phantom{-}\frac{1}{8}  \rho_1 \left[   |{\cal F}_0 |^2  +\frac{3}{2} \sin^2 \theta_K  \{  |{\cal F}_\parallel |^2 +  |{\cal F}_\perp |^2 \} \right]  \, , \\ \nonumber
I_2 & =- \frac{1}{8}  \rho_1 \left[   |{\cal F}_0 |^2  -\frac{1}{2} \sin^2 \theta_K  \{  |{\cal F}_\parallel |^2 +  |{\cal F}_\perp |^2 \} \right] \, , \\ \nonumber
I_3 & = \phantom{-}\frac{1}{8}  \rho_1   \left[ |{\cal F}_\perp |^2  -  |{\cal F}_\parallel |^2   \right]   \sin^2 \theta_K \, ,\\ \nonumber
I_4 &= - \frac{1}{4}  \rho_1\, {\rm Re}({\cal F}_0 {\cal F}_\parallel^*)  \sin \theta_K \, , \\
\label{eq:IopeSM}
I_5 &=  \phantom{-}\rho_2\, {\rm Re}({\cal F}_0 {\cal F}_\perp^*)  \sin \theta_K \, , \hspace{4cm} (\mbox{SM basis})\\ \nonumber
I_6&= -   \rho_2 \,{\rm Re}({\cal F}_\parallel {\cal F}_\perp^*) \sin^2 \theta_K \, , \\ \nonumber
I_7&=   \phantom{-}\rho_2 \,{\rm Im}({\cal F}_0 {\cal F}_\parallel^*)  \sin \theta_K \, ,\\ \nonumber
I_8&= - \frac{1}{4} \rho_1\, {\rm Im}({\cal F}_0 {\cal F}_\perp^*)  \sin \theta_K \, ,\\ \nonumber
I_9&=  \phantom{-}\frac{1}{4} \rho_1\, {\rm Im} ({\cal F}_\perp {\cal F}_\parallel^*) \sin^2 \theta_K  \,, \nonumber
\end{align}
where
\begin{align} \label{eq:SMrho}
\rho_1 \equiv \rho_1^\pm=2 {\rm Re} \rho_2^- \, , \quad \rho_2 \equiv {\rm Re} \rho_2^+ =\delta \rho \, , \quad
{\rm Im} \rho_2^\pm=0\, . \hspace{1cm} (\mbox{SM basis})
\end{align}

\section{The $\bar{B}\to\bar{K}^*\ell\ell$ form factors} 
\label{app:B2Kst}
The $\bar{B}\to\bar{K}^*$ vector transversity form factors are defined as \cite{Bobeth:2010wg} 
\begin{equation}
\begin{aligned}
  f_{\perp} (q^2)& ={\cal N }_{K^*} \frac{\sqrt{2\, \lambda_{K^*}}}{m_B + m_{K^*}} V(q^2)\,, 
\\
  f_{\parallel} (q^2) & = {\cal N}_{K^*} \sqrt{2}\, (m_B + m_{K^*})\, A_1(q^2)\,,
\\
  f_{0} (q^2)& ={\cal N}_{K^*} \frac{(m_B^2 - m_{K^*}^2 - q^2) (m_B + m_{K^*})^2 A_1(q^2)
   - \lambda_{K^*}\, A_2(q^2)}{2\, m_{K^*} (m_B + m_{K^*}) \sqrt{q^2}}\,, 
\end{aligned}
\label{eq:Kstff}
\end{equation}
where $\lambda_{K^*} \equiv \lambda(m_B^2, m_{K^*}^2, q^2)$, and the normalization factor is
\begin{align}
  \label{eq:NKstar}
  {\cal N}_{K^*} & = G_F   V_{tb}^{}V_{ts}^{*} \alpha_e\,
    \sqrt{\frac{q^2  \sqrt{\lambda_{K^*}}}{3 (4 \pi)^5\, m_B^3}}\,.
\end{align}
We follow Ref.~\cite{Das:2014sra} for the numerical values of $V,A_{1,2}$. Specifically,
the form factors are taken from \cite{Ball:2004rg} as compiled in  \cite{Bobeth:2010wg}, and we
employ an uncertainty estimate for the ratios $V/A_1$ of $8\%$ and $A_2/A_1$ of $10\%$ from \cite{Hambrock:2013zya}.

\section{The $\bar{B}\to\bar{K}\pi\ell\ell$ form factors} 
\label{app:B2Kpi}
The $\bar{B}\to\bar{K}\pi$ transversity form factors read\
\begin{align} \nonumber 
F_0 &= \frac{{\cal N}_{nr}}{2}  \bigg[  \lambda^{1/2 }w_+(q^2,p^2,\cos \theta_K)+\frac{1}{p^2}\{(m_K^2-m_\pi^2)\lambda^{1/2}-(m_B^2-q^2-p^2) \lambda^{1/2}_{p}\cos \theta_K\} w_-(q^2,p^2,\cos \theta_K) \bigg]\,,\\
\label{eq:Fi}
F_\parallel &= {\cal N}_{nr}  \sqrt{ \lambda_p \frac{q^2}{p^2}} \, w_-(q^2,p^2,\cos \theta_K)\,, \qquad
F_\perp = \frac{{\cal N}_{nr}}{2}\sqrt{ \lambda \lambda_p \frac{  q^2}{p^2}} \, h(q^2,p^2,\cos \theta_K)\,,
\end{align} 
where ${\cal N}_{nr}$ is a normalization factor  \cite{Das:2014sra}.
The HH$\chi$PT expressions of the form factors $w_\pm$ and $h$ to the lowest order in $1/m_b$ are given as
\begin{align} \nonumber
w_\pm &= \pm \frac{gf_B}{2 f^2} \frac{m_B}{v\cdot p_\pi+\Delta} \, ,\\
h &= \frac{g^2 f_B}{2f^2} \frac{1}{ 
[v\cdot p_\pi+\Delta][v\cdot p+\Delta+\mu_s]} \, , \label{eq:ffinput}
\end{align}
where $v=p_B/m_B$,
$\Delta=m_{B^*}-m_B=46$ MeV and $\mu_s= m_{B_s}-m_B=87.3$ MeV \cite{PDG}.
Here, $g$ is the HH$\chi$PT coupling constant and $f_B$ is the decay constant in the $SU(3)$ limit
assumed in this work.  We further use $f^2=f_\pi f_K$. The values of these parameters used in our numerical analysis are given in Table~\ref{tab:input}. 
\begin{table}[ht]
\centering{
\begin{tabular}{c|c|c}
Parameter           &     Value & Source \\\hline
$|V_{ts}^* V_{tb}|$ & $0.0407 \pm 0.0011$ &  \cite{Charles:2004jd}\\
$\Gamma(B_0)$ & $(4.333 \pm 0.020) \cdot 10^{-13}$ GeV & \cite{PDG} \\
$f_\pi $ & $130.4 \pm 0.2$ MeV & \cite{PDG} \\
$f_K $ & $156.2 \pm 0.7$ MeV &  \cite{PDG}$^\dagger$ \\
$f_{B_d} $ & $188 \pm 4$ MeV &  \cite{Dowdall:2013tga}\\
$g$ & $0.569 \pm 0.076$ & \cite{Flynn:2013kwa}$^\dagger$ \\
$r_{BW}$ & $2.1 \pm 0.7$ GeV$^{-1}$ &  \cite{delAmoSanchez:2010fd}$^\dagger$ 
\end{tabular}
}
\caption{\label{tab:input}
Numerical input used in this work. $\Gamma(B_{0})$ denotes the mean total width. $^\dagger$Uncertainties added in quadrature.}
\end{table}

\section{Generic  finite width considerations \label{app:epsilon}}
Using
\begin{align}
\pi \delta(x)=\lim_{\epsilon \to 0} \frac{\epsilon}{x^2+\epsilon^2}
\end{align}
implies for the zero width approximation $\Gamma_{K_J} \to 0$:
\begin{align}
\pi \delta (p^2 -m_{K_J}^2) = \frac{m_{K_J} \Gamma_{K_J}}{(p^2-m_{K_J}^2)^2 +m_{K_J}^2 \Gamma_{K_J}^2} \, .
\end{align}
We consider the BW lineshape at amplitude level,
\begin{align}
\frac{\sqrt{\epsilon}}{x +i \epsilon}  =\frac{ \sqrt{\epsilon} x }{x^2+\epsilon^2}
-i   \frac{ \sqrt{\epsilon}^3  }{x^2+\epsilon^2} \, .
\end{align}
The limit $\epsilon \to 0$ does not exist: 
\begin{align}
\lim_{\epsilon \to 0} \frac{\sqrt{\epsilon}}{x +i \epsilon}  =
-i  \,  \lim_{\epsilon \to 0} \frac{ \sqrt{\epsilon}^3  }{x^2+\epsilon^2} \, .
\end{align} 
The real part vanishes; to show this investigate $x \neq 0$ and $x=0$.
The imaginary part vanishes for $x \neq 0$, too. For $x=0$ it diverges as $1/\sqrt{\epsilon}$, but it does not yield the delta distribution,
because the integral vanishes as $\sqrt{\epsilon}$ for $\epsilon \to 0$.

To discuss interference with the BW amplitude with finite width, let $\delta_{K^*}$  be the relative phase%
\footnote{The imaginary  part makes a "phase rotation" in $\delta_{K^*}$ between these and our previous works \cite{Das:2014sra}.}
\begin{align}
\frac{\sqrt{\epsilon}}{x +i \epsilon}  \exp[i \delta_{K^*}]=\frac{ \sqrt{\epsilon}  }{x^2+\epsilon^2} \[
x \cos \delta_{K^*} + \epsilon \sin \delta_{K^*} + i (-\epsilon \cos\delta_{K^*} + x \sin \delta_{K^*}) \] \, .
\end{align}
For Re-type observables  we expect the following dependence on the strong phase:
\begin{align}
\frac{\sqrt{\epsilon}}{x} \cos \delta_{K^*} & \quad \mbox{for} \quad  |x| \gg \epsilon \quad (\mbox{outside signal window})\,, \nonumber \\
\frac{1}{\sqrt{\epsilon}}\sin \delta_{K^*} & \quad \mbox{for} \quad |x| \ll \epsilon \quad (\mbox{signal window})\,.
\end{align}
For Im-type observables, such as $I_{7,8,9}$ in the SM basis,  we expect the following dependence on the strong phase:
\begin{align}
\frac{\sqrt{\epsilon}}{x} \sin \delta_{K^*} & \quad \mbox{for} \quad  |x| \gg \epsilon \quad (\mbox{outside signal window})\,, \nonumber \\
- \frac{1}{\sqrt{\epsilon}} \cos \delta_{K^*} & \quad \mbox{for} \quad |x| \ll \epsilon \quad (\mbox{signal window})\,. \label{eq:imregions}
\end{align}
Note in both Re- and Im-type observables the dependence on ${\rm sign}(x)$ for $ |x| \gg \epsilon $, that is, a sign flip between below $(x<0)$ and above $(x>0)$ the $\bar K^*$-resonance.
Numerically, $\epsilon(\bar K^*) =0.05$ and  $\epsilon(\phi)=0.004$.

\end{document}